# Quantum Nanophotonics with Energetic Particles: X-rays and Free Electrons

Xihang Shi[1†], Wen Wei Lee[2†], Aviv Karnieli[3], Leon Merten Lohse[4], Alexey Gorlach[1], Lee Wei Wesley Wong[2], Tim Salditt[4], Shanhui Fan[3], Ido Kaminer[1], Liang Jie Wong[2]*

[1]*Solid State Institute and Faculty of Electrical and Computer Engineering, Technion – Israel Institute of Technology, Haifa 3200003, Israel*

[2]*School of Electrical and Electronic Engineering, Nanyang Technological University, 50 Nanyang Avenue, Singapore 639798, Singapore*

[3]*E. L. Ginzton Laboratories, Stanford University, 348 Via Pueblo, Stanford, CA USA*

[4]*Institute for X-ray Physics, University of Göttingen, Friedrich-Hund-Platz 1, 37077 Göttingen, Germany*

Corresponding Author: liangjie.wong@ntu.edu.sg

**Abstract**

Rapid progress in precision nanofabrication and atomic design over the past 50 years has ushered in a succession of transformative eras for moulding the generation and flow of light. The use of nanoscale and atomic features to design light sources and optical elements – encapsulated by the term nanophotonics – has led to new fundamental science and innovative technologies across the entire electromagnetic spectrum, with substantial emphasis on the microwave to visible regimes. In this review, we pay special attention to the impact and potential of nanophotonics in a relatively exotic yet technologically disruptive regime: the high-energy particles such as X-ray photons and free electrons – where nanostructures and atomic design open the doors to unprecedented technologies in quantum science and versatile X-ray sources and optics. As the practical generation of X-rays is intrinsically linked to the existence of energetic free or quasi-free-electrons, our review will also capture related phenomena and technologies that combine free electrons with nanophotonics, including free-electron-driven nanophotonics at other photon energies. In particular, we delve into the demonstration and study of quantum recoil in the X-ray regime, the study of nanomaterial design and free-electron waveshaping as means to enhance and control X-ray radiation, examine the free-electron generation enabled by nanophotonics, and analyze the high-harmonic generation by quasi-free-electrons. We also discuss applications of quantum nanophotonics for X-rays and free electrons, including nanostructure waveguides for X-rays, photon pair enhanced X-ray imaging, mirrors, lenses for X-rays, etc. Our review highlights the uniqueness of the X-ray regime and the enormous potential of quantum nanophotonics to revolutionize it through tailored interactions between photons, free electrons, and the bound electrons of atoms and nanostructures.

## 1 Introduction

Advancements in X-ray technologies have entered the quantum realm, driven by breakthroughs in free-electron-driven nanomaterial technologies that enable investigations into spatial-temporal dynamics down to sub-nanometre and attosecond scales [1]. Practically, the generation of X-rays necessitates the use of energetic free or quasi-free electrons, making the manipulation and control of free electrons themselves also germane to any discussion on X-ray innovation. Research on free-electron-driven light emission and the shaping of free electrons has advanced swiftly in recent decades, from nanoengineering at atomic scales [2–6] and attosecond electron pulses [7–13], to the precise control of free-electron-light interactions [14–26]. The recent surge in interdisciplinary research leveraging these advances has spurred the emergence of new fields, such as novel coherent compact light sources from near-infrared to X-rays [27–35] and free-electron quantum optics [14,15,36,16,17,37,18–21,38–46,22,47–58,24,59–65,26,66–70]. This review captures (1) emerging quantum and classical phenomena in X-ray generation and manipulation through the interaction between free electrons and nanomaterials; (2) developments in the shaping of free electrons, as well as the interaction between free electrons and nanomaterials, of relevance to the design of future X-ray as well as optical sources, and (3) existing and emerging applications of X-rays and free electrons enabled by nanomaterials and quantum science.

A major chapter in the modern history of photonics arguably began with the conception of photonic crystals by Eli Yablonovitch [71] and Sajeev John [72]. Their ideas leveraged nanofabrication to create nanoscale periodic structures, mimicking the periodicity of actual atoms in crystalline materials. These nanoscale periodic structures, and combinations of them, function effectively as artificially designed materials with unique capabilities in manipulating visible and longer wavelength light [73–77], as well as their interactions with free electrons [78,27,79,29,80–85] – giving rise to the field of nanophotonics. Rapid progress in the fabrication of well-defined structures with ever-higher resolutions has made nanophotonic concepts applicable to progressively smaller wavelengths, not surprisingly entering the X-ray regime. With the ability to structure materials at the atomic scale for the manipulation of correspondingly short wavelengths, the concept of photonic crystals has in a sense come full-circle – back to the periodic arrays of actual atoms that initially inspired them. Indeed, crystalline van der Waals materials and heterostructures have proven to be promising platforms for tunable X-ray emission, arising from the coherent interaction between free electrons and nanocrystals [28,31,86,32–34,87,88]. In addition, high-resolution and scalable nanoengineering at atomic scales [2–6,89] has facilitated the development of X-ray waveguides [90–97], lenses [98–100], and other X-ray optics [100,101]. The ability to manipulate light at longer wavelengths is also relevant to X-ray scintillators, which convert incident X-ray photons to visible photons. As such, conventional nanostructures have enormous potential to enhance the X-ray detection process, conferring benefits such as better spectral and directional control and improved temporal resolution [102–107]. Figure 1 presents an overview of free-electron-light interactions across various nanophotonic regimes.

The intimate connection between free electrons and X-rays can be appreciated from both classical and quantum perspectives. Classically, when light is described as a wave, free electrons act as sources of broad-spectrum electromagnetic radiation, which contain significant portions of the X-ray range. Processes such as bremsstrahlung, undulator radiation, and Smith-Purcell radiation couple these broad-spectrum, evanescent fields into propagating X-ray waves. From a quantum mechanical perspective, where light is described as particles (photons), the law of conservation of energy dictates that to produce an X-ray photon (or photon in any other electromagnetic spectrum), the particle involved must have kinetic energy at least equal to the photon's energy. Free electrons, along with quasi-free electrons (which are not fully free but behave similarly), are uniquely essential in this context because they have the distinction of being the most practical and widely accessible particle being accelerated to the high energies needed for X-ray generation.

Experiments to extract light from electrons must be designed to overcome the discrepancy between the relatively large momentum of free electrons compared to that of the emitted photons, at any given energy, and also to overcome the fact that intrinsic cross-sections can be relatively low for higher-order processes [108]. Nanostructures can boost the interaction cross-section by facilitating momentum matching through the smallness of their feature sizes [27–34,109,110], or through highly confined polaritons [28,109,111]. Nanomaterials also open up the quantum regime of free-electron-photon interactions [112]: for example, the quantum recoil phenomenon [113–115]–the shift in the emitted photon energy from classical predictions due to the particle nature of light. Quantum recoil can also manifest when soft X-ray photons and lower energies photons are emitted through interactions with nanomaterials [110,116,117].

The underlying physics beneath the quantum recoil phenomena is the entanglement between the radiating electrons and the emitted photons. The entanglement has also become the foundational principle of the field of free-electron quantum optics [14,15,36,16,17,37,18–21,38–46,22,47–58,24,59–65,26,66–70]. One important aspect of free-electron quantum optics is generating quantum light [54,50,58,55,65,56] and measuring the quantum features of materials with electrons [42,47,48,118,119]. The treatment of electrons as wavepackets (and not just particles) also implies that different eigenmodes of a single electron could interfere with each other [120]. Following this principle, engineered electron wavefunction [121,122], which could be achieved by electron holographic structure [123–125], has been theoretically proposed to enhance the single electron radiation and manipulate the radiation direction [66,69].

Advancements in nanomaterials over the past decade have unveiled novel facets of free-electron-photon interactions in the ultrashort pulse regime. One field that has consequently emerged is that of dielectric laser accelerators (DLAs) [83,126–139], in which a train of ultrashort electron bursts are synchronized with an incident laser pulse to achieve acceleration gradients on the order of GeV/m. Another area of active research is the generation of attosecond photoelectrons, driven by femtosecond light interacting with nanoscale metal emitters [8,13,140,141]. Furthermore, the quantum characteristics of light can be imparted to electrons during their interaction with coherent [14–16,22], squeezed [142], and Fock-state light [70]. As a result, energy comb electrons [9,17,24,143–145], heralded electrons [142,146,147], and non-Poisson distributed electrons have been developed [22,142,146,147]. It should be noted that the investigation of light generation from these novel electron sources has only begun in the domain of long-wavelength light and is pending exploration in the X-ray regime.

The arrangement of the review is as follows: Section 2 discusses recent works on quantum recoil and light emission enabled by nanomaterials. Section 3 presents tunable X-ray emission from shaped free electron wavefunctions. Section 4 examines the quantum electrodynamics processes studied during free electron interacting with nanomaterials, including attosecond electrons and electron correlations. Section 5 highlights high-harmonic generation as a promising compact ultrashort-wavelength source. Section 6 explores X-ray optics developed through nanotechnology. Finally, Section 7 addresses applications of X-rays and quantum electrons.

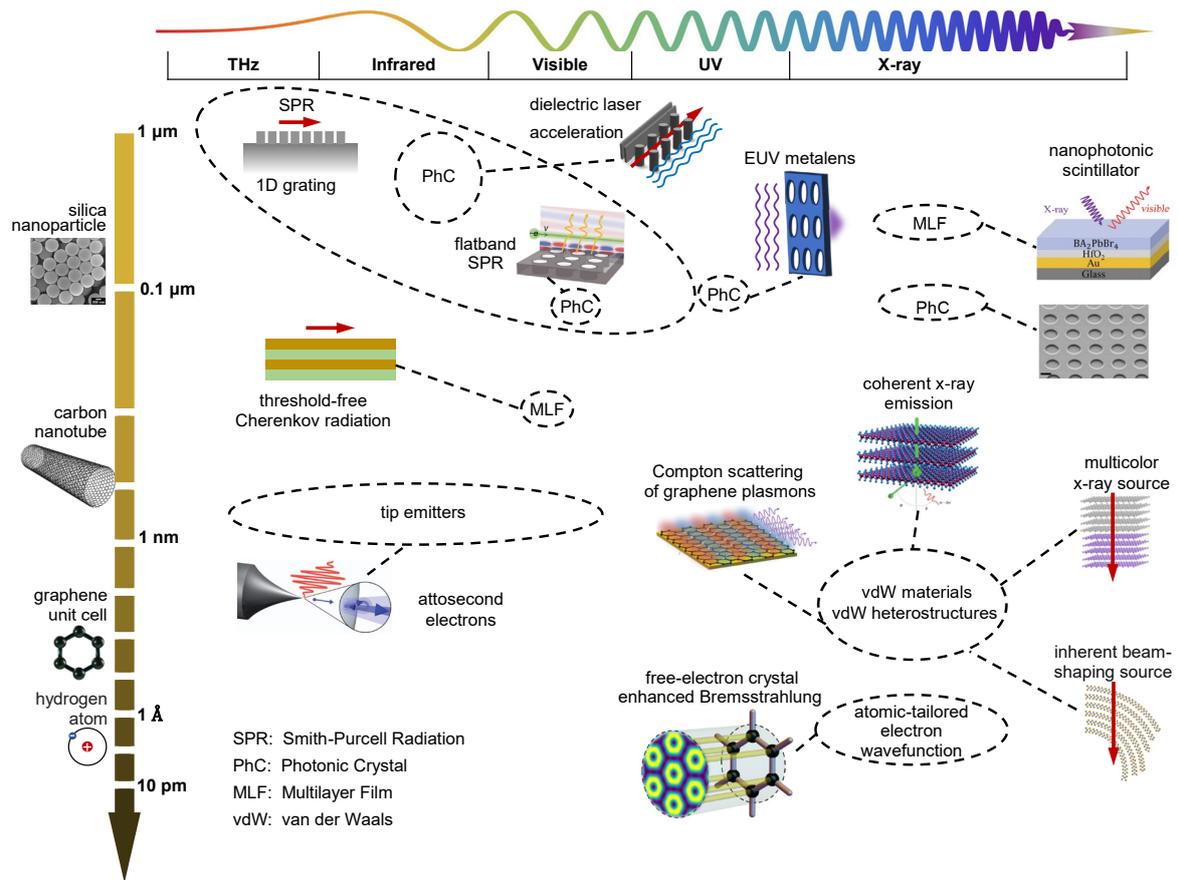

**Fig. 1 Overview of free-electron-light-matter interactions across various nanostructure periodicities and electromagnetic wavelengths.** Smith-Purcell radiation is emitted when free electrons travel through or in the vicinity of periodic structures. The radiation wavelength is tied to the structure's periodicity and the electron velocity, resulting in a versatile platform for generating light over a broad frequency range from the terahertz to ultraviolet regimes [30,148,149]. The emission rate could be further increased through emerging nanophotonic designs such as the photonic flatbands [85]. Free electrons can also gain net energy from light, such as through inverse Smith-Purcell radiation and dielectric laser accelerators [83,126–138]. Van der Waals materials are multilayer two-dimensional (2D) materials with atomic periodicities between 0.1 nm and 1 nm, making them pertinent platforms for generating both soft and hard tunable X-rays from free electrons. The wavelengths of the emitted X-rays can be tuned by adjusting the electron energies and the periodicities of van der Waals materials [28,31,32,69]. Van der Waals materials can be combined into heterostructures to realize customizable X-ray sources with further unique properties [87,33,34,88]. Short-wavelength optical components like extreme ultraviolet metalenses [150] have also been realized with nanomaterials. Nanophotonics can also be used to enhance and control the output of scintillators, which convert high-energy particles like X-rays to low-energy particles like visible photons [102–107]. In devices like tip emitters [8,141], nanophotonics enable us to enhance and control the conversion of energy from light to free electrons. Illustration for "flatband SPR" is reprinted with permission from Y. Yang *et al.*, Nature **613**, 42–47(2023) [85]. Copyright 2023 Springer Nature. Illustration for "attosecond electrons" is reprinted with permission from M. Krüger *et al.*, Nature **475**, 78–81(2011) [8]. Copyright 2011 Springer Nature. Illustration for "Compton scattering of graphene plasmons" is reprinted with permission from L.J. Wong *et al.*, Nat. Photonics **10**, 46–52(2016) [28]. Copyright 2016 Springer Nature. Illustration for "coherent X-ray emission" is reprinted with permission from

N. Talebi, Nat. Photonics **17**, 213–214(2023) [151]. Copyright 2023 Springer Nature. Illustration for "nanophotonic scintillator" is reprinted with permission from C. Roques-Carmes *et al.*, Science **375**, eabm9293 (2022) [104] and W. Ye *et al.*, Adv. Mater. **36**, 2309410 (2024) [107]. Copyright 2020 AAAS and 2024 John Wiley and Sons.

## 2 Quantum Phenomena in Free-Electron-Driven X-ray Generation from Nanomaterials

### 2.1 Quantum Recoil in Spontaneous Emission from Free Electrons

Quantum recoil refers to the shift in emitted photon energies compared to predictions made using the classical nonrecoil approximation for free electrons [113]. It is closely tied to the particle nature of light, which was recognized following the successful explanation of the photoelectric effect by Albert Einstein in 1905 [152]. The particle nature of light soon found its application in light-mater interaction in 1922 which is now known as Compton scattering [153,154]. Later, in 1940, Vitaly Ginzburg discovered in theory that the particle nature of light manifests in shifting the spectrum of free electron radiation, a phenomenon named as quantum recoil [113,114]. In 1954, Julian Schwinger investigated the electron radiation in an external field by incorporating the particle nature of light and obtained the quantum correction for radiation power [155–157].

In the following sections, we will review the quantum effects arising from the particle nature of photons in electron radiation, especially the quantum recoil phenomenon in free-electron radiation.

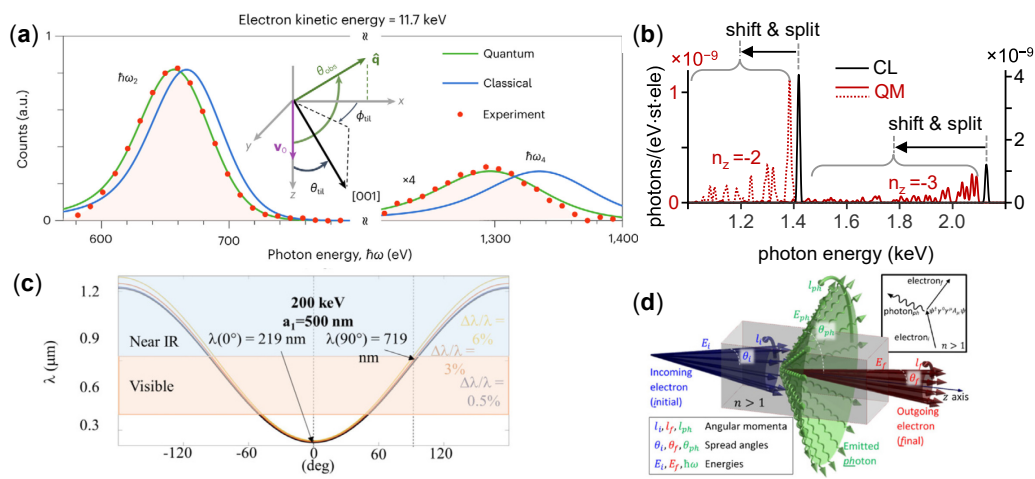

**Fig. 2. Quantum recoil in spontaneous emission from free electrons**. (**a**) Experimental measured quantum recoil from parametric X-ray radiation. Reprinted with permission from S. Huang, *et al.*, Nat. Photonics **17**, 224–230 (2023) [117]. Copyright 2023 Springer Nature. (**b**)

The transverse recoil effects shift and split the spectra of coherent X-ray emission. Reprinted with permission from X. Shi, *et al.*, Nat. Commun. **15**, 7803 (2024) [110]. Copyright 2024 Authors, licensed under a Creative Commons Attribution (CC-BY-NC-ND) License. (**c**) Quantum recoil effects in Smith-Purcell radiation. Reprinted with permission from S. Tsesses *et al.*, Phys. Rev. A **95**, 013832 (2017) [116]. Copyright 2017 the American Physical Society. (**d**) Illustration of twisted photons emitted from a vortex electron. The angular momentum is conserved during the photon emission. Reprinted with permission from I. Kaminer, *et al.*, Phys. Rev. X **6**, 011006 (2016) [115]. Licensed under a Creative Commons Attribution (CC BY 3.0) License.

### *2.1.1  Quantum Recoil in Parametric X-ray Radiation and Smith-Purcell Radiation*

Quantum recoil becomes prominent when nanoscale structural periodicities are involved in the generation of high-energy photons like X-rays. This occurs because the electron can experience momentum recoil due to the inverse reciprocal lattice vectors, in addition to the emitted photon wavevector. It has been theoretically predicted [113–116] and experimentally observed by directing free electrons at periodic structures [117], such as van der Waals atomic sheets and nanoscale gratings. In classical theory, the emitted radiation spectrum is calculated directly from the dynamics of point electrons [158], i.e., the position $\mathbf{r}(t)$ and the velocity $\mathbf{v}(t)$, without considering any energy exchange between free electron and the emitted photons. These classical theory calculations form an essential portion of classical electrodynamics, capturing radiation of free electron through theory such as the Lienard-Wiechert potential [159,160] and Thomson scattering [161]. While valid in a broad range of cases, it is not true for all regimes, especially when the emitted photon energy are high and structural feature sizes are small– as is the case of X-ray photon generation from electrons scattering off a structure with atomic-scale periodicity, or even in the case of slow electrons emitting lower energy photons[58,65].

Quantum recoil in the soft X-ray regime that is close to the water window was measured in a parametric X-ray radiation experiment [117]. Parametric X-ray radiation is emitted by free electrons traversing periodic crystal structures. It was first developed in theory in the 1970s [162–165] and realized in experiments in 1985[166,167], and features high radiation intensity in a narrow spectral interval. Recent studies show that the van der Waals materials provide a good platform for tunable parametric X-ray radiation [31,86,32,34,33,88]. Parametric X-ray radiation is a type of Smith-Purcell radiation [87,168]; however, it requires the electron to penetrate through bulk crystals, unlike the typical setup in general Smith-Purcell radiation, where the electron travels above the surface [169].

The recent quantum recoil experiment was conducted by semi-relativistic electrons from a scanning electron microscope passing through graphite and hexagonal boron nitride films. Compared to the classical nonrecoil prediction, the measured radiation peaks are red-shifted by around 10 eV and 50 eV for emitted photons of energies of 700 eV and 1.3 keV, respectively, as shown in Fig. 2b. Theoretical studies show that the energy shift could be enormous, by tuning the electron energy, interlayer spacing (by using different materials), and the sample tilt angles [117].

When a free electron interacts with materials, the electron undergoes both longitudinal and transverse momentum changes, following energy-momentum conservation. Consider an electron moves along the z-axis, interacting with a crystalline structure and emitting a photon. The energy-momentum conservation writes as

$$q_{iz} + g_z = q_{fz} + k_z$$
$$\mathbf{g}_\perp = \mathbf{q}_{f\perp} + \mathbf{k}_\perp \quad (1)$$
$$E_i = E_f + \hbar\omega,$$

where $q_{iz}$ and $q_{fz}$ are the longitudinal components of the incident and final electron wavevectors, respectively, $\mathbf{q}_{f\perp}$ are the transverse component of the final electron wavevector, $g_z$ and $\mathbf{g}_\perp$ are the longitudinal and transverse components of one reciprocal lattice vector, $k_z$ and $\mathbf{k}_\perp$ are the longitudinal and transverse components of the emitted photon, $E_i$, $E_f$ and $\hbar\omega$ are the energies of the incident electron, the final electron and the emitted photon, respectively.

When adopting the linearized electron dispersion approximation, only the first-order changes in the electron wavevector are considered. Such approximation generates the classical radiation dispersion,

$$\omega = \frac{\hbar q_{iz}}{\gamma m}(q_{iz} - q_{fz}) \approx v(k_z - g_z), \quad (2)$$

where $\gamma$ is the Lorentz factor, and $v$ is the velocity of the incident electron.

However, when adopting the squared electron dispersion, by retaining up to the second order of electron wavevector changes, we obtain the quantum-corrected radiation spectra with the frequency

$$\omega \approx v(k_z - g_z) - \frac{\hbar}{2\gamma m}(k_z - g_z)^2 - \frac{\hbar}{2\gamma m}(\mathbf{k}_\perp - \mathbf{g}_\perp)^2, \quad (3)$$

The quantum recoil correction on the photon energy could be obtained by comparing Eqs. (2) and (3). Though the quantum recoil correction could be obtained in the context of momentum-energy conservation, a full quantum electrodynamic analysis is required to obtain the radiation intensity [110]. Eq. (3) shows that both longitudinal and transverse momentum changes would shift the photon energy. When the electron interacts with certain van der Waals materials, such as graphite and hexagonal boron nitride, the radiation peaks corresponding to $\mathbf{g}_\perp \neq 0$ dominates. Therefore, one could only detect the spectrum shift following Eq. (3), as shown in Fig. 2a.

When an electron interacts with bulk materials, such as silicon, different $\mathbf{g}_\perp$ components can significantly affect the radiation intensity [110]. Consequently, the classically predicted angular frequency in Eq. (2) would be corrected by different values according to different $\mathbf{g}_\perp$ in Eq. (3). It should be noted that different $\mathbf{g}_\perp$ components negligibly recoiled the electron energy, but affected the movement directions of the final electrons, i.e., the transverse recoils on the electrons. Due to the entanglement between the final electrons and the emitted photons, transverse recoils would manifest in the radiation spectra, i.e., shift the split of the classically predicted spectra (Fig. 2b). The transverse recoil effects cannot be explained by the classical model with quantum corrections applied, and are still awaiting experimental demonstration.

Smith-Purcell radiation is the radiation excited by free electrons moving along periodic gratings [169]. It is a natural extension of Cherenkov radiation, eliminating the velocity threshold of the latter. The studies of Smith-Purcell radiation generally assume the electron velocity is constant along a linear trajectory [27,30]. However, careful studies following energy-momentum conservation find a quantum correction to the emitted photon's wavelength by an extra term proportional to the de Broglie wavelength of the electron [116], whereas the emission power at the quantum recoil corrected wavelength requires further theoretical analysis and experimental demonstration. Experimental demonstrations of quantum recoil effects in Smith-Purcell radiation for lower energy photons (such as visible and infrared light) are being pursued, with recent advancements in the demonstration of Smith-Purcell radiation from record-low-energy electrons[170,171] providing promise for such experiments.

*2.1.2 Quantum Recoil in Cherenkov Radiation*

Cherenkov radiation was discovered by P.A. Cherenkov in 1934 [172,173]. It was the first type of radiation recognized as being emitted by free electrons without being accelerated. This phenomenon occurs when free electrons move uniformly inside a medium at a speed faster the speed of light in that medium [174,175]. In 1940, Vitaly Ginzburg discovered in theory that the particle nature of light manifests in shifting the spectrum of free-electron radiation, a phenomenon named as quantum recoil [113,114], which has been studied in subsequent works [116,176–178]. However, Ginzburg himself claimed that "The quantum condition for radiation is different from the classical condition, but practically coincides with it for radiation in the visible and ultra-violet regions, in which we are interested." [113]. Nevertheless, Ginzburg also claimed that the recoil effects remain significant in studying the radiation reaction, especially under the condition of superluminal emitters [114,179].

Ginzburg and Landau believed the quantum recoil in Cherenkov radiation is proportional to $\omega\omega_C^{-1}$, where $\omega$ is the frequency of the emitted photons and $\omega_C = mc^2\hbar^{-1} = 0.5$ MeV $\hbar^{-1}$ is the Compton frequency [113,114]. Therefore, the quantum recoil is negligible for radiation in the visible and ultra-violet regions. Meanwhile, the refractive indexes of materials approach unity at high frequency, significantly increasing the electron velocity threshold for generating radiation. Therefore, it is challenging to observe quantum recoil effects in Cherenkov radiation. Recently, Ref. [115] has shown that quantum recoil also manifests in radiation from electrons with orbital angular momentum (OAM). Besides energy and longitudinal momentum conservation in previously studied Cherenkov radiation by a plane wave electron, angular momentum is also conserved [115,180]. Therefore, twisted radiation can be generated by a vortex electron beam following a post-selection process on electrons, as illustrated in Fig. 2d. Besides, in classical theory, the frequency cutoff appears at extremely high frequency - a trivial reason being that a charged particle cannot emit a photon with energy higher than it originally had, making the cutoff less significant. However, quantum analysis shows that taking the electron velocity arbitrarily close to the Cherenkov radiation threshold would shift the frequency cutoff all the way to zero [115]. Therefore, in the quantum framework, the Cherenkov radiation has a sharp frequency cutoff due to the recoil correction, irrespective of material dispersion.

### 2.1.3 *Inelastic Scattering from Free Electrons*

The underlying physics for the quantum recoil in free-electron radiation comes from the particle nature of light, which was recognized following the successful explanation of the photoelectric effect by Albert Einstein in 1905 [152]. Later, in 1923, Arthur Compton's experiments provided further evidence of the particle nature of light, by showing that photons could transfer energy and momentum to other particles like stationary electrons [133,134,135,136], resulting in the scattered photon experiencing a wavelength shift. The shift in wavelength is not captured by Thomson's classical scattering theory, which assumes the X-rays are continuous classical electromagnetic waves [161]. By adopting the particle nature of X-rays and recognizing the electron recoil, Arthur Compton obtained the famous Compton scattering equation $\lambda_{\text{sca}} - \lambda_{\text{inc}} = \lambda_c(1 - \cos\theta)$, where $\lambda_{\text{sca}}$ and $\lambda_{\text{inc}}$ are the wavelengths of the scattered and incident light, $\theta$ is the scattering angle. $\lambda_c = hm_e^{-1}c^{-1} = 2.43 \times 10^{-12}$ m is the Compton wavelength of the electron, where $h$ is the Plank constant, $m_e$ is the electron mass, and $c$ is the vacuum light speed. That this shift is quantum in nature can be seen from the fact that it is directly proportional to the Plank constant $h$, recognized as the signature of the quantum regime.

This shift is typically very small, necessitating the use of very high-energy photons for an observable shift to be measured. For example, 17.4 keV photons were used in the first experiment by Arthur H. Compton [181], resulting in an increase in the scattered wavelength of about 0.02 Å. Nonlinear inverse Thomson or Compton scattering, in which multiple low-energy photons are converted into one high-energy photon [182,183], can occur with low-energy photons but requires high-intensity lasers due to their relatively small cross-sections. For example, the pioneering nonlinear Thomson scattering experiment and nonlinear Compton scattering experiments used lasers with intensities of $10^{14}$ W/cm$^2$ [184] and $10^{18}$ W/cm$^2$ [185,186], respectively.

In the case of an electron not initially at rest, as is the case in general Compton scattering and inverse Compton scattering experiments, the resulting wavelength shift is no longer entirely a quantum effect. In fact, most of the wavelength shifts in these cases typically result from the classical relativistic Doppler shift, due to the presence of a moving source (the electron). Inasmuch as quantum recoil is a direct result of the particle nature of light, quantum recoil can be understood as the spontaneous emission counterpart of the photon energy shift observed in the rest frame of the initial electron during Compton scattering. However, the above discussion reveals several important differences between quantum recoil and the wavelength shift in Compton scattering:

1. Whereas quantum recoil in spontaneous emission is entirely a quantum effect, the wavelength shift in Compton scattering (between input and output photons), is entirely quantum only when the initial electron is at rest – in which case the Compton shift also represents the difference between classical and quantum predictions of outgoing photon wavelength [153,154].

2. The maximum wavelength shift in Compton scattering is limited to twice the Compton wavelength $\lambda_c = h m_e^{-1} c^{-1}$. In contrast, the wavelength shift in quantum recoil (between classical and quantum predictions) can be arbitrarily large [117,110].

3. Being a first-order quantum electrodynamics process, spontaneous emission tends to have a larger cross section than scattering processes like Compton scattering, a second-order process. This makes spontaneous emission an exciting and complementary platform for exploring quantum effects and realizing quantum technologies.

In free-electron radiation under strong external field, the particle nature of light reduces the emission power and the maximum energy of the emitted photons. The quantum correction become significant only in the presence of a strong external field and when the energy of the emitted photons is large, such as exceeding 0.01 of the electron's total energy [155–157].

## 2.2 Coherent X-ray Emission from Nanomaterials

Coherent photon emission from nanomaterials through free-electron interaction can be classified into two categories according to whether the electron moves uniformly or is accelerated [187]. In the first category, the electrons move uniformly through or near the medium, exciting the polarization of the medium, which in turn generates coherent radiation. It includes Cherenkov radiation [113,114,188], transition radiation [114], Smith-Purcell radiation [169], and parametric X-ray radiation (also named as diffracted X-ray radiation) [163,164]. In the second category, the electrons are accelerated by the atomic potential within the medium, generating radiation during the acceleration. It includes channeling radiation [189] and coherent Bremsstrahlung [190,191].

In the first category, from the classical perspective, polarization currents excited by free electrons can, in turn, generate coherent radiation. From the quantum perspective, free electrons emit dressed photons, which are then transferred to free photons. Readers interested in Cherenkov radiation, transition radiation, and Smith-Purcell radiation and their quantum-mechanical descriptions are recommended to refer to the review papers [114,192–194]. Parametric X-ray radiation is emitted by polarization currents induced by the incident-free-electrons on the bound electrons, which was first proposed in theory [163,164] and demonstrated in experiments later [166,195,196] with ultra-relativistic electrons. More theoretical and experimental studies of parametric X-ray radiation in the past can be found in the review [187].

In the second class, from the classical perspective, the electron's trajectory is altered by the atomic Coulomb potential or vector potential from an external magnetic field, and radiation is emitted during the electron acceleration and deceleration. From a quantum perspective, free electrons emit photons as they transition from high-energy states to low-energy states, with the photon energy equal to the difference between these states. One specific case of the potential-influenced electron trajectory is channeling, where the electron's trajectory is guided by the crystal axis or plane, when incident electrons are nearly parallel to the crystal axis or plane, respectively, with the incident angle smaller than the Lindhard angle [189]. In this case, the electrons do not feel the discreteness of the lattice field and are guided by the average potential field. Coherent Bremsstrahlung dominates when the incident angle is larger than the Lindhard angle [190,191]. Coherent Bremsstrahlung was the first effect that proved the crucial role of coherence length(also known as the formation zone), which is the distance the electron interacts coherently with the medium, in the generation of coherent radiation [162,187]. In Coulomb scattering processes, atoms, which are more widely distributed than the wavelength, exhibit coherent behavior within the coherence length, leading to the generation of coherent Bremsstrahlung.

It should be noted that the above-mentioned radiation process may occur simultaneously. But they could manifest separately in different frequency ranges and radiation angles, as shown in Fig. 18 of Ref. [187]. They could also share the same photon energy and radiation direction and, therefore, interfere with each other. Previous studies on the interferences were reported in Refs. [187,197–201].

*2.2.1 Tunable X-ray Emission from Nanostructures*

The present table-top X-ray sources rely on X-ray tubes [202], which give rise to incoherent Bremsstrahlung and characteristic X-rays. The emitted radiation is unidirectional and covers a broad frequency range. Therefore, the emission processes are inefficient as only a small part of the emitted radiation is exploited in application. Besides, the radiation is spatially and temporally incoherent, inhibiting the imaging resolution. Recently, advances in nanophotonics inspired the studies of tunable coherent X-ray emission. The tunability can be realized by controlling the electron velocity, the nanostructure, the radiation direction, etc.

One mechanism for creating spatially coherent X-rays is inverse Compton scattering, including the undulator and wiggler radiation [203]. Generally, ultrarelativistic electrons are required to scatter much longer wavelength optical laser or static wiggler field into ultrashort wavelength X-rays [204]. This is because the optical wavelength or wiggler period seen by the moving electron is scaled by the Lorentz contraction factor ($1/\gamma$). Quasi-particles, such as graphene plasmons, are accompanied by a much smaller wavelength that could reach one of hundreds of the free-space wavelength [205]. Therefore, the ultrashort wavelength graphene plasmons could facilitate the generation of X-rays with semi-relativistic electrons in inverse Compton scattering. It was shown in theory [28] that 3.7 MeV electrons, interacting with graphene plasmons at the frequency of 100 THz, could excite 20 keV hard X-rays (Fig. 3a). The use of plasmonic metasurfaces have been proposed to spectrally and spatially shape the output X-ray radiation [206]. Multilayer graphene metamaterials, such as 50 layers in Ref. [207], provide a relatively large interaction area, and can therefore generate high-intensity X-rays from a larger current. The generation of high-energy photons through free electrons scattering off plasmonically enhanced vacuum modes has also been studied, opening the possibility of plasmon-based X-ray sources that do not require external light sources [208].

X-rays can also be excited by free electrons bombarding materials. One widespread application is X-ray tubes, in which incoherent Bremsstrahlung is excited. However, coherent interaction also occurs when the electrons traverse crystal structures. Two mechanisms, parametric X-ray radiation and coherent Bremsstrahlung, contribute to the coherent X-ray excitation. Parametric X-ray radiation and coherent Bremsstrahlung share the same radiation dispersion $\omega = (\mathbf{k} + \mathbf{g}) \cdot \mathbf{v}$ under the nonrecoil approximation, where $\omega$ is the radiation angular frequency, $\mathbf{k}$ and $\mathbf{g}$ are the photon wavevector and reciprocal lattice vector, respectively, $\mathbf{v}$ is the electron velocity. The radiation dispersion should be corrected considering the recoil, which is discussed in Section 2.1.2. The radiation dispersion reveals that the photon energy and the radiation direction could be tuned through the electron velocity $\mathbf{v}$ and the reciprocal lattice vector $\mathbf{g}$. Therefore, one can demonstrate the tunable X-ray emission by van der Waals materials irradiated by semi-relativistic electrons available, for example, from an electron microscope. The electron velocity $\mathbf{v}$ can be precisely tuned in a scanning electron microscope and transmission electron microscope, and the reciprocal lattice vector $\mathbf{g}$ could be adjusted by exploiting van der Waals materials with varying lattice constants [31] (Fig. 3c). The reciprocal lattice vector $\mathbf{g}$ could also be rotated by tilting the crystal structure, bring more flexibility to create tunable X-ray emission [32].

The advance of two-dimensional (2D) materials, especially van der Waals heterostructures, can create nanostructures with feature sizes comparable with X-ray wavelength. Meanwhile,

X-ray optics rely on bulky and/or inefficient optical elements that are challenging to fabricate. Therefore, it is beneficial to create X-ray sources that emit shaped X-rays directly, eliminating subsequent X-ray optical elements. These sources come from the coherent interaction between free electrons with aperiodic nanostructures. Such as in Ref. [33], chirped periodicity is approximated by a multilayer van der Waals heterostructure stacked by different kinds of van der Waals materials (Fig. 3e). A focused X-ray beam with a diffraction-limited focal spot can be generated by a free electron passing through. X-ray caustics, such as an Airy-beam, are excited by a free electron passing through a cylindrically bent van der Waals material (Fig. 3d) [34]. Curved crystal structures can be produced from strained van der Waals materials. These proposals await experimental demonstrations, though the radiation from chirped aperiodic structures have already been realized in the optical domain [84], providing promise for similar demonstrations in X-rays. A recent prototype in the experiment [87] [209].shows that two-color X-ray emission could be achieved from a graphite/MoS$_2$ van der Waals heterostructure (Fig. 3b).

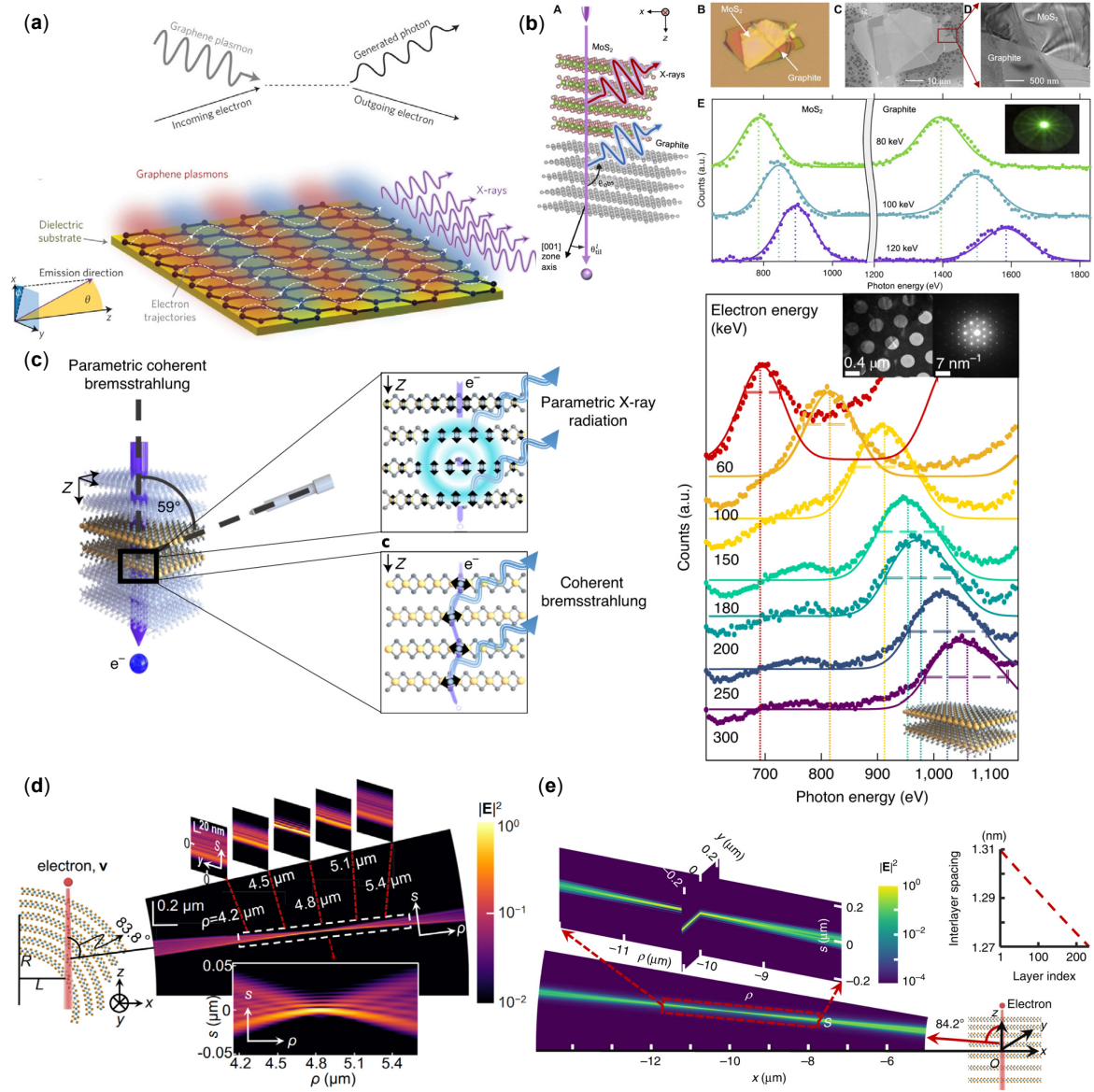

**Fig. 3 Advancements in X-ray generation nanophotonic innovations and tunable emission techniques.** (**a**) Inverse Compton-scattering from graphene plasmons to X-rays. Reprinted with permission from L.J. Wong *et al.*, Nat. Photonics **10**, 46–52(2016) [28]. Copyright 2016 Springer Nature. (**b**) Two-color X-ray emission from graphite/MoS$_2$ van der Waals heterostructure. Reprinted with permission from S. Huang *et al.*, Sci. Adv. **9**, eadj8584 (2023) [87]. Licensed by a Creative Commons Attribution NonCommercial License 4.0 (CC BY-NC). (**c**) Tunable X-ray emission by changing the incident electron energy. Reprinted with permission from M. Shentcis *et al.*, Nat. Photonics **14**, 686–692 (2020) [31]. Copyright 2020 Springer Nature. (**d**) X-ray caustics generated by curved van der Waals materials. Reprinted with permission from X. Shi *et al.*, Optica **10**, 292-301(2023) [34]. Copyright 2023 The Optical Society. (**e**) Focused X-ray beam from periodicity chirped van der Waals heterostructure. Reprinted with permission from X. Shi, *et al.*, Light Sci. Appl. **12**, 148 (2023) [33]. Licensed under a Creative Commons (CC BY) license.

*2.2.2 Tunable Low-Energy Photons with Nanostructures*

Low-energy photons could also be generated from free electrons interacting with nanomaterials, in terms of Cherenkov radiation [113,114,188], transition radiation [114], Smith-Purcell radiation [169], etc. Cherenkov radiation is generally excited by free electrons moving uniformly inside homogeneous materials [114]. The studies of Cherenkov radiation trace back to 1934 [188], the first time people found that free electrons could emit radiation without being accelerated. The emergence of nanophotonics spurs the investigation of 2D-Cherenkov radiation in the context of various types of polaritons in 2D materials [210–213,213–217]. The experimental demonstration of the 2D-Cherenkov radiation is from Ref. [218], identifying the quantized Cherenkov radiation emission through electron energy loss spectroscopy (EELS) (Fig. 4a).

Among all the 2D polaritons, graphene plasmons are a well-known and extensively studied example. Graphene plasmons are extremely difficult to excite using optical methods due to the significant momentum mismatch with free photons [205]. Free electrons moving in the vacuum can be regarded as a source of evanescent supercontinuum light [219]. Therefore, free electrons become a pertinent tool to excite graphene plasmons by moving parallel to [115,220] and across a monolayer graphene sheet [221]. The former generates a 2D caustic graphene plasmon wake, an exemplary case of a free-electron-excited Kelvin-Mach wake in a 2D Fermi sea [222]. The latter is similar to the deep-water hydrodynamic splashing phenomenon, with a picosecond of charge concentration formed analog to the "Rayleigh jet" in hydrodynamic splashing (Fig. 4b). The graphene plasmons oscillate in the THz regime and can be coupled to free-space THz radiation by grating structure. Based on this mechanism, free-electron-driven THz sources have been proposed [223] and realized in the experiment [224], which works as a novel on-chip THz emitter.

Graphene plasmons also stand out as the unique platform to study superluminal phenomena. The phase velocity of highly confined graphene plasmons is significantly slower, often only a fraction of the speed of light in vacuum, typically one out of several hundred. [205]. Thus, it is readily easier to study superluminal phenomena with semi-relativistic electrons. Ref. [225] predicted a novel superlight inverse Doppler effect, where forwarding propagating waves by superluminal sources could experience a redshift (Fig. 4c).

The nanomaterials can also lift the electron velocity thresholds in Cherenkov radiation, which limits the application of Cherenkov radiation-based detectors in low-energy electrons. One solution uses resonant transition radiation from free electrons passing through periodic stacked photonic crystals [226]. The radiation is similar to Cherenkov radiation, with the radiation angle sensitive to electron velocity but not bound by the Cherenkov radiation threshold. Another way is using hyperbolic metamaterials (HMMs), which have opposite sign permittivity along different directions, to eliminate the Cherenkov radiation threshold (Fig. 4d) [29]. In type I HMM, where $\varepsilon_x < 0$ and $\varepsilon_{y,z} > 0$, there is no velocity threshold. In type II HMM, where $\varepsilon_x > 0$ and $\varepsilon_{y,z} < 0$, the condition to generate CR by an electron propagating along the $x$ direction is $v_{\text{ele}} < c/\sqrt{\varepsilon_x}$. In Ref. [29], low-energy electrons (0.25–1.4 keV) excite radiation in the wavelength range of 500–900 nm from a type II HMM.

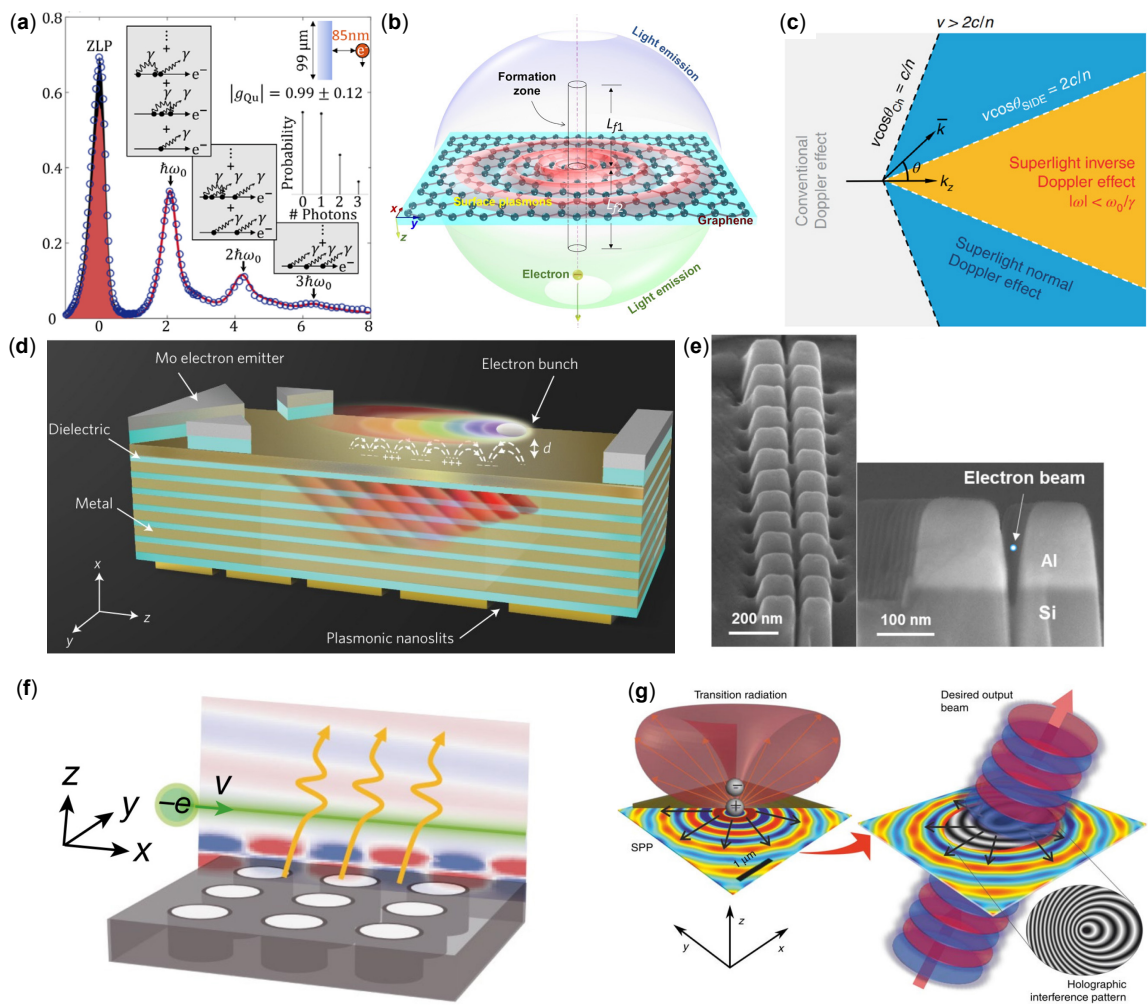

**Fig. 4 Tunable low-energy radiation by free electrons interacting with nanomaterials. (a)** Electron energy loss spectroscopy (EELS) spectrum of quantized Cherenkov radiation. Reprinted with permission from Y. Adiv *et al.*, Phys. Rev. X **13**, 011002 (2023) [218]. Licensed under a Creative Commons Attribution 4.0 International license. **(b)** Transition radiation from a monolayer graphene. Reprinted with permission from X. Lin *et al.*, Sci. Adv. **3**, e1601192 (2017) [221]. Licensed under a Creative Commons Attribution-NonCommercial license. (**c**) Superlight inverse Doppler effect. Reprinted with permission from X. Shi *et al.*, Nat. Phys. **14**, 1001–1005 (2018) [225]. Copyright 2018 Springer Nature. (**d**) Threshold eliminated Cherenkov radiation from hyperbolic metamaterials. Reprinted with permission from F. Liu *et al.*, Nat. Photonics **11**, 289–292 (2017) [29]. Copyright 2017 Springer Nature. (**e**) Deep-ultraviolet Smith-Purcell radiation. Reprinted with permission from Y. Ye *et al.*, Optica **6,** 592–597 (2019) [30]. Copyright 2019 The Optical Society. (**f**) Free-electron radiation at photonics flatband resonances. Reprinted with permission from Y. Yang *et al.*, Nature **613**, 42–47(2023) [85]. Copyright 2023 Springer Nature. (**g**) Holographic free-electron light source. Reprinted with permission from G. Li *et al.*, Nat. Commun. **7,**13705 (2016) [79]. Licensed by a Creative Commons Attribution 4.0 International License.

Smith-Purcell radiation is generated by electrons passing over grating structures [169]. It has no velocity threshold and can be excited by low-energy electrons [170,171]. The radiation process is strongly correlated with the grating structure, making it a versatile platform for tunable coherent light sources. With the grating fixed, the emission wavelength could be tuned by changing the electron velocity [27]. By designing the grating structure, we could manipulate the radiation directly from the sources, generating structured light, such as a focused beam [80,84], Airy beam [84,227], and vortex light [81,228]. More interestingly, the efficiency of Smith-Purcell radiation could be increased with inverse design [83]. One significant achievement is the 2D flatband grating structures, which were experimentally demonstrated to enhance the Smith-Purcell radiation by two orders of magnitude [85].

Generating short-wavelengths light with Smith-Purcell radiation, like in the deep-ultraviolet regime, relies on reducing the gap between the electron and the grating. Ref. [30] reported in the experiment the shortest yet wavelength of 230 nm generated by letting a highly focused electron beam (with spot size ~10 nm) pass through a nano-slot, corresponding to zero-gap Smith-Purcell radiation (Fig. 4e). The detection of Smith-Purcell radiation could be achieved in an scanning electron microscope equipped with a light-collection system comprising a small parabolic mirror [27,30] or an objective lens [229]. Smith-Purcell radiation could also be upgraded by using magnetic nanograting [230], which is motivated by advances in fabricating nanopatterned ferromagnets. This is a compact version of undulator radiation, in which X-rays could be excited by semi-relativistic electrons.

The collective mode of the nanomaterials could be used to enhance and manipulate free-electron radiation [78,79,231]. A localized excitation could generate a delocalized response over a larger area beyond the electron spot size and light wavelength. For example, collective modes in a metasurface [78] and epsilon-near-zero materials [231] are excited by electron beams over a small spot size, which is converted to coherent radiation. More interestingly, we can customize the metasurface as a collective mode converter, transforming SPPs into the desired propagating light [79]. The pattern is designed following holographic principles, where the interferences between the point-generated SPPs pattern and the propagating light pattern are utilized (Fig. 4g).

## 2.3 Quantum X-ray States from Nanomaterials

### 2.3.1 High-Energy Photon Regime

The generation of quantum X-ray states using spontaneous parametric down-conversion (SPDC) was first reported in 1969 [232–234]. To this day, this method remains the primary approach for producing quantum X-ray states. However, the conversion efficiency of X-ray SPDC is notably low, commonly measured around $10^{-11} - 10^{-14}$ [235–240], motivating research on generating quantum X-rays directly in free-electron lasers. Indeed, the exploration of quantum X-ray states can be traced back to initial studies of free-electron laser [241–243].

Glauber correlation functions and the statistics of the occupation number of the available energy states are good tools to investigate the quantum state of free-electron laser [244,245]. For free-electron laser operating in the self-amplified spontaneous emission (SASE) process (Fig. 5a), the radiation is spatially highly coherent in first-order (with the degree of spatial coherence reaching 80%[246–248]) from experimental measurement. However, since the radiation in SASE process is produced stochastically by electron shot noise, the temporal structure of the pulse fluctuates strongly from shot to shot. Therefore, higher-order correlation measurements of the radiation show the light behaves as chaotic sources [249–251].

It should be noted that sub-Poissonian distribution of spontaneous harmonic radiation [252] was also measured in free-electron laser. Sub-Poissonian distribution implied that the radiation could be squeezed light, which was predicted in earlier theoretical studies [253,254], which contained classical descriptions as an approximation. However, J.W. Park et al. argued that the experimental results could be explained using classical free-electron laser theory, and therefore, a more definitive measurement of the photon statistics is needed in improved experiments [255,256]. Besides, a fully consistent quantum theory for free-electron laser in the SASE process predict that the radiation in SASE process is similar to thermal statistics [245] or chaotic statistics [257], since the radiation originates from the random fluctuations in electrons.

However, when a seeding laser is incident with the electron beams, it is expected that the coherent features of the laser could be transferred to the electrons (Fig. 5b). This conjecture can find experimental demonstration that the electron wavefunction could be affected by the photon statistics during electron-light interaction [22], such that the radiation could carry certain degree of coherence as the seeding laser. Experimental measurements of the radiation with a seeding laser in free-electron laser show second-order intensity correlation close to unity[258,259], indicating lasering characteristics. Theoretical studies based on quantum theory show that the radiation follow a displaced thermal distribution, close to the statistics of a coherent laser [245].

To date, based on our current understanding, there has been no direct generation of quantum light from free-electron laser. The research on the statistics of seeded-FELs spurred the investigation of whether the quantum seeding light could induce the quantum X-ray [257], especially considering the recent experimental discoveries that the quantum statistics of light could affect the quantum features of electrons [22,260] and theoretical studies on electrodynamics driven by bright squeezed vacuum [261].

Other attempts have been made to generate quantum light from free-electron lasers. For example, recent theoretical analyses have shown that entangled X-ray pairs could be produced in the SASE process as a second-order effect. Consequently, X-ray pairs can be obtained by post-selecting the emitted electrons in specific energy states (Fig. 5c) [237].

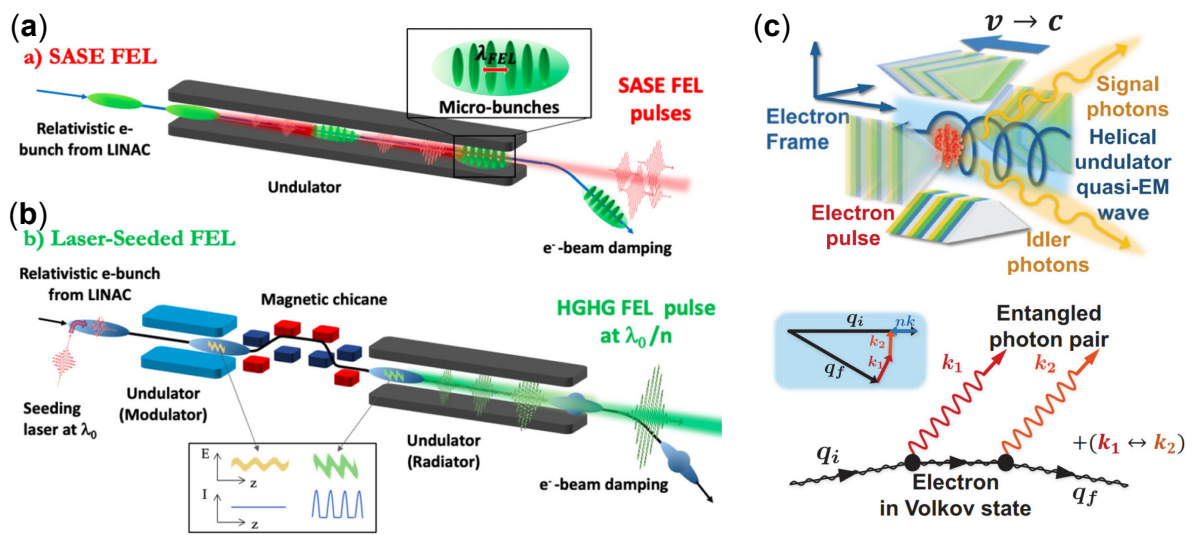

**Fig. 5 Quantum X-ray states generation in free-electron lasers (a,b)** Illustrations of the free-electron lasers working in the self-amplified spontaneous emission (SASE) and seeded modes. Reprinted with permission from F. Benatti *et al.*, Opt. Express 29, 40374-40396 (2021) [245]. Copyright 2021 The Optical Society. (**c**) Entangled X-ray pairs are generated in free-electron lasers as a high-order process. Reprinted with permission from L. Zhang *et al.* Phys. Rev. Lett. **131**, 073601 (2023) [237]. Copyright 2023 American Physical Society.

### 2.3.2 Low-Energy Photon Regime

The quantum nature of the interaction between free-electrons and low-energy photons has been gradually unveiled over the past decades, with experiments unravelling the importance of the quantum description of the coherent electron wavefunction as well as of the quantization of the electromagnetic field. The first observation of the photon-induced near-field electron microscopy (PINEM) effect [14] showed that a free electron wavefunction interacts with a strong laser field (of frequency $\omega$) in a quantized manner, gaining or losing energy of $\hbar\omega$.

The full scope of coherent electron wavefunction effects were shortly after predicted by theory [15,16], and several ensuing experiments demonstrated the importance of the quantum wave nature of the free electron [9,17,19–21,51,144,262,263]. In particular, Ref. [17] has demonstrated the first observation of free-electron Rabi oscillations which have also been interpreted as a quantum walk of the free electron on an infinite energy ladder, driven by a coherent light source. In this picture, the electron is hopping between equidistant energy levels, with a symmetric hopping amplitude $g = \frac{e}{\hbar\omega}\int_{-\infty}^{\infty} dz\, e^{-\frac{i\omega}{v}z} E(z)$, the electron-laser coupling. A semiclassical scattering operator $S = \exp(gb - g^*b^\dagger)$ can be used to describe this quantum walk, where $b \equiv e^{-\frac{i\omega}{v}z}$ is a free-electron hopping operator, satisfying $[b, b^\dagger] = 0$. Starting at $E = E_0$, the probability of finding the electron at energy level $E = E_0 + n\hbar\omega$, $n$ an integer, is then $J_n^2(2|g|)$ [15,16], where $J_n(x)$ is the $n$-th order Bessel function. Since the demonstration of Rabi oscillations and quantum walks on the energy ladder, several other experiments employing, for example, multiple laser frequencies [9,262] have been demonstrated, extending the range of possibilities to shape and manipulate free-electron wavefunctions in space and in time using low-energy photons [264–269].

The importance of the quantum nature of light in free-electron—photon interactions has been only recently explored. The semiclassical quantum walk theory breaks down when one considers interactions driven by nonclassical light. This is now known as the quantum PINEM effect, first introduced theoretically by Kfir [18] and Di Giulio et al. [37]. There, the quantum scattering operator takes the form of $S = \exp(g_Q ba^\dagger - g_Q^* b^\dagger a)$, where $g_Q = \frac{e}{\hbar\omega}\int_{-\infty}^{\infty} dz\, e^{-\frac{i\omega}{v}z} E_{vac}(z)$ is the quantum coupling to the vacuum field amplitude and $a, a^\dagger$ are the photonic annihilation and creation operators. The modulus square of $g_Q$ represents the number of photons emitted by a single electron, and $|g_Q| = 1$ is the convention for the separation between strong and weak coupling regimes [18].

During the quantum interaction, the electron makes discrete energy transitions (Fig. 6a). However, now the probability of finding an electron at energy level $E_0 + n\hbar\omega$ is strongly dependent on the quantum state of the incident light [40,45,270]. In fact, it was shown experimentally that a transition from a quantum to classical random walk is possible when the photon statistics of the incident light are tuned between a Poissonian and thermal statistics [22]. On the other hand, for an initial empty cavity in the vacuum state (spontaneous photon emission), the emitted light displays a Poissonian distribution, wherein the electron energy spectrum becomes strongly asymmetric: no energy gain for $n > 0$ and a Possonian energy loss for $n < 0$ [18,37,45]

This series of breakthroughs laid the grounds for the now emerging field of free electron quantum optics. The main conceptual advancement is the realization that radiating free electrons become entangled with the emitted photons via the laws of conservation of energy and momentum. This has led to a wealth of studies leveraging this entanglement to realize quantum light sources based on free electrons [18,45,53,56,65,36,49,54,55,57,58,60,67,68].

The efficient generation of quantum light states, such as large-number Fock states, squeezed states, cat states and Gottesman-Kitaev-Preskill (GKP) states is a long-standing goal for the general quantum optics community. Recent studies have theoretically proposed that the entanglement between electron energy and photon number could be harnessed for the generation of Fock states(Fig. 6). Fock states of light may be heralded by post-selecting the associated free electron energy [36]. The heralded single-electron and Fock-state photon sources have been experimentally demonstrated in Ref. [50], as shown in Fig. 6b, where high-energy electrons excite photons in a fiber-coupled $Si_3N_4$ micro-resonator. Through coincidence detection of the electrons and photons [271], one could isolate the physical scattering events from background noise. Such correlation-enhanced measurement demonstrates the high-fidelity generation of electron-photon pairs.

Post-selection of electron energy renders such light generation non-deterministic and may hamper its speed, but this need not be a fundamental limitation: the use of inherent nonlinearities of cavity quantum electrodynamics [67] or the intrinsic nonlinearity of low-energy electrons [38,39,52,58,64,272,273] (Fig. 6c) that gives rise to unique vacuum Rabi oscillations [58,273] can allow for ultrafast and deterministic schemes of quantum light generation (i.e., without post-selection). These nonlinear deterministic interactions have further been shown to construct a universal gate set for gate-based photonic quantum computation with free-electron ancilla qubits [67] . The two-photon Fock state, or photon pairs, represents a higher-order process and, consequently, has a lower probability of occurrence than the one-photon Fock state. Nevertheless, by selectively post-selecting the electrons under specific conditions, such as zero deflection and a defined energy loss, one could find scenarios where the probability of generating a two-photon state becomes predominant [68]. Many-photon states of quantum light can be achieved through consecutive photon-electron interactions [45]. As illustrated in Fig. 6d, the incident photon is in a vacuum or Fock state, and the incident electrons are in monochromatic energy states. By directly measuring the electron after each interaction, one could obtain a pure Fock –state of light. The number of photons in the Fock-state increases, though stochastically, with the number of interactions.

More advanced quantum light states could be excited by pre-shaped electrons, such as electrons prepared in the so-called energy-comb state (referred to as 'comb' electrons hereafter), which could be obtained from free electrons interacting with the near-field in nanomaterials (Section 4.4). In the quantum optics terminology, the comb electron implements a displacement operation on the electromagnetic field, generating a coherent state of light from the vacuum state [45]. Being an eigenstate of the electron ladder operator, the state of a comb electron is retained (i.e., it does not become entangled with the light) even after emitting photons, and subsequent interactions with several comb electrons can amplify the displacement value (Fig. 6e). When the energy separations of the comb electrons are multiples of the photon energy, as illustrated in Fig. 6f, the comb electrons could be used to generate macroscopic non-Gaussian quantum light states [54,57,67] such as optical cat states as well as Gottesman-Kitaev-Preskill (GKP) states. This could be done through consecutive interaction with a resonant cavity initially in the vacuum (or squeezed vacuum) state [54] or in a coherent state of light [55]. These protocols were later shown to be compatible with continuous variable quantum information processing [57], offering a universal quantum-optical gate set for continuous-variable schemes, where the electron-light interaction acts as a conditional displacement operation.

While the generation of quantum light by free electrons can be realized in the weak coupling regime, either by post-selection or consecutive interactions [45,50,54,56], generation and control over macroscopic quantum light states, such as large Fock states, cat and GKP states mentioned above, can be made much more efficient in the strong-coupling regime [18], where $|g_Q| \geq 1$. The state of the art value of $|g_Q|$ in experiments so far is $|g_Q| = 1$ from free electrons interacting with lossy plasmonic waveguides [218], and $|g_Q| \sim 0.1$ in lossless dielectric cavities [50,59].

Many theoretical attempts to increase $|g_Q|$ are currently underway, such as optimizing the photonic structures and impact parameters [274,275,56,62,63] and increasing the interaction length [60,61]. In addition to achieving strong coupling, increasing interaction lengths can also enhance single photon nonlinearities due to recoiled free electrons[38,52,58,60,64,65,273]. Other strategies such as free electron cavities [276] are promising for increasing the interaction length without increasing the size of the optical system.

It should be noted that many previous experimental strategies [27,29,30,85,170] to increase the free-electron radiation power could be adopted for enhancing the coupling, though these past experiments did not measure the value of $|g_Q|$ directly.

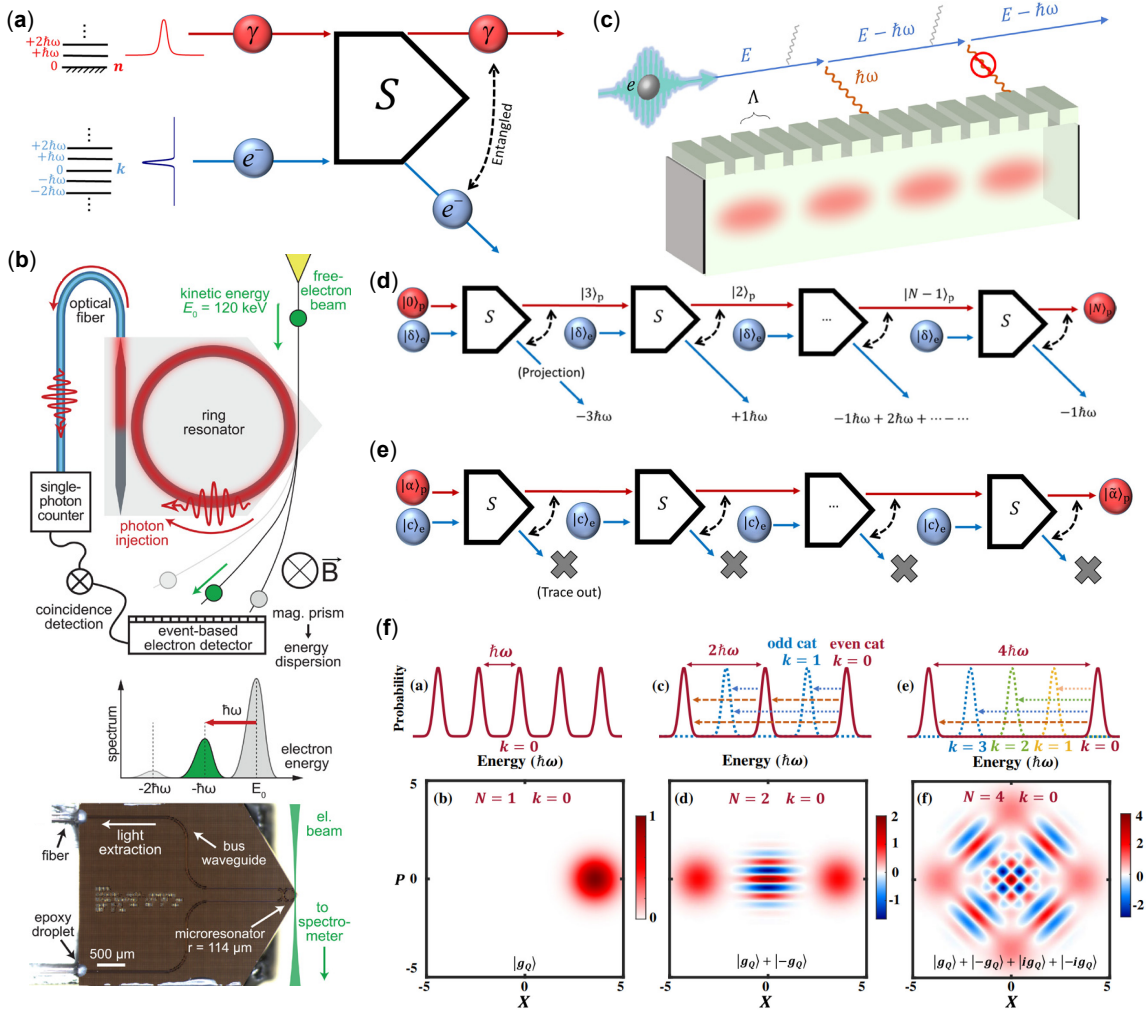

**Fig. 6 Advancements in quantum light generation through electron-light interaction. (a)** The electron is described in its energy ladder space, and the photon is described with the harmonic oscillator spectrum. Free electrons experience discrete energy transitions upon emitting or absorbing photons. Reprinted with permission from A. Ben Hayun *et al*., Sci. Adv. **7**, eabe4270 (2021) [45]. Licensed under Creative Commons Attribution NonCommercial License 4.0 (CC BY-NC). (**b**) Entangled electron-photon pairs are generated through spontaneous emission into a photonic chip. Reprinted with permission from A. Feist *et al.*, Science **377**, 777–780 (2022) [50]. Copyright © 2022, AAAS. (**c**) Deterministic nonlinear single photon generation by low-energy free electrons. Reprinted with permission from A. Karnieli *et al.*, Sci. Adv. **9**, eadh2425 (2023) [58]. Licensed under a under a Creative Commons Attribution NonCommercial License 4.0 (CC BY-NC). (**d,e**) Large-number Fock-state and large displacement amplitudes are generated by consecutive interactions with monochromatic electrons and comb electrons, respectively. Reprinted with permission from A. Ben Hayun *et al*., Sci. Adv. **7**, eabe4270 (2021) [45]. Licensed under Creative Commons Attribution NonCommercial License 4.0 (CC BY-NC). (**f**) Photonic cat state and GKP states are generated by higher-harmonic comb electrons and post-selection. Reprinted with permission from R. Dahan et al., Phys. Rev. X **13**, 031001 (2023) [54]. Licensed under a under a Creative Commons Attribution NonCommercial License 4.0 (CC BY-NC).

# 3 Enhancing X-ray Emission via Free-Electron Waveshaping

## 3.1 *Electron Spatial Wavefunction Engineering in Nanophotonics*

The wave–particle duality of electrons was developed in 1925 by de Broglie [112], and demonstrated experimentally by studying the interference of electrons traversing a crystal structure [120]. Therefore, in principle, it is possible to engineer electrons in a manner similar to how we manipulate other types of waves [277,278].

One electron could be coherently split into multiple spatial modes through a phase plate, generating a superposition electron state. The most often adopted phase plate is a bi-prism [279], which could be made of a uniformly magnetized straight needle [280] (Fig. 7a). The electron beam undergoes different phase shifts when passing through the two sides of the bi-prism due to the Aharonov–Bohm effect. Recently, controllable phase plates could be realized by using an array of programmable electrostatic elements [121,281,122] and phase holograms [282], enabling the introduction of adaptive optics techniques to the field of electron microscopy (Fig. 7b).

Another type of spatial modulation of electron wavefunctions is the electron vortex beam [283], which has spiralling wavefronts that give rise to the orbital angular momentum (OAM) around the propagation direction. Similar to the generation of optical vortex beams, spiral phase plates are required to alter the phase of the electron beam. Although fabrication of a spiral phase plate for electron (minimum $2\pi$ phase shift in transverse plane) is very challenging, it is demostratred that one can realize a functioning spiral-like phase plate by stacking atomic-thin graphite films [125]. Later, people found that nanofabricated diffraction holograms can provide more freedom to create electron vortex beams with OAM up to $100\hbar$ [123,124] (Fig. 7c).

More complex structured electron wave packets like Bessel beams [284] and Airy beams [285] could be generated by letting free electrons pass through nanoscale holograms. The experiment detects the parabolic trajectory of the highest-intensity lobes and the self-healing feature of the electrons after passing an obstacle (Fig. 7e). More interestingly, the self-accelerating electrons could be created by engineering the initial electron wavefunction [286], without needing any external potential. The engineered electron wavefunction evolves following an acceleration trajectory. It mimics an electron wavepacket that accumulates the Aharonov-Bohm phase under the influence of an external potential. Besides, an engineered wavepacket could extend the lifetime of unstable particles [286].

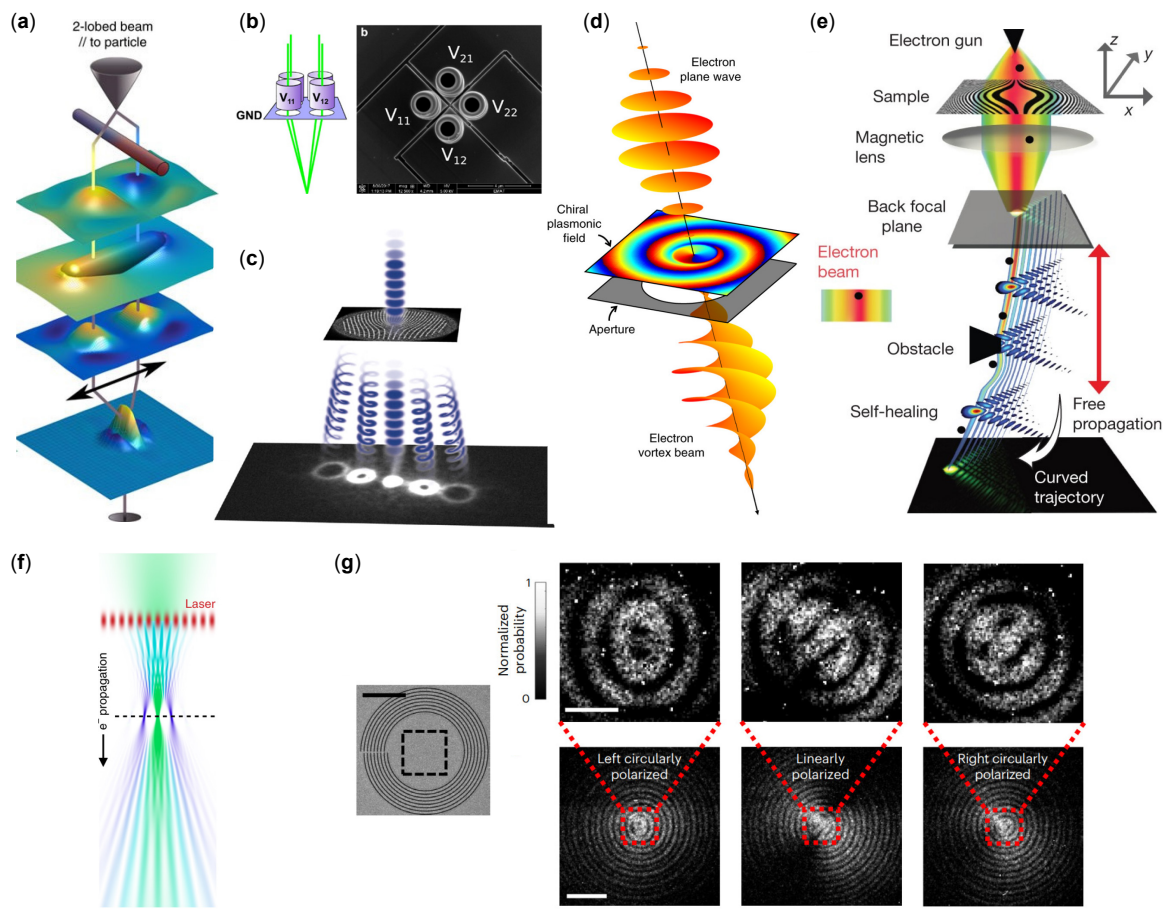

**Fig. 7 Exploring quantum electrodynamics through superposition states, phase plates, and advanced wavefront shaping in electron microscopy.** (**a**) Illustration of splitting one electron into a superposition state, coherently interacting with plasmons, and then interfering at the detector plane. Reprinted with permission from G. Guzzinati *et al.*, Nat. Commun. **8**, 14999 (2017) [280]. Licensed under a under a Creative Commons (CC BY) license. (**b**) Programmable phase plate from electrostatic elements. Reprinted with permission from J. Verbeeck *et al.*, Ultramicroscopy **190**, 58–65 (2018) [121]. Copyright 2018 Elsevier B.V. (**c**) Multiple electron vortex beams by a phase plate. Reprinted with permission from B.J. McMorran *et al.*, Science **331**, 192–195 (2011) [123]. Copyright 2011 AAAS. (**d**) Schematic representation of OAM transfer to electrons from the chiral plasmonic field. Reprinted with permission from G.M. Vanacore *et al.*, Nat. Mater. **18**, 573–579 (2019) [264]. Copyright 2019 Springer Nature. (**e**) Electron Airy beam generated from a hologram. Reprinted with permission from N. Voloch-Bloch *et al.*, Nature **494**, 331–335 (2013) [285]. Copyright 2013 Springer Nature. (**f**) Laser phase plate from standing light wave for electron microscopy. Reprinted with permission from O. Schwartz *et al.*, Nat. Methods **16**, 1016–1020 (2019) [287]. Copyright 2019 Springer Nature America, Inc. (**g**) Spatial modulation of the free electrons by the polarization of the incident light. Reprinted with permission from S. Tsesses *et al.*, Nat. Mater. **22**, 345–352(2023) [269]. Copyright 2023 Springer Nature.

Electron-light interactions also provide an efficient method to shape the electron spatial wavefunction [287–290]. Specifically, the standing light wave could work as a phase plate [287,288] for electrons (Fig. 7f), similar to how crystals work for electrons. Besides interacting with free-space light, it has been shown that the interaction with near-field or evanescent light reduces the requirement for high laser intensity [38,291], as demonstrated in photon induced near-field electron microscopy (PINEM) (Section 4.4). For example, electron vortex beams could also be produced by plane wave electrons interacting with chiral plasmonic fields [264] (Fig. 7d). Recently, a more complex electron pattern could be achieved [269] in electron microscopy through free electrons interacting with externally controlled standing surface plasmon waves (Fig. 7g). Additional degrees of freedom, such as the polarization of the incident light, can alter the spatial distribution [269] or momentum modulation [1] of the electron wavefunction.

*3.2 Quantum Electrodynamic Control by Shaped Electron Wavefunctions*

A fundamental question exists regarding the interpretations of the electron wavefunction in spontaneous radiation. One possible interpretation is that the electron wave function squared $|\psi|^2$ is equivalent to the continuous spread of the electron chargein space (middle panel of Fig. 8a). Another is that $|\psi|^2$ is the probability of finding a point electron in space, while the electron charge is always localized in space (right panel of Fig. 8a). The first-principle calculation from quantum electrodynamics (QED) and experimental results supported the second interpretation [293], complementing earlier results derived for electron energy loss spectroscopy (EELS) [219,294]. Similar QED derivations predicted the independence of spontaneous emission on the longitudinal electron wavefunction [295–297]. These results clearly deviate from the semiclassical analysis of such phenomena [293,296–300]. However, as we shall see below, this interpretation breaks down when we consider the nonparaxial shaped wavefunctions, quantum recoil, and postselection, which can all lead to strong dependence of spontaneous light emission on the electron wavefunction.

Under paraxial approximation, the spontaneous radiation from a single electron can be described by treating the electron as a point charge, apart from a weighted average (incoherent summation) of the probability density $|\psi|^2$ along the transverse dimension [168,293]. However, beyond paraxial conditions, by pre-shaping the incident electron in a superposition state, the electron wavefunction could coherently affect the radiation process [66,301]. Consider an incident electron shaped as a superposition of two transverse spatial modes $|i_1\rangle + |i_2\rangle$, the cross section of the interaction is $\int df \left| \delta_{i_1 f} M_{i_1 f} + \delta_{i_2 f} M_{i_2 f} \right|^2$, where the integral is carried out over the final states, $\delta_{i,f}$ is the delta function representing the momentum-energy conservation, and $M_{i,f}$ is the scattering amplitude. Under paraxial conditions, the states $i_1, i_2$ are closely spaced, and in this scenario one can define a transverse envelope for the electron: $\psi(r_T)e^{ik_z z}$, where $\psi(r_T) = \langle r_T|i_1\rangle + \langle r_T|i_2\rangle$. Owing to the paraxial approximation, the two transverse states have the same contribution to energy conservation (only dictated by the change in the longitudinal momentum $k_z$), and it can be shown that the trace-out of the final momentum state of the electron then leads to a spatial delta function $\delta(r_T - r'_T)$ which in turn leads to the incoherent summation over $|\langle r_T|i_1\rangle|^2$ and $|\langle r_T|i_2\rangle|^2$. If further the near-field is approximately uniform in the transverse direction within the extent of the wavefunction, then this leads to nullifying the contribution of the cross term $\text{Re}\{\langle i_1|i_2\rangle\}$ between the two initial states (or the coherent contributions from the two scattering amplitudes $\text{Re}(M_{i_1 f}M^*_{i_2 f})$ to the same final state). However, beyond the paraxial approximation, for example when the two transverse states are two large-angle plane wave fronts, or when the near-field is strongly nonuniform, the above reasoning no longer applies. When two different transitions between $i_1, i_2$ to the same final state $f$ are tailored to satisfy momentum and energy conservation for the emission of the same photon, $\text{Re}(M_{i_1 f}M^*_{i_2 f})$ can have a strong effect on the emission pattern. This effect was studied in detail for the case of X-ray emission from non-paraxial shaped electron wavefunctions, for bremsstrahlung and undulator radiation patterns [66,301].

In a further step, Ref. [69] showed that by shaping the electron wavefunction to periodically overlap with the crystal structure of monolayer two-dimensional (2D) materials, one could coherently enhance and shape the radiation emission profile. The emission rate increases coherently with the number of plane wave states in the pre-shaped electron superposition state and the number of atoms involved, as illustrated in Fig. 8b. For bulk material, we have shown in Section 2.1.3 that by preparing the incident electron wavefunctions according to different transverse recoil pathways [110], we could coherently enhance the radiation.

The underlying reason that electron wavefunction could affect the radiation is due to the entanglement between the emitted photons and the post-emission electrons [18,22,37,44–49,54,66,118]. One direct consequence of the entanglement is that the coherence could be transferred from electron to photon [46,302], or from photon to photon via an electron mediator [53] . The coherence transfer could affect detectable properties of the radiation, like pulse duration and coherent range, as shown in Fig. 8c. Conversely, by detecting the coherence of the radiation, we could retrieve the coherent length of the electron wavepacket [46].

Apart from studying the engineered single electrons, quantum-correlated electrons, also known as electrons in many-body quantum states, could also coherently affect the radiation [168,303,304]. These ideas are complementing the well-establsihed results of free electron superradiance due to classical spatial bunching of electron beams [305]. One exciting direction is that the quantum-correlated electrons could generate electron-induced superradiance in bound-electron emitters [303,306,307], as illustrated in Fig. 8e.

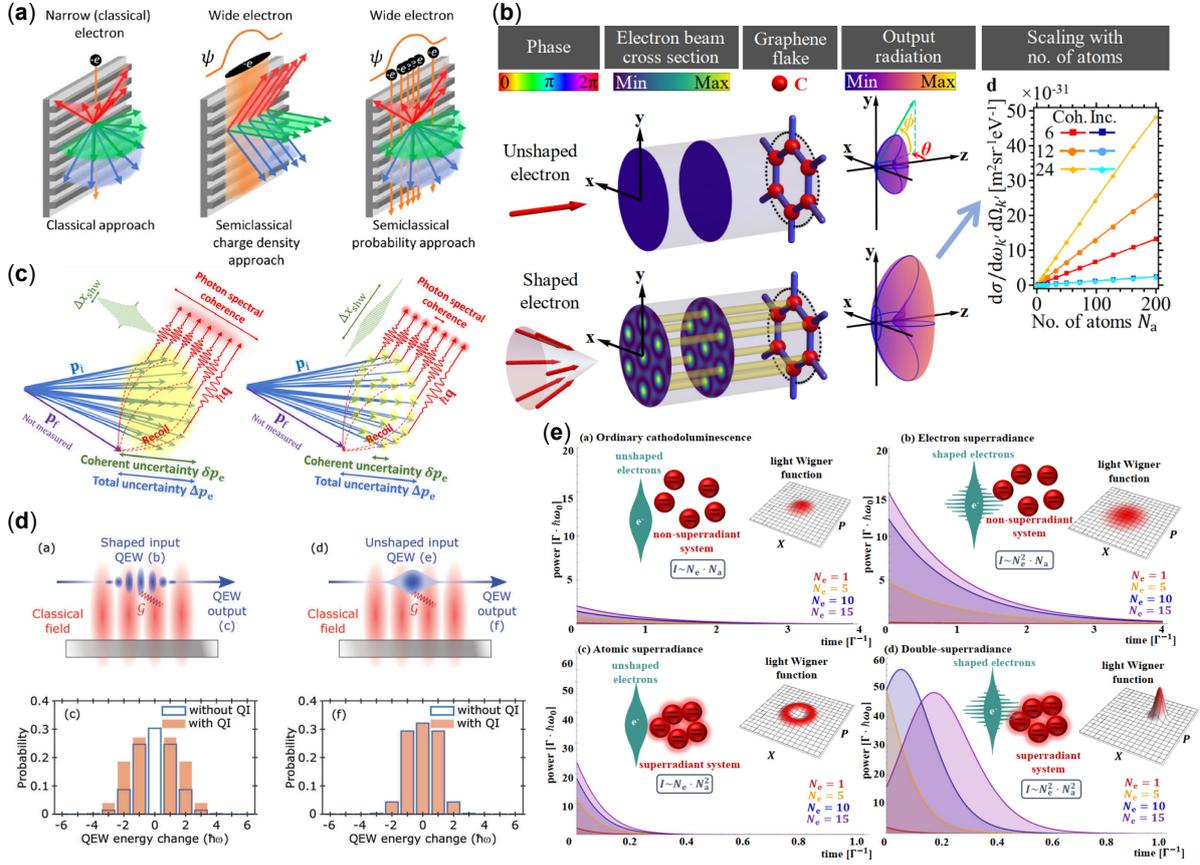

**Fig. 8 Visualization of electron wavefunction interactions and radiation outcomes.** (**a**) Illustration of radiation pictures from a point electron, charge-density electron, and free electrons with undefined trajectory. Reprinted with permission from R. Remez *et al.*, Phys. Rev. Lett. **123**, 060401 (2019) [293]. Copyright 2019 American Physical Society. (**b**) Enhanced radiation from an electron in a superposition state. Reprinted with permission from L.W.W. Wong *et al.*, Light Sci. App. **13**, 29 (2024) [69]. Licensed by a Creative Commons Attribution 4.0 International License. (**c**) Radiation pulse duration and coherent range are tied to the momentum uncertainties of the incident electron. Reprinted with permission from A. Karnieli *et al.*, Sci. Adv. **7**, eabf8096 (2021) [46]. Licensed by a Creative Commons Attribution NonCommercial License 4.0 (CC BY-NC). (**d**) Shaped electron wavefunction results in the elimination of the zero-loss peak due to quantum interference. Reprinted with permission from J. Lim *et al.*, Adv. Sci. **10**, 2205750 (2023) [301]. Licensed by a Creative Commons (CC BY) license. (**e**) Double-superradiant cathodoluminescence. Reprinted with permission from A. Gorlach *et al.*, Phys. Rev. A 109, 023722 (2024) [307]. Copyright 2024 American Physical Society.

# 4 Generating and Manipulating Free Electrons with Nanophotonics

The ability to generate on-demand, coherent, and indistinguishable single electrons has long been a desire for advancing quantum applications, including quantum computing [308,309]. Recent electrodynamic interactions with nanomaterials could potentially affect the generation of X-rays with features inaccessible by classical theory. It is worth noting that the electron spatial wavefunction engineering discussed in Section 3.1 could be used to generate electron superposition states and tune the phase-front of free electrons. In this section, we will review the generation and manipulation of free electrons using nanomaterials, covering topics like attosecond electrons and electron quantum statistics.

## 4.1 Attosecond Electron Pulse Control and Photoemission with Nanoscale Emitters

The near-field enhancements at nanotips, nanoparticles, and other nanostructures open the door to strong-field photoemission with field strengths excessing $1\text{ V Å}^{-1}$ [141,310]. The most general photoemission is from metal tips with radii from tens of nanometer [8] down to several atoms [311] at the apex, as shown in Fig. 9a. The nanometer-scale tip not only amplifies the field intensity by tens-fold but also ensures that the field intensity remains nearly homogeneous across the apex. Therefore, the photoemission shows a strong light-phase sensitivity [312,313] that will affect the quantum electrodynamics, which will be discussed in the following two sections. Except for metal tips, metallic nanoparticles supporting localized surface plasmon resonances [314] and nanotubes [315] also work as tip emitters. These nanostructures introduce additional parameters, such as particle geometry and plasmon resonance, which influence the light-phase sensitive response of photoemission. For the fundamental physics underlying photoemission, i.e., photoelectric effects, please refer to the Refs. [312,313,316,317].

The tip emitters could provide ultrashort electron pulses with high brightness [140,141,310,318] and high coherence [286–288]. However, the current from one tip is limited. Experimental demonstrations of a nanotip array (Fig. 9b) [322] and a nanoparticle array [314], which enhance the current, make a step toward chip-scale light-phase sensitive electron sources.

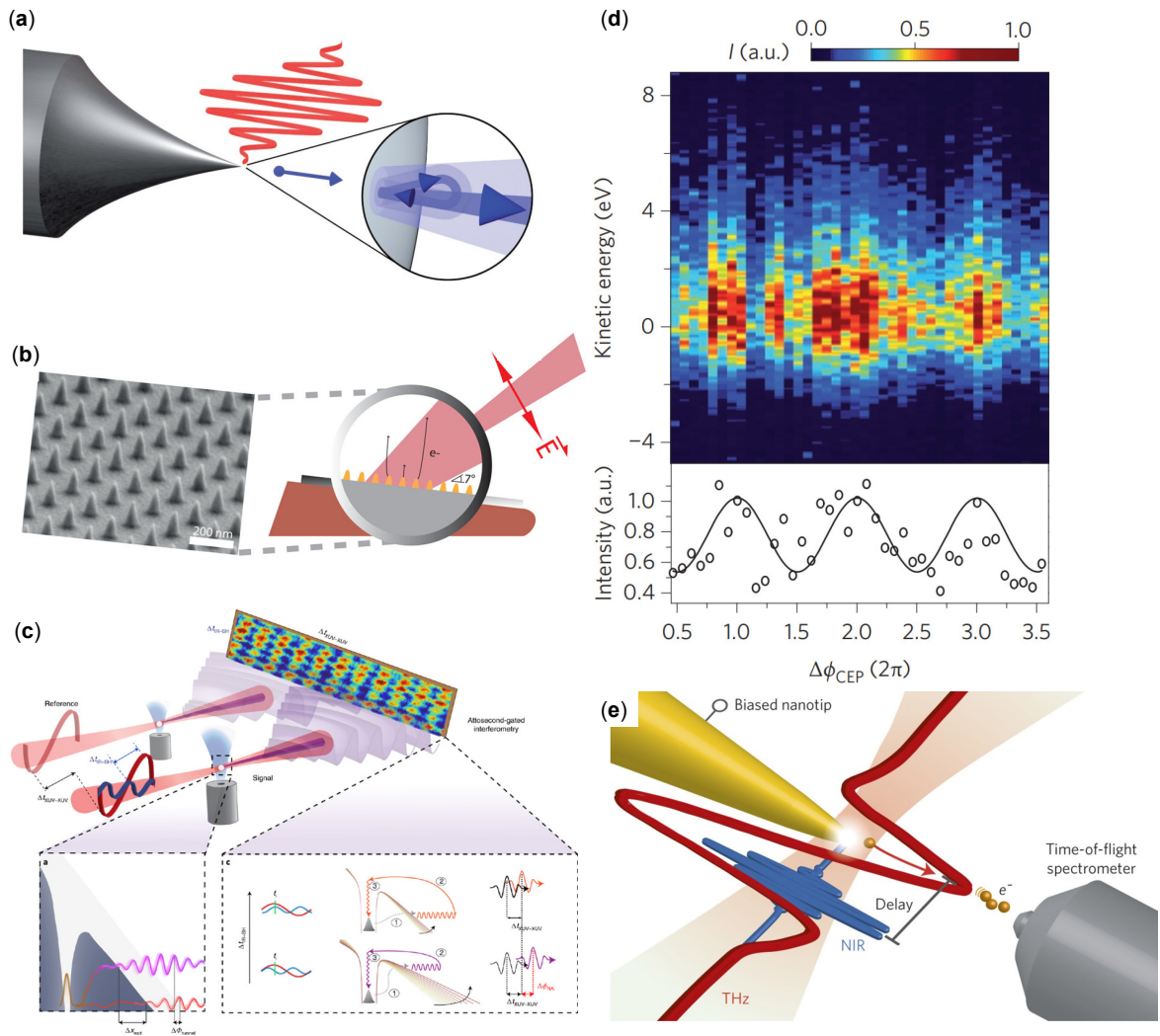

**Fig. 9 Enhanced photoemission and attosecond electron pulse generation from nanoscale emitters.** (**a**) The illustration of the photoemission process induced by a few-cycle incident pulse incident on a metal tip. Reprinted with permission from M. Krüger *et al.*, Nature **475**, 78–81(2011) [8]. Copyright 2011 Springer Nature. (**b**) Photoemission from gold tip array. Reprinted with permission from L. Brückner *et al.*, Nano Lett. 24, 5018–5023 (2024) [322]. Copyright 2024 American Chemical Society. (**c**) Two-color interferometry of the electron tunnelling dynamics. Reprinted with permission from O. Kneller *et al.*, Nat. Photonics **16**, 304–310 (2022) [10]. Copyright 2022 Springer Nature. (**d**) Electron yield and energy spectra from carrier-envelope phase variation. Reprinted with permission from B. Piglosiewicz *et al.*, Nat. Photonics **8**, 37–42 (2014) [323]. Copyright 2014 Springer Nature. (**e**) Control of nanotip photoemission with terahertz pulses. Reprinted with permission from L. Wimmer *et al.*, Nat. Phys. **10**, 432–436 (2014) [324]. Copyright 2014 Springer Nature.

### *4.2 Attosecond Electron Pulse Generation from Nanoscale Emitters*

The paramount advantage of strong-field interactions with nanometer-scale emitters lies in generating ultrashort electron pulses [140,141], reaching down to the attosecond scale [8,13].

The current record stands at 53±5 attoseconds, achieved from a single-cycle optical pulse (with a duration of about 1.9 femtoseconds) and a carrier frequency of 1.8 eV incident on a tungsten tip [13]. The physical process could be intuitively explained in classical and quantum physics aspects. From the classical perspective, when a strong pulse with only a few cycles strikes the metal tip, liberated electrons are driven back to the tip, undergo elastic re-scattering, and gain additional energy, as illustrated in Fig. 9b. From the quantum perspective, the strong electric field enhancement at the tips reduces the binding Coulomb barrier, facilitating tunnelling, as illustrated in Fig. 9c.

The photoemission process exhibits sensitivity to the carrier-envelope phase of the incident pulse. Although directly measuring the temporal profiles of the electron pulses presents challenges, the variations in the spectral interference patterns of the electrons suggest that the electron dynamics are indeed influenced by the carrier-envelope phase [8]. Carrier-envelope phase also affects other parameters such as electron yield and spectra of the electrons [323] (Fig. 9d), the current autocorrelation [325], etc. Readers interested in the development of the attosecond science are referred to the review [316,325–327].

The dynamical control of the electrons could be realized by dual-frequency excitation on nanotips [324]. The photoelectron emission by the higher-frequency light is gated and streaked by the lower-frequency light (Fig. 9e) [324]. Indeed, this is more often used to measure the spectro-temporal features of the electrons, known as streaking spectroscopy [328,329]. The two lights can have a controllable time delay [13] or differ in frequency [10,12,330], typically with one being a harmonic of the other. For example, in Ref. [13], a weak replica of the pump pulse is generated by a piezo state. Variation of the time delay permits the reconstruction of the tunnelled electron wavefunction. Similarly, by controlling the phase of the dual-frequency incident lights [10,12,330], one could measure the sub-optical-cycle emission/tunnelling dynamics, control the emission rate and the kinetic energy, as shown in Fig. 9e. Recently, it has been proposed that the electron's wavefunction in energy space could also be reconstructed by splitting one electron into a superposition state, with the spectrum of one-half shifted by interacting with light, and then allowing them to interfere [331].

Note that photon-induced near-field electron microscopy (PINEM), which will be discussed in Section 4.4, could also produce attosecond electron pulse trains [9,17,24,143–145] from free electrons. Single-cycle electron pulses could also be produced in PINEM when the incident light induces single-cycle modulation on the electrons [332].

## 4.3  *Electron Correlations and Quantum Statistics in Free Electron Beams*

Electron-electron correlations are investigated in different avenues [70,146,147,260,333], demonstrating the potential for heralded electron sources in the field of free-electron quantum optics.

Fermions would allow one particle to be in one quantum state at most. As a result, electrons from one tip source tend to anti-bunch from each other [334]. In addition to the exclusion principle, antibunching also results from the repulsive Coulomb repulsions between electrons [333]. Coulomb interactions in free electron beams are generally detrimental in ultrafast electron microscopy, particle accelerators, and free-electron lasers. To name a few, it could decrease the electron beam's brightness and spatial coherence [335], and result in spectral broadening [336].

Recently, the electron correlations from Coulomb repulsions, including the anti-bunching phenomenon, have been quantitatively characterized with event-based electron spectroscopy [146,147] (Fig. 10a). The investigation is based on photoemission from tip emitters [337] at two vastly different electron energy regimes, 200 keV [146] and 40 eV [147]. With event-based electron spectroscopy, the pair, triple, and quadruple states of free electrons by each laser pulse are recorded, revealing a strong correlation in kinetic energy (Fig. 10b). The anti-bunched electrons with sub-Poissonian distribution are also analyzed under different extraction voltage of the Schottky field emitter [146].

Without considering the high-order Coulomb-correlated pairs [146,147], in general, Poissonian statistics of the electron number by coherent excitation light can be expected [338]. The electron statistics by non-classical quantum light was first demonstrated in Ref. [260]. The emitted electrons follow the multi-photon processes induced by a bright squeezed vacuum light of peak intensities of $3 \times 10^{12}$ W cm$^{-2}$. This work recorded instances of electron bunching (Fig. 10c), identifying extreme statistical events where up to 65 electrons were emitted from a single light pulse, with an average emission rate of 0.27 electrons per pulse.

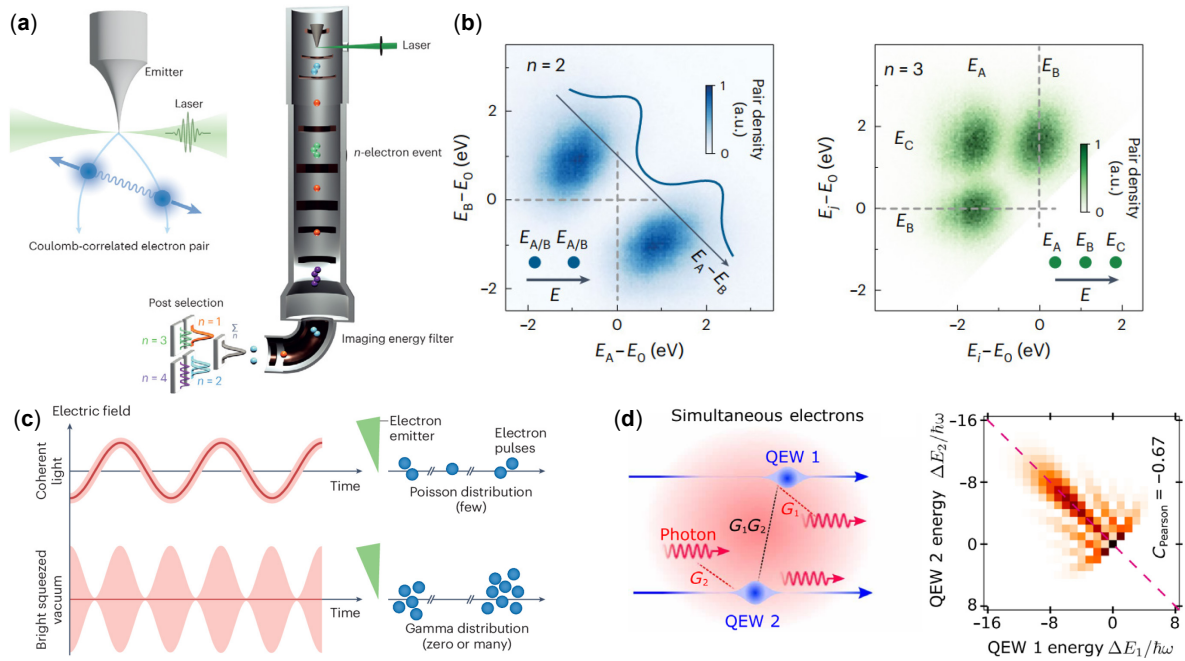

**Fig. 10 Free-electron statistics and coulomb correlation.** (**a**) Coulomb-correlated electron pairs are post-selected and measured. Reprinted with permission from J.W. Simonaitis *et al.*, Nat. Phys. **19**, 1382–1383 (2023) [339]. Copyright 2023 Springer Nature. (**b**) The energy correlation of n=2 and 3 electrons. Reprinted with permission from R. Haindl *et al.*, Nat. Phys. **19**, 1410–1417 (2023) [146]. Licensed by a Creative Commons (CC BY) license. (**c**) Poisson distribution and Gamma distribution of the electrons, which are excited by coherent and bright squeezed vacuum light. Reprinted with permission from A. Leitenstorfer *et al.*, Nat. Phys. **20**, 890–891 (2024) [142]. Copyright 2024 Springer Nature. (**d**) Strong linear energy correlation of two electrons interacting with photons simultaneously. Reprinted with permission from S. Kumar *et al.*, Sci. Adv. **10**, eadm9563 (2024) [70]. Licensed by a Creative Commons Attribution NonCommercial License 4.0 (CC BY-NC).

Except for inheriting quantum properties from the light, the emitted electrons are also affected by the tips' quantum properties. When the tip is small enough, approximately 1 nm at the apex, or is deposited by a single molecule, the emitted electrons are affected by single-molecule molecular orbitals [340]. Sub-Poissonian electron distribution could also be achieved through energy filtering, with the second-order correlation $g^{(2)} = 0.34$ measured in the experiment.

Electron correlations can also develop through momentum changes via photons [70]. This process relies on multiple electrons interacting with photons simultaneously, such that photons emitted by one electron are absorbed by another. This interaction results in strong correlations with a large Pearson correlation coefficient, which quantifies the degree of linear correlation between two quantities (Fig. 10d). To build a strong linear correlation, it is necessary to have a small number of initial photons and a strong coupling constant between free electrons and photons. This process has the potential to establish strong correlations among a large number of electrons, compared to those relying on Coulomb interaction.

*4.4 Quantum Electrodynamics through Light Interaction*

In Section 2.1, we discussed the quantum effects on electron radiation, induced by the electron recoil. The electron recoil from electron-light interactions can be investigated directly using electron energy loss spectroscopy (EELS). The readers interested in how EELS is used to analyze the optical response of structured materials are referred to the reviews [219,341]. The energy-momentum correlation could also be resolved in far-field electron scattering distribution [289].

The elastic scattering of electrons by light was first studied by Kapiza and Dirac in 1933, which is known as the Kapiza-Dirac effect [342]. They predicted that free electrons could be elastically scattered by the potential of standing-wave light, similar to electron diffraction by crystal lattices [120]. But the Kapiza-Dirac effect is a ponderomotive-type effect that relies on a strong-field laser. Recently, the Kapiza-Dirac effect from a standing wave pulse was experimentally demonstrated (Fig. 11g) [26]. Tuning the time delay of the two pulses generates a time-dependent diffraction pattern and can be used to retrieve the temporal dynamics of the electron wavefunctions. Readers interested in the Kapiza-Dirac effect are referred to the review [343].

Studies in recent decades found that evanescent light fields offer an efficient way to enhance electron-light interaction, generating a new field in electron microscopy: photon-induced near-field electron microscopy (PINEM) [14–16]. In PINEM, femtosecond photoelectrons are used as probes to detect the optical modes excited by a pump laser, as illustrated in Fig. 11a. The crucial point is synchronizing the femtosecond photoelectrons and the optical modes by dividing a light pulse into two parts [344]. One part of the light, after harmonic generation, excites femtosecond electron pulses from a photocathode, while the second part, passing through an optical delay, excites the optical modes of a nanomaterial. By controlling the optical delay between the two lights, PINEM can be used to measure the field distribution and temporal dynamics of surface plasmon polaritons [345] and photonic cavity modes [19,20] (Fig. 11b). Furthermore, the ultrashort electron pulse promises direct access to the spatiotemporal imaging of two-dimensional (2D) polariton [346] (Fig. 11c) and the nanoscale charge dynamics in materials [347].

What is more appealing is that PINEM can be used to study the quantum electrodynamics in the interaction with external light. The most fundamental dynamics are the electron energy loss and gain, which manifest in discrete peaks in the electron spectrum [14]. The discrete peaks represent the integer number of photons absorbed/emitted by free electrons. By controlling the laser intensity and interaction length, hundreds to thousands of photons could be absorbed/emitted by free electrons (Fig. 11d) [21,24]. To detect the discrete electron spectrum, one needs fine detection resolution and narrow energy bandwidth. For example, in Ref. [139], it is not sufficient to detect the quantum nature of the interaction, i.e., the discrete energy peaks, when the detection resolution is above 0.5 eV, and the energy bandwidth of the incident electron is above 1.5 eV.

Through the PINEM interaction, one can realize the quantum control of the free electron energy spectrum. The PINEM electron becomes an energy comb with multiple sidebands separated by the photon energy, as shown in Fig. 11d. As a result, in the spatiotemporal space, the electron evolves into a train of attosecond electron pulses (Fig. 11f) [9,17,24,143–145]. The attosecond electron pulses could find applications in attosecond imaging, which will be shown in Section 4.5. Apart from electron combs, the PINEM process could also be used to reduce the energy bandwidth of electrons [348] if the light has significantly longer wavelength.

Since the free electron interchanges an integer number of photon energies, we could define an energy ladder space for electrons [37,45]. In the energy ladder space, the electron distribution demonstrates multilevel Rabi oscillations as a function of drive field strength (Fig. 11e) [17,349]. Additionally, the evolution of electrons in the energy ladder space is guided by the quantum statistics of the photons (Fig. 11h) [22]. When the incident laser is coherent light, the electron undergoes a quantum walk, while when the incident light is a thermal light, the electron performs a classical walk.

PINEM electrons are not only inelastically scattered, i.e., gain/lose energy, but also be elastically scattered. Interacting with plasmon standing waves, the free electrons undergo both longitudinal and transverse recoils [38,289,291]. Unlike the Kapiza-Dirac effect, which is dominated by the $\mathbf{A}^2$ terms in the Hamiltonian, PINEM is mainly affected by the $\mathbf{P} \cdot \mathbf{A}$ term for semi- relativistic electrons [289,291], where $\mathbf{A}$ is the vector potential of the light and $\mathbf{P}$ is the momentum operator for electrons. However, under conditions of low-energy electrons, in the range of hundreds of electron volts and below, or in the presence of a strong external field [38], both the linear and quadratic terms of the vector potential become essential. Free electrons interacting with standing plasmon waves have been used to modulate the electrons' spatial profiles [269].

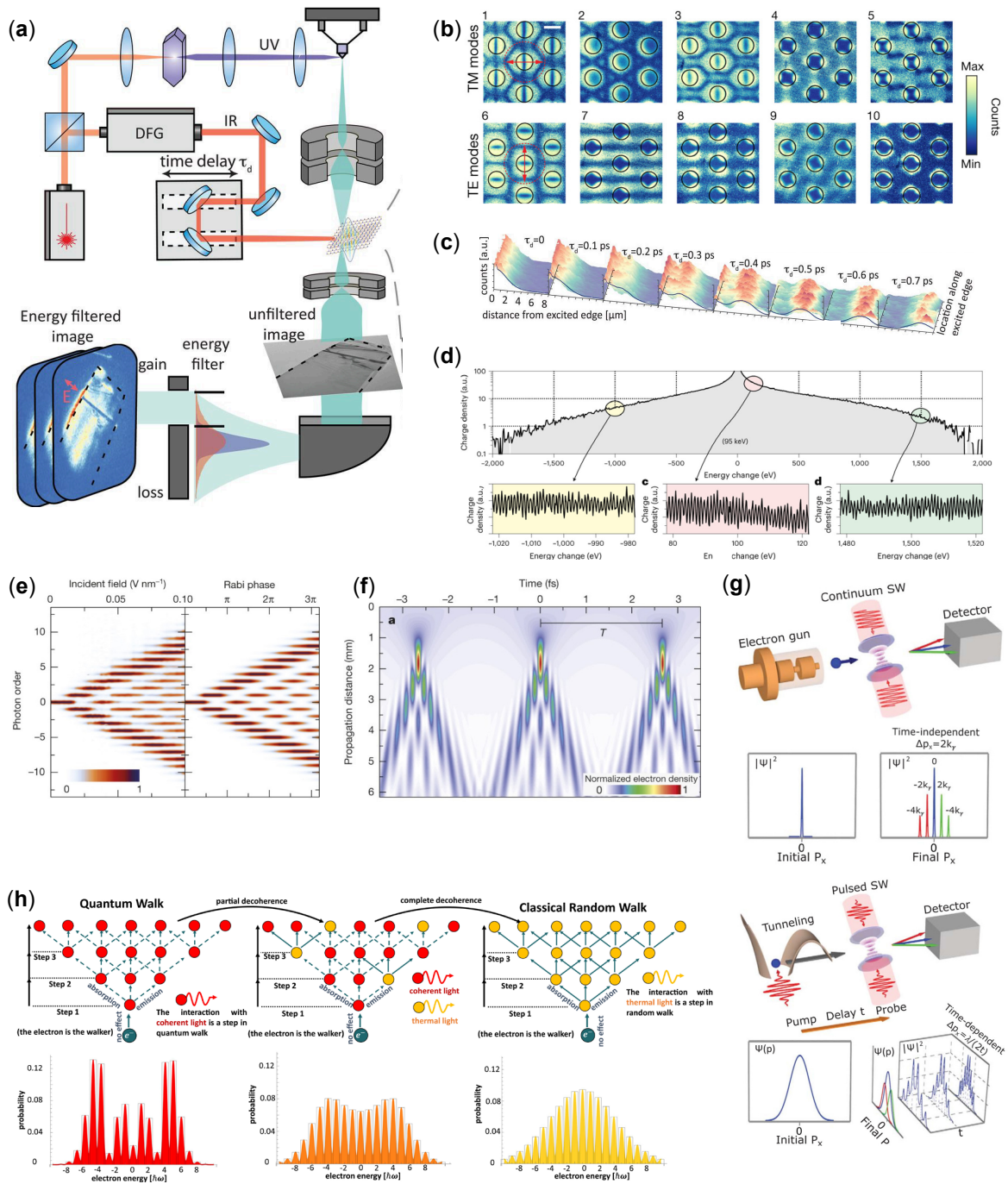

**Fig. 11 Quantum electrodynamics through light interaction.** (**a**) Illustration of PINEM. The femtosecond photoelectrons and the optical modes are synced. Reprinted with permission from Y. Kurman *et al.*, Science **372**, 1181–1186 (2021) [346]. Copyright 2021 AAAS. (**b**) Bloch modes of a photonic crystal measured by PINEM. Reprinted with permission from K. Wang *et al.*, Nature **582**, 50–54 (2020) [20]. Copyright 2020 Springer Nature. (**c**) The experimentally measured propagation dynamics of a phonon-polariton wavepacket. Reprinted with permission

from Y. Kurman *et al.*, Science **372**, 1181–1186 (2021) [346]. Copyright 2021 AAAS. (**d**) Measured electron energy spectrum of free electrons interacting with two-color lasers. Reprinted with permission from M. Tsarev *et al.*, Nat. Phys. **19**, 1350–1354 (2023) [24]. Copyright 2023 Springer Nature. (**e,f**) Rabi oscillation of the electron energy distribution as a function of the incident field, and development of an attosecond electron pulse train by PINEM electrons. Reprinted with permission from A. Feist *et al.*, Nature **521**, 200–203 (2015) [17]. Copyright 2015 Springer Nature. (**g**) The ultrafast Kapitza-Dirac effect from free electrons interacting with a pulsed standing light wave. Reprinted with permission from K. Lin *et al.*, Science **383**, 1467–1470 (2024) [26]. Copyright 2024 AAAS. (**h**) Quantum statistics of electrons in the energy ladder space imprinted from the quantum statistics of photons. Reprinted with permission from R. Dahan *et al.*, Science **373**, eabj7128 (2021) [22]. Copyright 2021 AAAS.

## *4.5 Quantum Measurements with PINEM-Shaped Electron Beams*

Mapping the spatial distribution [350] and temporal dynamics [144] of fields is always pursued to develop nanomaterials. PINEM (Section 4.4), which has been developed in the recent decade, has shown the ability to combine subwavelength spatial resolution and sub-optical-cycle temporal resolution. PINEM can detect near-field distributions with subwavelength spatial resolution due to the localized spot size of the electron beams [20,132,345,351]. The temporal resolution depends on the duration of electron bunches, which is generally in the sub-picosecond range. Therefore, PINEM can be used to detect ultrafast dynamics within sub-picosecond timescales [347].

It has been shown that PINEM could produce attosecond electron trains [9,17,24,143,144,262,352]. Therefore, recently, setups using the PINEM electrons to detect temporal dynamics with sub-optical-cycle resolution have been developed [144,353,354]. When the pulse duration is sub-optical-cycle, and the pulses are separated by one optical-cycle, Refs. [352,353] demonstrated the attosecond optical field movies of surface plasmon polaritons (Fig. 12a). Apart from using attosecond electrons, sub-optical-cycle dynamics of optical, electronic, or structural processes can also be observed through the interference of electron beams modulated by a sample and amplitude or phase-controllable light interaction [355–357], which is named as free-electron homodyne detection (Fig. 12b).

Among the models that characterize the quantum features of materials, a two-level system describing the bound electrons is generally exploited [118,43,42]. The full quantum analyses show that energy-modulated electron wavefunctions could facilitate quantum measurements in several aspects [47,48,52,301,358,359]. First, the contrast of the EELS peaks corresponding to the coherent interactions is enhanced for weak free-electron–bound-electron interaction.

Second, the shaped electrons could measure the qubit state of the two-level system and the decoherence time. Third, the fast emission rate in superradiation could be detected via the EELS detection of the shaped electrons.

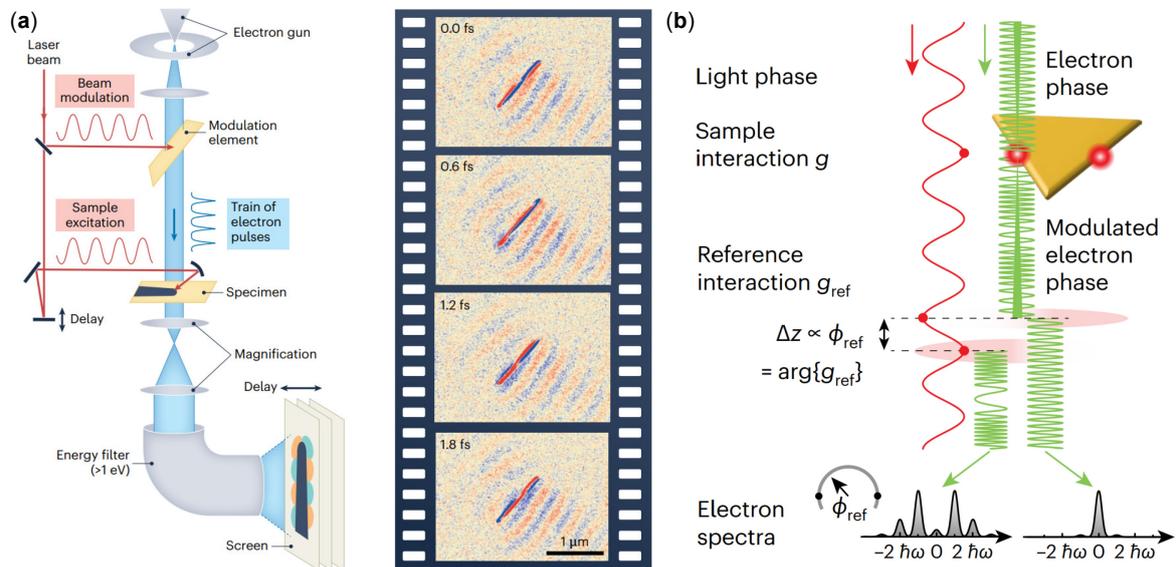

**Fig. 12 Attosecond electron microscopy.** (**a**) The measured plasmonic field within one optical cycle from an attosecond electron microscopy. Reprinted with permission from D. Nabben *et al.*, Nature **619**, 63–67 (2023) [353]. Copyright 2023 Springer Nature. (**b**) Free-electron homodyne detection. Reprinted with permission from J.H. Gaida *et al.*, Nat. Photonics **18**, 509–515 (2024) [356]. Licensed by a Creative Commons Attribution 4.0 International License.

# 5 High-Harmonic Generation of Extreme Ultraviolet and X-rays

## 5.1 High-Harmonic Generation as Extreme Ultraviolet Light Sources

The first experiments in high-harmonic generation (HHG) [360–362] showed that the gas of atoms strongly driven by the electromagnetic field can emit multiples of frequencies of the driving field. The HHG in atoms could produce more than twenty harmonics, reaching extreme ultraviolet and X-ray frequency ranges. Therefore, HHG stands as a viable and compact source of extreme ultraviolet light. In the HHG process, the emitted harmonics are coherent with each other, resulting in the formation of short attosecond pulses [363]. Soon after the discovery of HHG, the attosecond pulses from HHG were experimentally demonstrated [364–366]. These works formed the basis of the attosecond science [316,367].

The theoretical model of the HHG process in a gas was built a few years after the first experiments [368]. The first model [368] considered the HHG process and the electron dynamics in three steps: (i) the electron tunnels out from the atomic potential suppressed by the intense driving field, (ii) the electron is consequently accelerated in the continuum by the driving field, and (iii) the electron returns to the ion and recombine, emitting a high frequency pulse.

A quantitative understanding of HHG was provided by the highly successful semi-classical theory [369], which describes the electron quantum mechanically while still treating the driving and emitted fields classically. This model considers the time-dependent Schrodinger equation for the electrons [370,371]. This time-dependent Schrodinger equation is usually numerically solved for one-dimensional (1D) cases, even though the full three-dimensional (3D) case can also be numerically solved. The time-dependent Schrodinger equation for the bound electron inside the atom in 1D is the following:

$$i\hbar \frac{\partial |\phi_g(t)\rangle}{\partial t} = \left( \frac{\hat{p}^2}{2m_e} + V(\hat{x}) - \hat{d}E_c(t) \right) |\phi_g(t)\rangle, \qquad (4)$$

where $\hat{d} = e\hat{x}$ is the dipole operator, $\hat{x}, \hat{p}$ are the position and momentum operators, $E_c(t)$ is the classical electromagnetic field, $V(\hat{x})$ is the atomic potential. The initial condition is $|\phi_g(t = 0)\rangle = |g\rangle$, where $|g\rangle$ is the ground state of the atom. Then, according to the semi-classical theory, the only parameter that defines the process is the average dipole moment which can be calculated as $d_{gg}(t) = e\langle\phi_g(t)|\hat{x}|\phi_g(t)\rangle$. All the characteristics of the emission should be found by solving the classical electrodynamic equation for the time-dependent dipole $d_{gg}(t)$, as for the example described in [158]. For example, the spectrum is described by the Larmor formula:

$$\frac{d\varepsilon}{d\omega} = \frac{\omega^4 |d_{gg}(\omega)|^2}{6\pi^2 \varepsilon_0 c^3}, \tag{5}$$

where $d_{gg}(\omega)$ is the Fourier transform of $d_{gg}(t)$, i.e., $d_{gg}(\omega) = \int_{-\infty}^{+\infty} d_{gg}(t) e^{i\omega t} dt$.

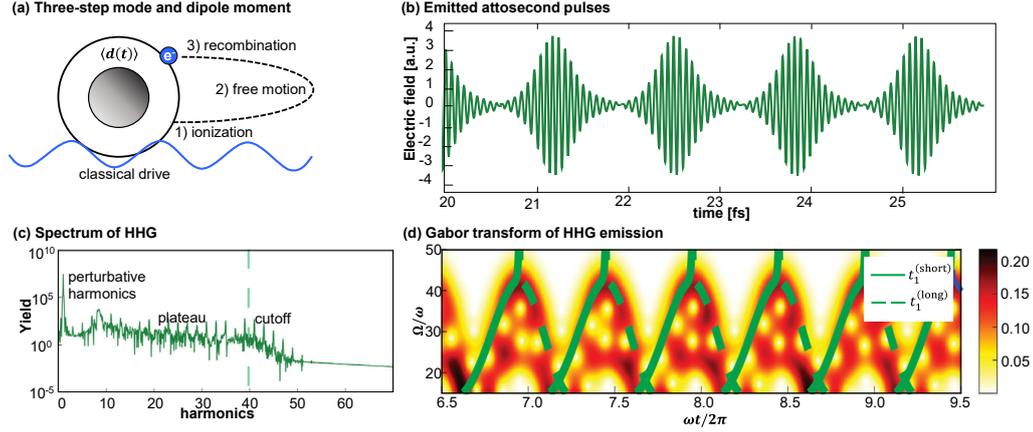

**Fig. 13 Classical theory of high-harmonic generation (HHG).** (**a**) The only parameter from the semi-classical theory is the average dipole moment $d_{gg}(t)$. All the other parameters can be found by solving classical electrodynamic equations for time-dependent $d_{gg}(t)$. (**b**) The example of attosecond pulses emitted during the HHG progress. (**c**) A typical HHG spectrum exhibits perturbative harmonics, a plateau with almost uniform intensities of harmonics, and a cutoff where the intensity of harmonics decreases exponentially. (**d**) The windowed-Fourier transform (Gabor) of time-dependent dipole moment. (b)-(d) are reprinted with permission from M. Even Tzur *et al.*, Nat. Photonics **17**, 501–509 (2023) [372]. Copyright 2023 Springer Nature.

*5.2 Nano-Optics in High-Harmonic Generation*

The field of nano-optics in HHG is rapidly growing. The early efforts to observe plasmon enhancement of HHG in the gases [373,374] faced challenges and yielded limited success. The observation of the HHG in solid bulk materials [375–380] gave new life for nano-optics in HHG. The solid structures can be designed with the appropriate morphology using different fabrication methods up to the nanometer level. Such nanostructures and structured surfaces can be used to enhance HHG yield, reduce the intensity threshold, and also focus the emitted X-ray and extreme ultraviolet light to a small spot, which is very complicated to do by other means.

Solid-state HHG was observed in ZnO crystal, starting the field of solid-state HHG [375]. Since then, HHG has been observed in multiple solid-state materials, including bulk crystals [375–380], amorphous solids [381], two-dimensional (2D) materials [382,383], and metamaterials [384].

One of the first demonstration of plasmonic enhancement [385] was shown in the array of bow-tie nanostructures. Later, HHG was shown in gold-coated sapphire cones [386], and in gold monopole nano-antennas [387,388], with the harmonic emission enhanced by one to several orders of magnitude. The HHG was also observed in 2D materials [382,383], enabling the production of highly localized harmonics. Ref. [389] demonstrated that conical ZnO nanostructures can be used to localize the driving field, and a Fresnel zone plate target in silicon can effectively focus the emitted light. The focusing of high harmonics was also done using nanostructured elements on the surface [390] as well as with vacuum guiding [150].

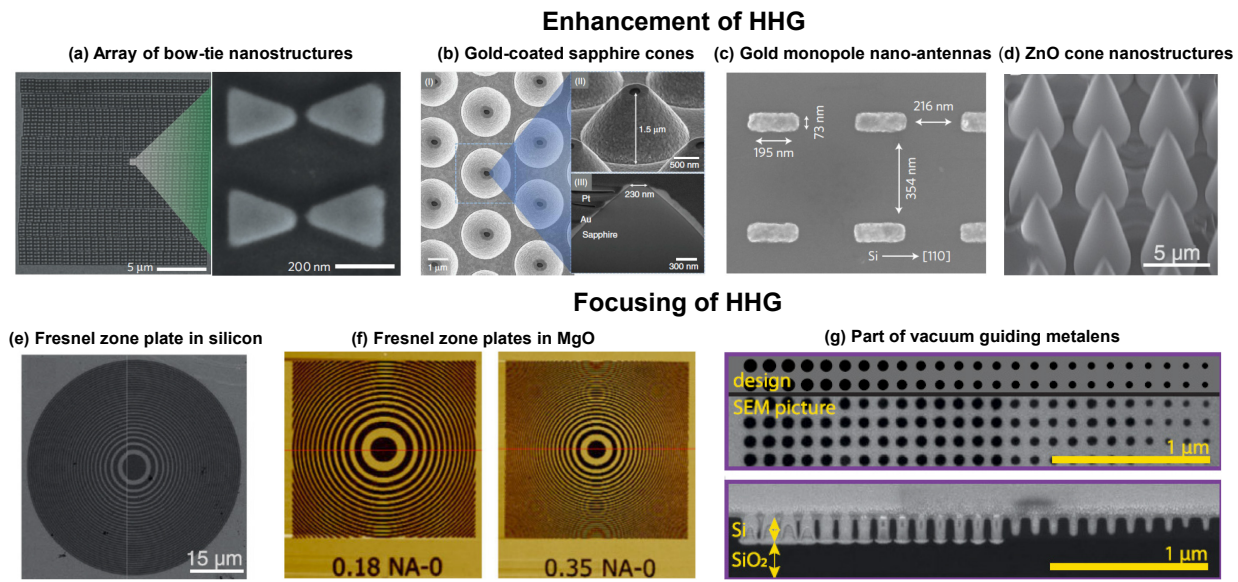

**Fig. 14 Nanostructures for enhancing and focusing HHG.** (**a**) An array of bow-tie nanostructures used for plasmon-enhanced HHG with a close-up view. Reprinted with permission from M. Sivis *et al.*, Nat. Phys. **9**, 304–309 (2013) [385]. Copyright 2013 Springer Nature. (**b**) An array of gold-coated sapphire cones that provide plasmon-enhanced HHG. Reprinted with permission from S. Han *et al.*, Nat. Commun. **7**, 13105 (2016) [386]. Licensed by a Creative Commons Attribution 4.0 International License. (**c**) An array of gold monopole nano-antennas on Si crystal that provide plasmon-enhanced HHG. Reprinted with permission from G. Vampa *et al.*, Nat. Phys. **13**, 659–662 (2017) [387]. Copyright 2017 Springer Nature. (**d**) ZnO cone nanostructures on ZnO crystal that provide enhanced HHG. Reprinted with permission from M. Sivis *et al.*, Science **357**, 303–306 (2017) [389]. Copyright 2017 AAAS. (**e**), (**f**), (**g**) Fresnel zone plate fabricated in silicon and MgO crystals to focus HHG emission. Reprinted with permission from M. Sivis *et al.*, Science **357**, 303–306 (2017) [389], M.

Ossiander *et al.*, Science **380**, 59–63 (2023) [150], and A. Korobenko *et al.*, Phys. Rev. X **12**, 041036 (2022) [390]. Copyright 2017 AAAS and 2023 AAAS and licensed by a Creative Commons Attribution 4.0 International license.

The control, enhancement, and focusing of HHG using nanostructures make the exploration of quantum effects in HHG more feasible, as discussed in the next section.

## *5.3 The Quantum Theory of High-Harmonic Generation*

### *5.3.1 Introduction*

Until recently, the theoretical framework of HHG was conventionally approached from semi-classical perspective. As elaborated in Section 5.1, the average time-dependent dipole moment of a single atom $d_{\text{gg}}(t)$, which can be calculated using time-dependent Schrodinger equation Eq. (4), defines all the characteristics of the emission through the classical electrodynamical equations [158]. Recently, the theory of HHG with the quantized electromagnetic field started to be considered in various works [372,391–404]. Here, we will consider and discuss the quantum theory of HHG in logical order but not in historical order. All the theory of quantum HHG from the gas of atoms can be summarized by a single Schrodinger equation:

$$i\hbar \frac{\partial |\Psi(t)\rangle}{\partial t} = \hat{H}|\Psi(t)\rangle, \tag{6}$$

where the Hamiltonian is $\hat{H} = \sum_{i=1}^{N} \hat{H}_{\text{A}}^{(i)} + \hat{H}_{\text{F}} - e\sum_{i=1}^{N} \mathbf{d}_i \cdot \hat{\mathbf{E}}(x_i)$. $\hat{H}_{\text{A}}^{(i)}$ is the Hamiltonian of the *i*-th atom, $\hat{H}_{\text{F}}$ is the free field Hamiltonian, $\hat{\mathbf{E}}(x_i) = i\sum_{k\sigma}\sqrt{\frac{\hbar\omega}{2V\varepsilon_0}}\left[\boldsymbol{\varepsilon}_{k\sigma}\hat{a}_{k\sigma}e^{ikx_i} - \boldsymbol{\varepsilon}_{k\sigma}^*\hat{a}_{k\sigma}^\dagger e^{-ikx_i}\right]$ is the quantized electric field, where $k$ and $\sigma$ represent the field mode and the polarization, $\hbar$ is the reduced Plank constant, $\omega$ is the angular frequency, $V$ is the quantization volume, $\varepsilon_0$ is the vacuum permittivity, $\boldsymbol{\varepsilon}_{k\sigma}$ is the polarization vector, $a_{k\sigma}^\dagger$ and $a_{k\sigma}$ are the photonic creation and annihilation operators. The behavior of the solution to Eq. (6) is strongly dependent on the initial conditions

$$|\Psi(t=0)\rangle = |\psi_{\text{light}}(0)\rangle \otimes |\Phi_{\text{atoms}}(0)\rangle, \tag{7}$$

where $|\psi_{\text{light}}(0)\rangle$ is the quantum state of the driving field, $|\Phi_{\text{atoms}}(0)\rangle$ is the initial state of the atoms. The primary objective of the quantum theory of HHG was firstly to find the classical regime of HHG, when the quantum theory coincides with the semi-classical theory, and secondly to find the regimes when the HHG emission can show quantum features. As was shown in the series of works [372,396,402,403], the properties of the high-harmonic emission arise from the initial state of the system described by Eq. (7). In the following subsections, we will discuss the works [396,402], which demonstrated that when the initial light state $|\psi_{\text{light}}(0)\rangle$ is a coherent state and the initial atomic state $|\Phi_{\text{atoms}}(0)\rangle$ is in the ground state, it corresponds to the classical regime of HHG. However, even with these initial conditions, as shown theoretically and demonstrated experimentally in the series of works [398,400,401] using post-selection on the emitted harmonics, it is possible to create the quantum kitten states for the fundamental driving mode. Then, according to the results of [402], we will discuss how many-body correlations in the atomic system $|\Phi_{\text{atoms}}(0)\rangle$ can transfer to the emission and lead to the quantum features in X-ray and extreme ultraviolet ray emission. As shown in [396], the quantum regime most commonly used is HHG from a single atom. In this case, it was shown that the emission is entangled with a single atom and the spectrum of the emission is completely different from the classical framework. Another way to get quantum features in the emission is to use the quantum driving field as shown in [403]. Different regimes of HHG are summarized in Fig. 15.

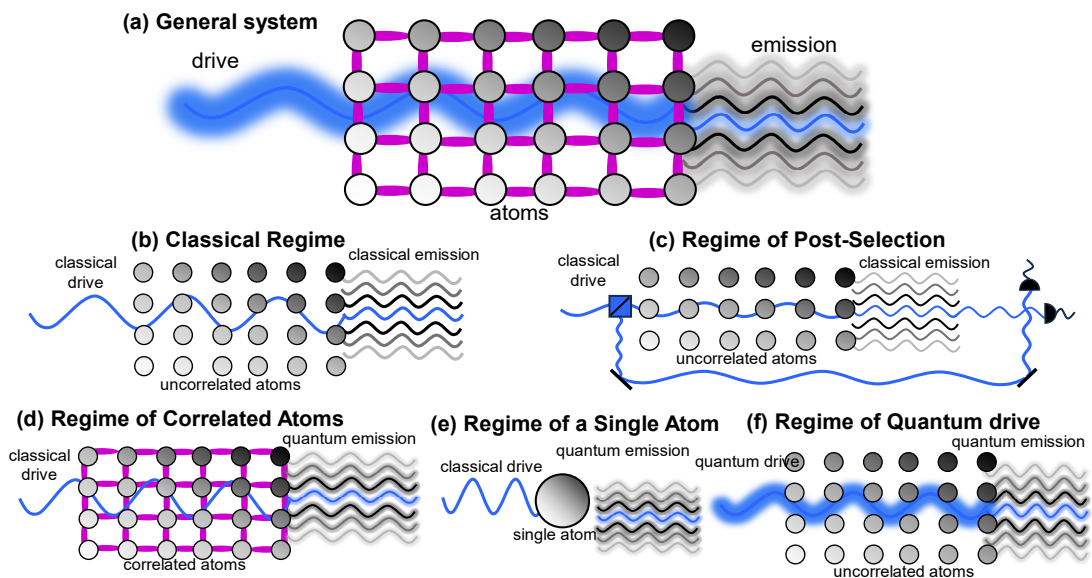

**Fig. 15 All the regimes of high-harmonic generation (HHG)**. **(a)** The general HHG system contains quantized driving fields, the atomic system that may have correlations between different atoms, and the emission that may have quantum features. **(b)** The classical regime of HHG corresponds to the classical (i.e., coherent state) driving field and uncorrelated many-body system. Then the emission is a coherent state without any quantum features. **(c)** Using post-selection schemes in the classical regime, it is possible to create from classical emission Schrodinger kitten states. **(d)** If the driving field is classical but the atoms are correlated, the emission can get quantum features transferred from the atomic system. **(e)** The regime of a single atom is the particular case of (d). **(f)** Uncorrelated initial states of atoms and the driving field with quantum features also lead to the quantum features in the emitted high harmonics.

### 5.3.2 Classical Regime of High-Harmonic Generation

Firstly, let us consider the classical regime of HHG, which corresponds to the semi-classical model described in Section 5.2. The correspondence between the semi-classical theory and the quantum theory was first schematically shown in [396]. In the recent paper [402], this correspondence was shown more explicitly. The initial condition in HHG that leads to the semi-classical theory corresponds to the classical coherent driving field and all atoms being in the ground state:

$$|\Psi(t=0)\rangle = |\alpha_d\rangle \otimes |ggg\ldots g\rangle, \tag{8}$$

where the state of light is the coherent state $|\alpha_d\rangle$ and all the atoms are initially in the ground state. It is important to note that this initial condition was explicitly correct for all the experiments done up to now. Here, we should emphasize that the semi-classical regime works only for many-body cases, when the number of atoms $N_{atoms}$ is large, $N_{atoms} \gg 1$. In the following subsections, we will show that even if the initial condition is the same, but the number of atoms is small (e.g., a single atom), then the semi-classical theory does not work and the emission has strong quantum features [396]. If the initial condition is described by Eq. (8), then it was shown that the emission also depends only on a single parameter $d_{gg}(t)$, similar to the semi-classical theory. As shown in [402], the emitted light is coherent and is unentangled from the atoms:

$$|\Psi(t)\rangle = \prod_i \left|\phi_g^{(i)}(t)\right\rangle \otimes |\alpha_d + \chi_{k_0\sigma_0}\rangle \prod_{(k\sigma)\neq(k_0\sigma_0)} |\chi_{k\sigma}\rangle, \tag{9}$$

where $|\phi_g(t)\rangle$ is the classical solution for the electron dynamics described in Eq. (4). All the emitted harmonics $(k, \sigma)$ are coherent states including the fundamental mode $(k_0, \sigma_0)$ $|\alpha_d + \chi_{k_0\sigma_0}\rangle$. Each mode is characterized by the wavevector $k$ and the polarization $\sigma$. The amplitude of the coherent emitted harmonics are $\chi_{k\sigma} = -i(2\omega V\varepsilon_0\hbar)^{-\frac{1}{2}}d_{gg}(\omega)\sum_i e^{ikr_i}$, where $d_{gg}(\omega) = \int_\infty^\infty e^{i\omega t}d_{gg}(t)dt$. Thus, all the emission is coherent, there are no quantum features, and the intensity of the emission coincides with classical theory Eq. (4). Interestingly, the quantum theory also includes the phase-matching condition that is described by the factor $\sum_i e^{ikr_i}$ inside the amplitude of the coherent states. The emission in the classical regime is shown in Fig. 17a.

### 5.3.3 Generation of Kitten States Using HHG

The modification of the fundamental mode of the driving field during HHG process was firstly considered in Refs. [393,398]. As Eq. (9) shows, the fundamental mode during the emission is the coherent state $|\alpha_d + \chi_{k_0\sigma_0}\rangle$ which is shifted from the initial driving field by $\delta\alpha = \chi_{k_0\sigma_0}$. Thus, the light is:

$$|\psi_{\text{light}}(t)\rangle = |\alpha_d + \chi_{k_0\sigma_0}\rangle \prod_{(k\sigma)\neq(k_0\sigma_0)} |\chi_{k\sigma}\rangle. \tag{10}$$

In the series of works [398,400,401], it was suggested and experimentally shown how to generate kitten state in the fundamental mode using the post-selection scheme (Fig. 16). The post-selection is done on whether any light was emitted:

$$\begin{aligned}|\tilde{\psi}_{\text{light}}(t)\rangle &= \left(1 - |\psi_{\text{light}}(0)\rangle\langle\psi_{\text{light}}(0)|\right)|\psi_{\text{light}}(t)\rangle \\ &= |\alpha_d + \chi_{k_0\sigma_0}\rangle \prod_{(k\sigma)\neq(k_0\sigma_0)} |\chi_{k\sigma}\rangle \\ &\quad - \xi_d|\alpha_d\rangle\,\xi_{\text{HHG}} \prod_{(k\sigma)\neq(k_0\sigma_0)} |0_{k\sigma}\rangle, \end{aligned} \tag{11}$$

where $\xi_{\text{HHG}} = \prod_{(k\sigma)\neq(k_0\sigma_0)}\langle 0_{k\sigma}|\chi_{k\sigma}\rangle$ and $\xi_d = \langle\alpha_d|\alpha_d + \chi_{k_0\sigma_0}\rangle$. Then, all the harmonics, except the fundamental one, should be measured using photodetectors. This corresponds to the projection of the state onto the high-harmonic emission $\prod_{(k\sigma)\neq(k_0\sigma_0)}|\chi_{k\sigma}\rangle$. After this projection, we get the kitten state:

$$\prod_{(k\sigma)\neq(k_0\sigma_0)}\langle\chi_{k\sigma}|\tilde{\psi}_{\text{light}}(t)\rangle = |\alpha_d + \chi_{k_0\sigma_0}\rangle - \xi_d|\xi_{\text{HHG}}|^2|\alpha_d\rangle. \tag{12}$$

This post-selection scheme is shown in Fig. 16a, and the experimentally generated kitten state similar to Eq. (10) is shown in Fig. 16b. These results are adopted from Ref. [398].

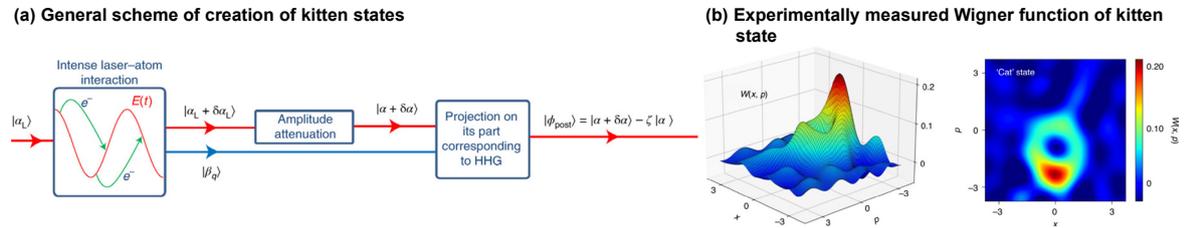

**Fig. 16 Generation of kitten states using HHG. (a)** The post-selection and projection of HHG emission scheme leads to the generation of kitten states in the fundamental mode. **(b)** Experimentally obtained kitten state. Reprinted with permission from M. Lewenstein *et al.*, Nat. Phys. **17**, 1104–1108 (2021) [398]. Copyright 2021 Springer Nature.

*5.3.4 Correlated Atoms*

In the previous two subsections, we considered uncorrelated initial state of the atoms: $|\Phi_{\text{atoms}}(0)\rangle = |ggg \ldots g\rangle$. However, the HHG can be modified if the initial matter state is correlated. This enables the measurement of matter correlations in solids using HHG, as demonstrated both theoretically and experimentally in Refs. [405,406]. Furthermore, matter correlations can be used to create quantum light, as theoretically shown in [402]. Let us restrict the initial condition to contain only ground $|g\rangle$ and first excited state $|e\rangle$. Then the emission will depend not only on $d_{gg}(t) = \langle g(t)|d|g(t)\rangle$, but also on the $d_{ge}(t) = \langle g(t)|d|e(t)\rangle$ and $d_{ee}(t) = \langle e(t)|d|e(t)\rangle$, where $|g(t)\rangle$ and $|e(t)\rangle$ are the solution of Eq. (4) for initial condition of $|g(0)\rangle$ and $|e(0)\rangle$ being the ground and the first excited state of the system. Fig. 17a,b shows the Wigner function of the emitted harmonics in the case (a) of uncorrelated atoms and in the case (b) of the squeezed initial state of atoms (b). It shows that the quantum state of the emitted light is not coherent light state as predicted by the classical theory and has quantum features as described in detail in [402].

The correlated system of atoms can be prepared in different ways. In the case of interaction-induced correlations, relevant systems include strongly interacting gases of atoms or molecules, whose interactions can be enhanced by trapping them in optical cavities. Such systems can realize the infinite-range (all-to-all) interactions [407]. Moreover, finite-range spin-spin interactions, which are ubiquitous in many hot-vapor [408] and cold-atom systems [409], can also realize spin-squeezed states and thus should also show variants of the quantum features we proposed here. Another promising platform for exploring these ideas is Rydberg atoms [410], offering the possibility of controllably generating correlated many-body atomic states [411,412].

*5.3.5 Single Atom Regime*

In section 5.3.1 we showed that the classical driving field and the atoms initially in the ground state lead to the classical emission. However, it is correct if there are multiple atoms. As it was shown in [396], the emission has strong quantum features if the atomic system has only a single atom. In this case, the emission does not have sharp harmonics and is more intense in a few orders of magnitude compared to the normalized many-body HHG (Fig. 17c). The quantum features of the emission are also completely different from the emission in the classical case. The emission during the HHG is entangled with the atom. Furthermore, the emission depends not only on the expectation value of the dipole moment $d_{gg}(t) =$ $\langle g(t)|d|g(t)\rangle$, but also on higher-order dipole correlations such as $\langle g(t)|d^2|g(t)\rangle$, $\langle g(t)|d^3|g(t)\rangle$, and $\langle g(t)|d^4|g(t)\rangle$. This demonstrates that, in the regime of a single atom, the entire electron dynamics—not just the average dipole moment—is imprinted on the emission. The main issue is that the emission from a single atom cannot be observed experimentally. The way to avoid it is to use nanostructures that can potentially enhance the emission and make this effect more feasible.

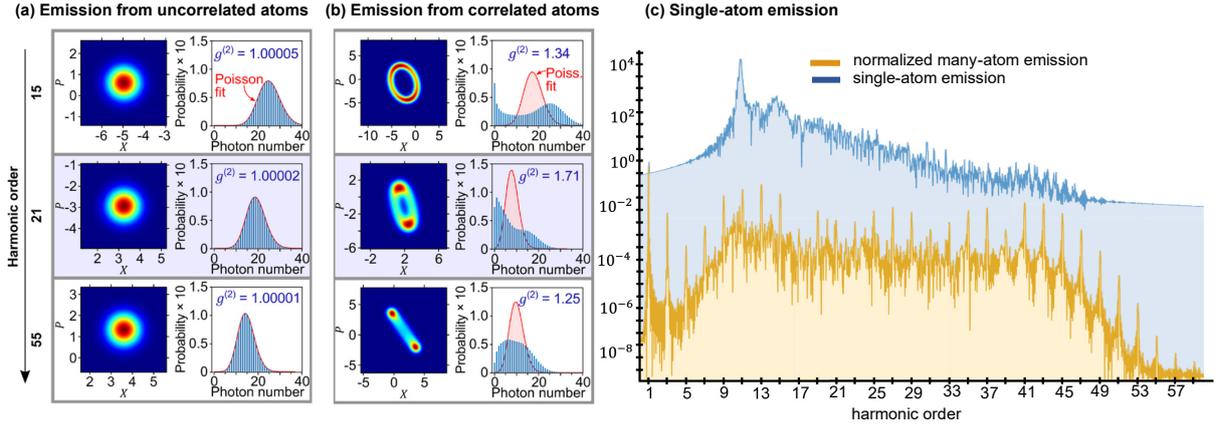

**Fig. 17 HHG for different atomic states.** (**a**) The Wigner function of the emitted light for 15$^{th}$, 21$^{st}$ and 55$^{th}$ harmonic in the case of initially uncorrelated atoms. (**b**) The Wigner function of the emitted light for 15$^{th}$, 21$^{st}$ and 55$^{th}$ harmonic in the case of initially correlated (squeezed) atoms. The Wigner functions are completely different from the classical regime considered in (a). (**c**) The comparison of normalized intensities (i.e., divided by the number of atoms squared) in a single-atom HHG and the classical regime of HHG. Reprinted with permission from A. Gorlach *et al.*, Nat. Commun. **11**, 4598 (2020) [396]. Licensed by a Creative Commons Attribution 4.0 International License.

### 5.3.6 *High-Harmonic Generation Driven by Quantum Lights*

All the previous subsections considered classical coherent driving fields. Let us now consider how the quantum features of the driving fields can affect high-harmonic emission. Refs. [372,403] investigated the emission from the atoms driven by quantum lights. It was shown that the spectrum as well as other quantum properties of the high-harmonic emission significantly depends on the quantum state of the driving field. The arbitrary quantum state of the driving field can be described by the Husimi function $Q(\mathcal{E}_\alpha)$ [413], which is the quasidistribution of the complex amplitude of the electric field $\mathcal{E}_\alpha$. Then the emission spectrum is:

$$\frac{d\varepsilon}{d\omega} = \frac{\omega^4}{6\pi^2 c^3 \varepsilon_0} \int d^2\mathcal{E}_\alpha \, Q(\mathcal{E}_\alpha) |d_\alpha(\omega)|^2, \qquad (13)$$

where $d_\alpha(t) = \langle g_\alpha(t)|d|g_\alpha(t)\rangle$ is the solution of Eq. (4) for the classical electromagnetic field with the complex amplitude $\mathcal{E}_\alpha$. According to Eq. (13), the shape of the spectrum is dependent on the Husimi distribution $Q(\mathcal{E}_\alpha)$ (Fig. 18a). While the coherent and Fock light states lead to approximately the same HHG spectrum corresponding to their narrow Husimi distributions, thermal and bright squeezed vacuum (BSV) states generate much higher harmonics for the same intensity. This feature shows that for HHG, broad Husimi distributions (e.g., thermal and BSV) are preferable over narrow ones (e.g., coherent and Fock) for the generation of high frequencies. This preference is surprising given their vanishing (average) electric field amplitudes, which are usually required to explain the dynamics in the three-step model [414]. We note that the predicted spectral broadening due to thermal and BSV drives is a microscopic effect that arises at the level of each single atom. This prediction can then be combined with the known macroscopic effects in HHG like phase-matching discussed in Section 5.3.1.

Importantly, even small deviations in quantum states significantly change not only the spectrum but also the temporal features of the emitted quantum pulses. As shown in [372], the temporal shape of the attosecond pulses changed significantly with changing the squeezing of the driving state (Fig. 18b). Fig. 18b shows that the coherent, amplitude, and phase squeezed driving lights (in the figure marked as dark red, red and orange respectively) lead to different emitted attosecond pulses. A phase-squeezed driving light results in horn-shaped attosecond pulses that significantly differ from standard attosecond pulses due to increased amplitude uncertainty. In contrast, an amplitude-squeezed driving light results in pulse-to-pulse envelope fluctuations for the attosecond pulses, and anharmonic (non-integer) spectral peaks in the HHG spectrum due to an increased phase uncertainty. Furthermore, the squeezed driving light leads to the squeezed high harmonics as shown in [415].

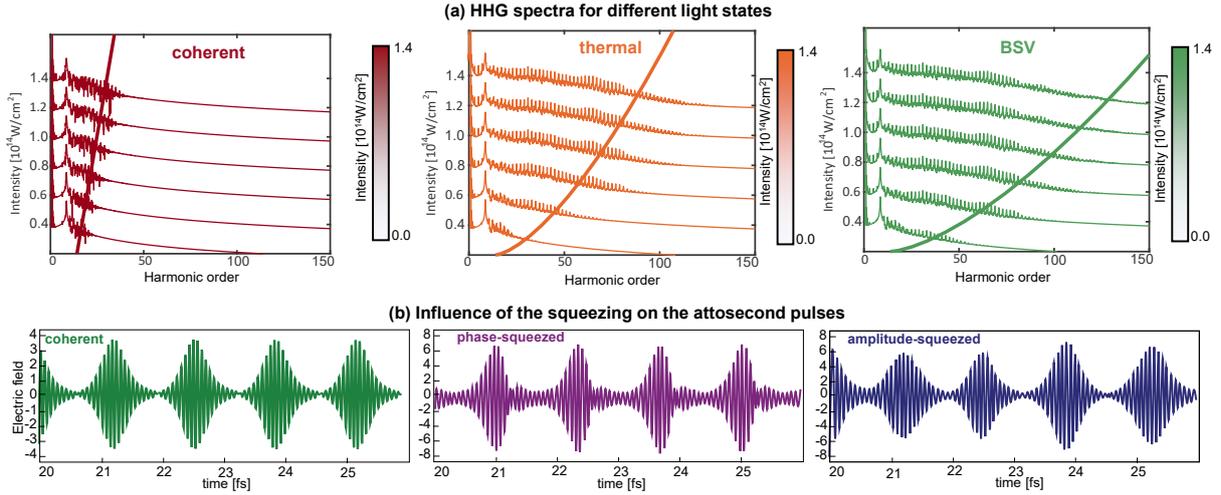

**Fig. 18 The emission from the atoms driven by quantum lights.** (**a**) Numerical simulations of high harmonic spectra according to Eq. (13) for several different intensities and the cutoff solid lines on the same plot for coherent, thermal, and BSV light states. Reprinted with permission from A. Gorlach *et al.*, Nat. Phys. **19**, 1689–1696 (2023) [403]. Copyright 2023 Springer Nature. (**b**) Exemplary attosecond pulses and HHG spectra, driven by coherent, phase-squeezed and amplitude-squeezed light with the same intensities. Reprinted with permission from M. Even Tzur *et al.*, Nat. Photonics **17**, 501–509 (2023) [372]. Copyright 2023 Springer Nature.

These predictions are within reach of current experimental capabilities in HHG, with several of them having been successfully demonstrated in experiments recently [404]. The spectrum behavior Eq. (13) for thermal and BSV light can be observed using classical measurements, such as conventional spectrometry of HHG. Classical femtosecond laser pulses of with energy as low as 200 nJ and a pulse duration of 30 fs were shown to be sufficient for driving HHG in optical fibers [416]. Current pulses of BSV generated through spontaneous parametric down-conversion (SPDC) [416–418] reach approximately the same intensity. Examples include 18 ps BSV pulses with energy of 10 μJ [417] and shorter femtosecond BSV pulses with energy of 350 nJ [418]. Intriguingly, even more intense pulses of non-coherent light can be generated by the amplification of weaker BSV pulses in solid-state or fiber amplifiers, which still maintain the Husimi distribution that extends the HHG cutoff.

### 5.3.7 *Outlook*

The experimental observation of predicted quantum effects in HHG requires fibers, solids and nanostructures to enhance the interaction coupling between light and matter, and to reduce the intensity threshold of HHG. Moreover, the measurement and post-selection of the high harmonics requires novel methods of focusing and collecting extreme ultraviolet rays and X-ray emission. Thus, we believe that the experimental developments in the field of quantum HHG go together with the development of nano-optics HHG.

## 6 X-ray Waveguide Nanophotonics

### 6.1 Waveguide Fundamentals

Waveguide optics and coupling of emitters to waveguide modes are essential for nanophotonics, and well established for visible and infrared light. In contrast, in the X-ray spectral range, waveguide optical effects and applications are much less prominent and common. The optical cross-sections and indices differ considerably, resulting in experimental regimes that are quite distinct from the optical counterparts. Yet, the basic principles of guided waves are identical, and a suitable index profile supports guided modes following the same principles, but in a different regime.

As a result of the X-ray index of refraction $n = 1 - \delta + i\beta$ with $\delta, \beta \ll 1$, an X-ray waveguide is formed by a thin film of low electron density surrounded by a cladding high electron density. With the angle of total reflection given by $\theta_c \leq \sqrt{2\delta}$ (with respect to the interface), waveguiding is limited to a small regime of grazing incidence $\theta \leq \theta_c$, and equivalently to propagation constants very close to the wavenumber $k$. At the same time, material absorption as governed by $\beta$ is significantly stronger in X-ray waveguides, due to photoelectric interactions in the cladding material. Fig. 19 illustrates the geometry of propagation and beam coupling for the two principal schemes, coupling from the front side (front coupling) or through a thinned cladding known as resonant beam coupling.

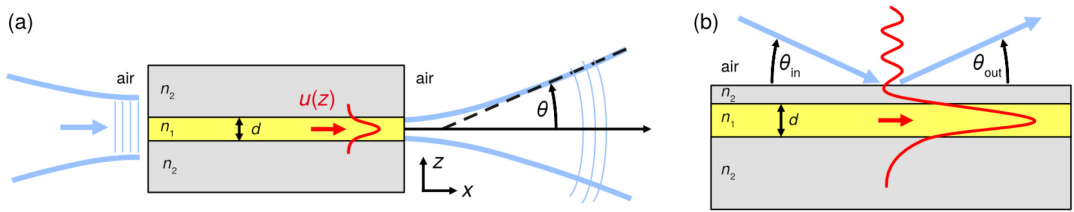

**Fig. 19 Coupling geometry of X-ray beam and waveguide.** (**a**) Front coupling scheme. A focused or collimated synchrotron X-ray beam is coupled into the front side of a waveguide. The excited mode with a given modal profile $u(z)$ propagates as guided waves and couples out into free space at the exit side with a divergence governed by the modal confinement. (**b**) Resonant beam coupling scheme. A collimated beam impinges under grazing incidence $\theta_{in}$ onto a thinned cladding, with the evanescent modes coupling into the resonantly enhanced leaky modes. The reflected beam at the exit angle $\theta_{out}$ shows a dip when modes are resonantly excited. Reprinted with permission from T. Salditt *et al.*, Nanoscale Photonic Imaging, Springer International Publishing, Cham, 2020 [419]. Licensed by a Creative Commons (CC BY) license.

While the first planar X-ray waveguides were demonstrated about thirty years ago [90–94,96], and the first two-dimensional (2D) channel waveguides about twenty years ago [95,97], applications, for example, in coherent imaging, X-ray spectroscopy and X-ray nano-optics developed very slowly, partly also due to the significant fabrication challenges [89]. With the advent of brilliant X-ray sources, progress in optical theory and simulation, nanofabrication, as well as novel experimental concepts, the situation is now changing, as we shall briefly review in this chapter. After a summary of the fundamental optics of X-ray waveguides, we address applications for coherent imaging, as well as waveguide fabrication. Finally, we review recent work in nano-optics and the coupling of waveguide modes with atomic emitters of radiation. Since X-ray waveguides distinctly alter the spatial modes of the electromagnetic field on the nanometer scale, the interaction of light with atomic emitters placed inside the waveguide is influenced in a rather subtle manner, giving rise to a variety of interesting phenomena.

## 6.2  Mode Structure

To illustrate the basics of X-ray waveguide optics, consider the simple 2D geometry of a planar waveguide. We loosely follow Ref. [420], sacrificing some rigor for brevity. In the simplest case, such a waveguide is a three-layer structure, consisting of a core of a material with low electron density and an outer cladding of a material with high electron density. The electromagnetic fields obey Maxwell's macroscopic equations and separate into transverse electric and transverse magnetic polarizations. Due to the weak contrasts in refractive indices, the two polarizations are approximately degenerate. Therefore, one can work with a single scalar field $U$, which corresponds to one of the transversal components of the electric (or magnetic) field. The field follows the Helmholtz equation

$$\{\partial_x^2 + \partial_z^2 + k^2 n(z)^2\} U(x,z,\omega) = 0, \tag{14}$$

where $k = \omega/c$ is the wave number and $n(z)$ is the refractive index. Here, we assume a quasi-monochromatic field of frequency $\omega$ and the field to be independent of $y$.

Equation (14) describes the scalar field under a (complex) potential well distributed along the $z$ direction and uniform along the $x$ axis. Its solutions can be expanded into a discrete set of resonant modes and a continuum of so-called radiation modes. The resonant modes can be found numerically as complex roots of a holomorphic function. One differentiates the resonant modes further into guided (resonant) modes, which are bounded, and leaky (resonant) modes, which are diverging for $|z| \to \infty$. Leaky modes are irrelevant for waveguides with thick cladding, but they play a central role in waveguides for evanescent coupling through a thinned top cladding. A general forward-propagating solution can be written as

$$U(x, z, \omega) = \sum_{m=1}^{N} c_m \, u_m(z) \, e^{ik\nu_m x} + \text{non-resonant} \tag{15}$$

where $N$ is the number of guided modes, $\nu_m$ are the so-called effective mode indices, $u_m(z)$ are the transversal mode profiles, and $c_m$ are the mode coefficients, which are determined by the boundary conditions. It should be noted that the mode profiles $u_m$ are complex-valued and not strictly orthonormal as a result of the presence of absorption, but they obey a bi-orthonormality relation which can even be naturally defined for the (unbounded) leaky modes [420]. For many practical purposes, however, one may consider them as small perturbations from the (orthonormal) modes for real refractive indices.

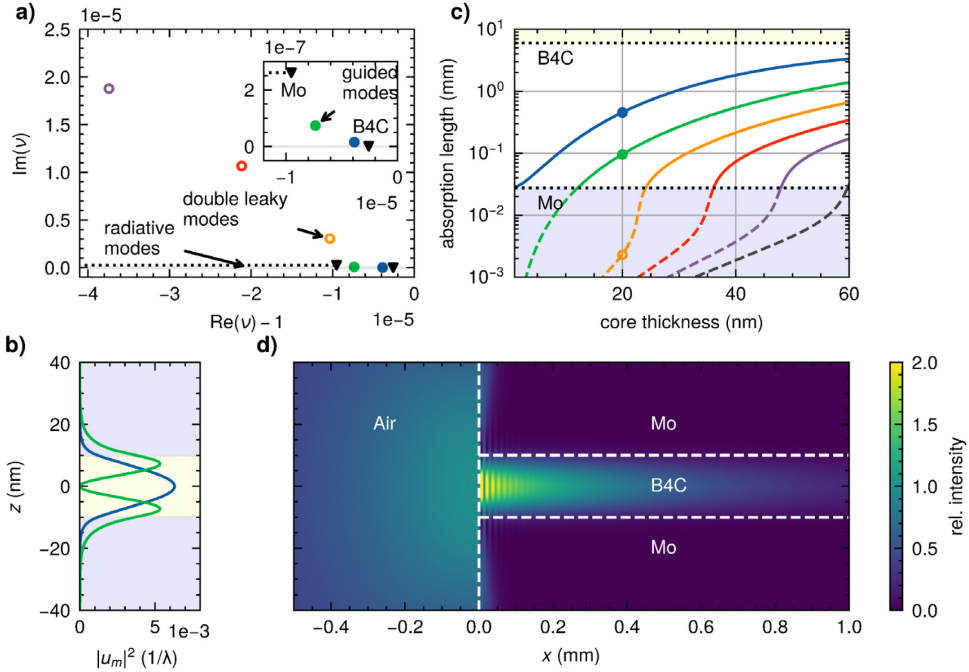

**Fig. 20 Mode structure of an X-ray waveguide at 13.8 keV photon energy.** (**a**) Effective mode indices $\nu_m$ of the guided and the leaky mode supported by the layer structure shown in (b), together with the mode profiles $u_m(z)$. (**c**) Absorption length of the waveguide modes for varying thickness of the $B_4C$ core-layer as well as the absorption lengths of cladding and core material. (**d**) Finite-difference solution of a Gaussian beam (full width at half maximum 100 nm) coupling into the waveguide. Reprinted with permission from L.M. Lohse *et al.*, Opt. Express **32**, 9518 (2024) [420]. Copyright 2024 The Optical Society.

Let us discuss the central properties of X-ray waveguides based on Eq. (15). Importantly, the effective indices of the guided modes are located in the complex plane between the refractive indices of core and cladding as shown exemplary in the inset of Fig. 20a. We can read off from Eq. (15) that the $1/e$ absorption length of a guided mode is $\Lambda_m = (2k\,\text{Im}\{\nu_m\})^{-1}$. Radiation modes, on the other hand, live in the cladding and are attenuated within the absorption length of the bulk cladding $\Lambda_\text{clad}$ or shorter. This generates the mode filtering property of X-ray waveguides: the non-resonant part of the field is greatly suppressed already after a few $\Lambda_\text{clad}$ (Fig. 20d), which can be two or more orders of magnitude shorter than the guided mode absorption lengths $\Lambda_m$. As shown in Fig. 20c, the absorption is lowest for the lowest mode, the fundamental mode. Thus, the higher modes are suppressed as well with increasing length. As an important consequence, the waveguide acts as a coherence filter [421,422], producing a highly (transversally) coherent wavefront at its exit, which is ideally suited for imaging. In the limit of only a single transmitted mode, the waveguide even emits a fully coherent field. The onset of waveguiding at the entrance of a front-coupled waveguide has also been studied [423,424].

The length scale to which X-rays may be confined in a waveguide has a fundamental lower limit $W = \lambda/(2\theta_c)$, where $\theta_c$ is the critical angle of total internal reflection [425]. This lower limit can be expressed as $W \approx \sqrt{\pi/4 r_0 \Delta\rho_e}$, where $r_0$ is the classical electron radius and $\Delta\rho_e$ is the difference in electron density between cladding and core [424]. Thus, it is independent of photon energy and takes values of, for example, $W \approx 10$ nm for molybdenum cladding and $\approx 20$ nm for silicon cladding (both assuming an air core). Consequently, typical mode confinement is at least two orders of magnitude larger than the typical wavelengths ($\lambda \lesssim 0.1$ nm) – X-ray waveguides are weakly guiding.

Next, a brief note on dispersion in X-ray waveguides. Synchrotron radiation prepared with crystal monochromators has typical bandwidths on the order of $\hbar\Delta\omega \sim 1$ eV. On this scale, refractive indices are nearly constant in the absence of absorption edges or atomic resonances. As a result, the mode profiles and mode indices in Eq. (15) are approximately constant, too, and the sole frequency dependency is in $k = \omega/c$. Multimode dispersion becomes relevant only for pulses shorter than 1 fs with typical parameters [420,426]. However, close to absorption edges or atomic resonances, the waveguide can cause significant dispersion [426], which has been studied numerically [427].

The mode coefficients $c_m$ in Eq. (15) can be computed via overlap integrals for given boundary conditions, such as coupling a beam into the front of a waveguide [420]. Computing the radiation modes in this analytic approach, however, is tedious. Given a field in the entry plane, it often makes sense to solve the entire field numerically in a different approach. A proven way to do this is the finite difference (beam) propagation method. It was first applied to X-ray waveguides by Fuhse [428] and has since then been further advanced to investigate the propagation of ultrashort pulses [427] and multilayer optics [429].

## 6.3 X-ray Waveguides for Coherent Imaging

Holographic illumination by the quasi-point source of X-ray waveguides is an important application of waveguide nano-optics. It takes advantage of the fact that modal propagation can be tuned, exactly calculated and designed in its properties, and hence used for coherence and spatial filtering.

### 6.3.1 Phase Contrast Imaging by Propagation

The dominating contrast mechanism of nanoscale X-ray imaging is phase contrast. Recall that at high photon energies $E$ of hard X-rays, the index of refraction $n(E, \mathbf{r}) = 1 - \delta(E, \mathbf{r}) + i\beta(E, \mathbf{r})$ leads to dominating phase interaction compared to absorption, quantified by the ratio $\delta/\beta \gg 1$, which is particularly high for low-$Z$ elements. Away from the absorption edges, $\delta(\mathbf{r})$ is proportional to the electron density. Contrast is created by the small phase shifts, which are imparted by inhomogeneous matter onto a propagating X-ray wavefront, based on density-dependent phase velocity. Given sufficient spatial coherence, these phase shifts are then transformed by self-interference into an intensity pattern in the detector plane [430], which can be inverted by phase-retrieval [431–433]. Compared to coherent diffractive imaging in the far-field, holographic imaging based on near-field interference results in a better posed mathematical structure of the phase problem [434]. For inline-holography or propagation imaging, the reference wave is formed by the enlarged primary wave (parallel or cone beam). Further, coherence requirements in near-field propagation imaging can be relaxed with respect to coherent diffractive imaging [435], and neither the sample nor the illumination field has to be compactly supported.

### 6.3.2 Challenges of Nanoscale X-ray Imaging

X-ray propagation imaging with nanoscale resolution requires nanoscale focusing of the X-ray beam and subsequent projection imaging with high geometric magnification. The divergent cone beam behind the focus illuminates the sample for holographic image formation. The properties of the illuminating wavefront are instrumental for holographic imaging. High-resolution and quantitative phase contrast require sufficient coherence, a large numerical aperture, and reduced wavefront aberrations [436–438]. Unfortunately, X-ray nano-focusing is associated with significant wavefront distortions, which violate the idealizing assumptions made on the probe in the course of image reconstruction (such as an ideal spherical wave). The standard procedure to correct this is to divide the hologram of the object by the empty (flat) image, which, however, can be associated with loss of resolution and image quality [439,440]. Solutions can be found either by optical filtering and wavefront cleaning, notably by X-ray waveguide optics [97,421,441], or – to some extent – by phase-retrieval schemes accounting for non-ideal illumination conditions, following the concept of ptychography for simultaneous probe and object reconstruction [442,443]. The disadvantage of these approaches is that multiple images with different sample translations have to be recorded for each projection angle, for example, when recording a tomographic scan. Further, they only work if the empty beam remains temporally stable.

### *6.3.3 Waveguide Optics for Imaging*

X-ray waveguides can be used for coherence and wavefront filtering, which is of importance since currently used sources often lack the spatial coherence needed for particular imaging experiments. In addition, they also serve in reducing or avoiding artifacts due to a non-ideal illumination. For this purpose, the waveguides are positioned in the focal plane of the focusing optics and are operated in front coupling mode. Note that different kinds of X-ray focusing optics are used, including reflective type optics notably curved mirrors such as the Kirkpatrick-Baez (KB) system, refractive optics notably by way of combining dozens to hundreds of lenses in form of compound refractive lenses (CRL), and diffractive optics, notably Fresnel zone plates (FZP), or multilayer Laue lenses (MZP). All types are limited by a rather low NA, and in many cases the spot size achievable with the available optics is insufficient. For this reason it can be useful to reduce the beam extension be further by waveguiding, which then results in increased numerical aperture and resolution. For imaging, two-dimensionally confining X-ray waveguides are required. As demonstrated in [437,444,445], the exit radiation of the waveguide then provides a highly coherent, well-controlled, smooth and quasi-point-like illumination for nanoscale X-ray imaging. Fig. 21 illustrates the basic principle and gives an example for a waveguide far-field intensity distribution. In this example, the FWHM of the waveguided beam is typically a factor of five smaller than the KB focus, resulting in corresponding coupling losses.

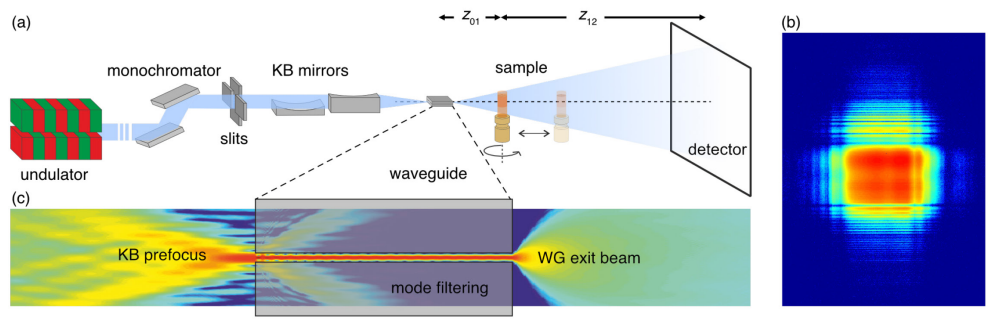

**Fig. 21 Coherent imaging with X-ray waveguide illumination.** (**a**) Monochromatized undulator radiation is focused by Kirkpatrick-Baez mirrors onto an X-ray waveguide, acting as a coherence and spatial filter. The sample is positioned at a distance $z_1$ behind the waveguide, and the holographic (near-field) diffraction pattern is recorded at distance $z_2$ behind the sample. (**b**) Far-field pattern of a waveguide, showing a smooth central cone corresponding to a small set of transmitted modes. High spatial frequency variations of the Kirkpatrick-Baez-mirrors are filtered, resulting in an advantageous illumination system for inline holography. Reprinted with permission from S.P. Krüger *et al.*, J. Synchrotron Rad. **19**, 227–236 (2012) [446]. Licensed under a Creative Commons Attribution 2.0 UK (CC BY 2.0 UK) license. (**c**) Finite difference calculation of the optical system, showing the Kirkpatrick-Baez focusing, propagation in the waveguide, and free space propagation of the waveguide exit beam. Reprinted with permission from T. Salditt *et al.*, Nanoscale Photonic Imaging, Springer International Publishing, Cham, 2020 [419]. Licensed by a Creative Commons (CC BY) license.

Fig. 22 shows an example of a cardiomyocyte cell, reconstructed from a holographic tomography scan in a waveguide beam [447]. Phase retrieval was performed by the nonlinear Tikhonov approach [448]. A total dose of $1.1 \times 10^5$ Gy was applied for a scan recorded at photon energy of 8 keV and comprising 720 projections. The cardiomyocyte was isolated from murine heart and prepared by freeze drying on a thin membrane. The magnification was adjusted to obtain a voxel size of 45 nm.

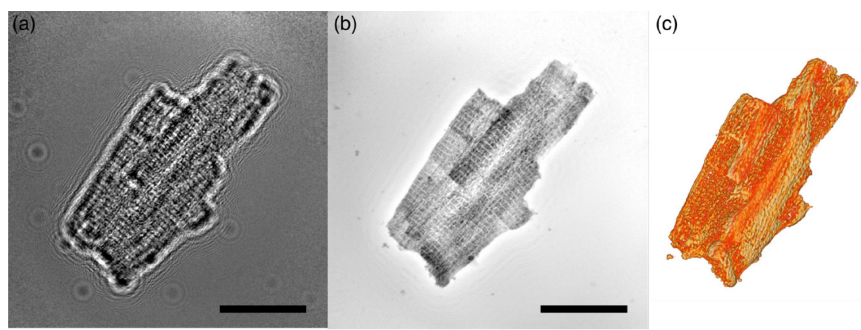

**Fig. 22 Waveguide holo-tomography of a biological cell.** (**a**) An exemplary projection and the (**b**) corresponding phase retrieval. (**c**) A volume rendering of the cell after tomographic reconstruction. Sub-cellular structures such as the nuclei, mitochondria, and myofibrils can be identified throughout the volume. Scale bar: 25 μm. Reprinted with permission from M. Reichardt *et al.*, Biophys. J. **119**, 1309–1323 (2020) [447]. Licensed under a Creative Commons Attribution 2.0 UK (CC BY 2.0 UK) license.

*6.3.4 c-Retrieval*

When using waveguide illumination and high magnification for nanoscale resolution, one deals with large Fresnel numbers $F \ll 1$; images are recorded in the so-called deep holographic regime. Here, the phase contrast transfer and, hence, phase sensitivity is highest, but standard phase retrieval based on the transport of intensity equation fails [449]. In this regime, inversion can be based on the contrast transfer function approach [431], which relies on linearization of the object's optical properties, and hence requires a slowly varying phase. In most applications, three or four different measurement planes with varying Fresnel numbers [450] are used, and a homogeneous (single-material) object is assumed. As a fast direct method, contrast transfer function (CTF)-based phase retrieval is well suited for tomography [451]. Iterative algorithms provide a more general phase-retrieval approach, based on alternating propagation between the object and measurement plane and the application of suitable constraints [452]. Assumption of a slowly varying phase, or a single material object, are not required and superior image quality compared to the CTF approach can be achieved.

To overcome errors associated with linearization, while maintaining the advantage of fast direct inversion, regularized inversion of the full non-linear forward operator has been described in [448] and is denoted as the nonlinear Tikhonov approach. A collection of phase retrieval and reconstruction algorithms has been published in [453]. While phase retrieval of inline holographic can be performed to some extent regardless of the illumination function used, the achievable resolution, the required dose and the phase sensitivity depend on the optical system, and have to be determined experimentally from different test patterns. The resolution for waveguide-based holography has reached the range of 30 nm–50 nm (half-period resolution), using conventional empty beam division, and even sub-15 nm for a phase retrieval formulation which used the waveguide probe and the compact size in the waveguide exit plain as a constraint [454]. Dose-resolution and coherence-resolution relationships for coherent X-ray imaging have been studied in [455–458]. More specific work has addressed finite partial coherence, multi-modal wavefield reconstruction [459], as well as the coherence-resolution relationship [435]. Altogether, experimental results and numerical simulations show that waveguide-based holography is suitable for achieving dose-efficient three-dimensional (3D) imaging close to the theoretic dose-resolution limits [435].

The properties of waveguide illumination, in particular, the compactness of the wavefield in the waveguide exit plane, can be advantageously exploited for coherent imaging. In the so-called super-resolution holography scheme [454], two images, one without and one with an object, are recorded as input for iterative phase retrieval, operating on the three planes: waveguide exit, object and detector plane. The constraint of a compact probe is sufficient to recover the phase of the empty beam and thereby fix the illumination in the object plane, which in turn provides a constraint to the phase shift induced by the object. Empty beam division is no longer required, and dose-efficient direct photon counting detectors can be used. Since information encoded in the tails of the waveguide beam is not lost, reconstruction can be achieved beyond the resolution limit given by the cone-beam numerical aperture. Fig. 23 shows the example of a test pattern imaged with a field of view of 5 μm × 5 μm and at a resolution of 11.2 nm, using a waveguide exit source size of 30 nm (full width at half maximum) [454].

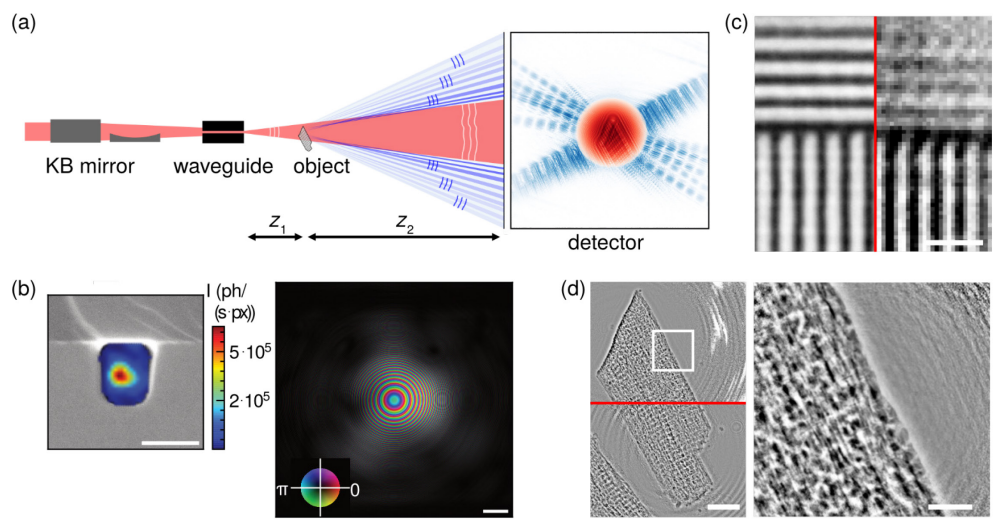

**Fig. 23 Super-resolution holography scheme.** Exploiting the central cone of a waveguide beam for holographic encoding together with information encoded in its tails, which is otherwise lost in the usual empty beam division. (**a**) Schematic illustrating simultaneous recording of holographic and diffractive signals. (**b**) Wavefield reconstructions, showing (left) intensity in the waveguide exit plane (superimposed with a scanning electron microscope image of the waveguide, and (right) the phase and amplitude in the object plane. (**c**) Test pattern with 50 nm lines and spacings. (**d**) Reconstruction of a cardiomyocyte cell recorded by a single super-resolution holography image without any need for scanning. Scale bars: (b) 100 nm (left) / 2 μm (right), (c) 250 nm, (d) 10 μm (left) / 2.5 μm (right). (**b-d**) are reprinted with permission from J. Soltau *et al.*, Optica **8**, 818 (2021) [454]. Copyright 2021 The Optical Society.

### *6.4 Fabrication of X-ray Waveguides and Advanced Geometries*

Planar waveguides for one-dimensional beam confinement can be conveniently fabricated by thin film deposition techniques, in particular by magnetron sputtering, with a plethora of possible thin film sequences including waveguide arrays [460]. Small guiding layers down to the fundamental limit of beam confinement can be fabricated [425]. Contrarily, challenges are sometimes encountered for larger film thicknesses when accumulating compressive or tensile stress. For front coupling, the thin films have to be capped with a second substrate wafer to block radiative modes. This can be achieved via soft alloy bonding [461].

Imaging applications, as discussed above, require 2D beam confinement, or channel waveguides. For multi-keV photon energy and waveguide cross sections of less than 100 nm, aspect ratios (length to width) of waveguide channels easily exceed $10^4$ to provide sufficient mode filtering, which imposes significant challenges in fabrication. Simple 2D waveguides can be formed by channels made of photoresist or low $Z$ materials patterned by e-beam lithography and then coated with metal or semi-conductor cladding [95,97]. Waveguides with air or vacuum guiding layers can be fabricated by dry etching of channels into silicon wafers and subsequent capping by wafer bonding [89,462]. An alternative fabrication scheme for 2D waveguides uses two crossed planar waveguides, which confine the beam in orthogonal directions. Combined in a crossed geometry, an effective 2D source is formed for holographic imaging [461]. This crossed 2D waveguide scheme is compatible with fabrication by thin film deposition techniques. Hence, smaller guiding layers, a wider range of materials, and more complex layer sequences can be realized, including a two-component cladding optimized for high transmission [101].

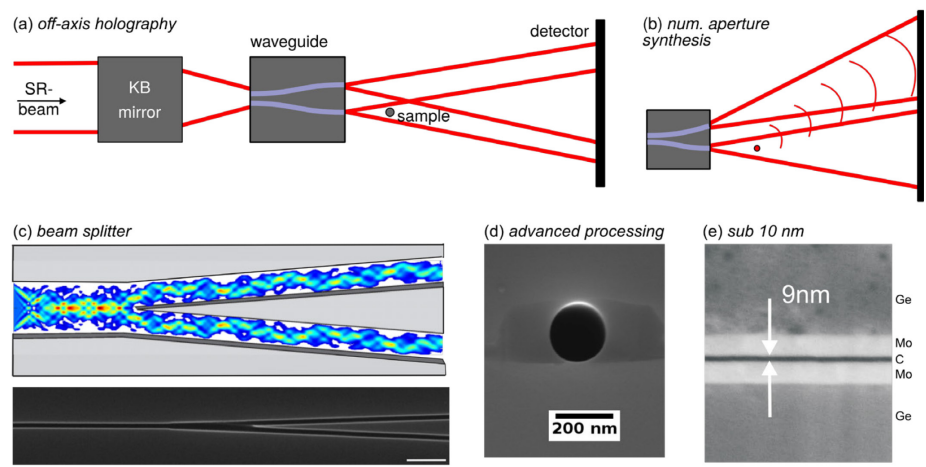

**Fig. 24 X-ray waveguides with multiple channels lithographically defined on a chip.** This could serve for example, (**a**) the encoding of absolute phase by off-axis holography, and (**b**) synthesis of the numerical aperture by tilted beams. (**c**) Finite difference simulation of a beam splitter and the corresponding lithographic structure (scanning electron microscope image). Reprinted with permission from S. Hoffmann-Urlaub *et al.*, Acta Cryst. A **72**, 515–522 (2016) [463]. Copyright 2016 IUCr. (**d**) Advanced fabrication and processing involving waver bonding of semiconductors (silicon and germanium). Surface diffusion during the bonding process, can result in very smooth and round cross-sections as shown here for a channel in germanium. (**e**) Guiding layers with $d \sim 10$ nm can be fabricated by thin film deposition. Reprinted with permission from S.P. Krüger *et al.*, J. Synchrotron Rad. **19**, 227–236 (2012) [446]. Licensed under a Creative Commons Attribution 2.0 UK (CC BY 2.0 UK) license.

Beyond simple quasi-point sources, waveguide optics enables a variety of optical functions, such as filtering, confining, guiding, coupling or splitting of beams. Advanced X-ray waveguides now begin to exploit such advanced functionalities, which again require lithographic fabrication techniques. Based on an array of waveguide channels, X-ray optics on a chip was demonstrated in [464]. Beam concentration by tapering [465], beam propagation in curved channels [464], and beam splitting for nano-interferometry [89] have also been realized. Multiplexed beamlets based on splitting and redirecting of beams in an X-ray waveguide chip can be exploited for optimized illumination systems for holography, for example, different view angles and/or synthesis of high numerical aperture. Apart from imaging, multiplexed X-ray beams generated on a waveguide chip could serve a variety of interferometric, nano-optical and quantum optical experiments.

### *6.5 Interaction with Emitters inside the Waveguide*

The majority of experiments have been performed using a class of waveguides that is slightly different from the ones presented in the last sections. By thinning the top cladding of a planar waveguide, a plane wave in grazing incidence can be evanescently coupled into the core layer, as first demonstrated by Spiller in 1974 [90]. For incidence angles $\theta_{\text{in}}$ that are compatible with one of the resonant modes ($\cos \theta_{\text{in}} \approx \text{Re}\{v_m\}$), the field inside the core layer is amplified due to the constructive interference between the downward-propagating field components and those that are reflected from the interface between the core and bottom cladding. This principle was later termed "resonant beam coupler" [91] and was applied to create bright and coherent secondary sources before being mostly superseded by front-coupling waveguides [466] due to the widespread availability of focusing optics. The thin cladding conversely allows the field from the core layer to leak out evanescently through the top. In the mode picture, the resonant modes of such a waveguide structure are not properly guided but leaky modes. The technical details have been discussed in Ref. [420].

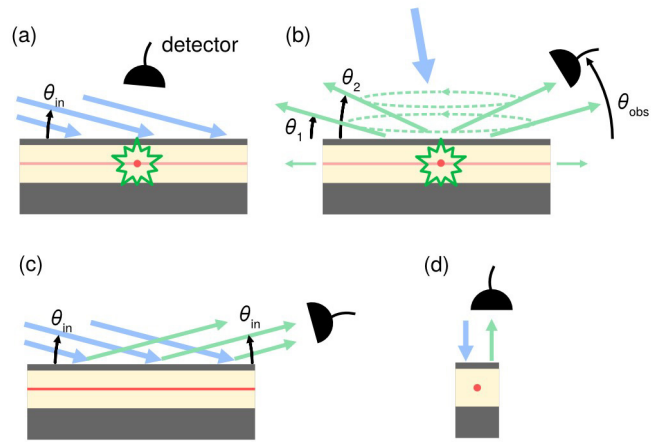

**Fig. 25 Types of experiments, where incident light (blue) excites emitters (red) which produce secondary radiation (green) inside an X-ray waveguide.** (**a**) Grazing incidence illumination at a resonant incidence angle $\theta_{\text{in}}$ enhances the fluorescence rate. (**b**) Incoherent emissions couple into waveguide modes, resulting in enhanced intensity at the resonant angles. (**c**) Resonant scattering preserves the phase of the illumination so that the emission is concentrated into the direction of specular reflection. The system can be interpreted as a Fabry-Pérot cavity resonator (d). Compared to front-coupling, grazing-incidence illumination has a number of practical advantages: (i) Fabrication is simpler, since no bonding or cutting is required. (ii) Experimental setups are simpler, because no (or only mild) focussing is required. (iii) Theoretical modelling is simpler, because the structures are fully translationally invariant in two dimensions and have no boundaries. A plethora of experiments involving emitters in such waveguides have been performed, which we broadly classify into two categories: (i) incoherent interactions such as fluorescence and characteristic radiation and (ii) coherent interactions with atomic or nuclear resonances.

### 6.5.1 Incoherent Emissions

The intensity at the maximum-positions of a resonant mode profile can be enhanced by 10 to 100-fold with respect to the incident field [467] without an additional focussing device if the waveguide is illuminated at a resonant incidence angle (Fig. 26a,b). Placing an atom at such a position consequently increases the photoionization rate and, hence, the fluorescence by the same factor [468,469]. The mode structure also provides some depth sensitivity. Optimizing layer structures to obtain the highest possible field strength has been studied in Ref. [467].

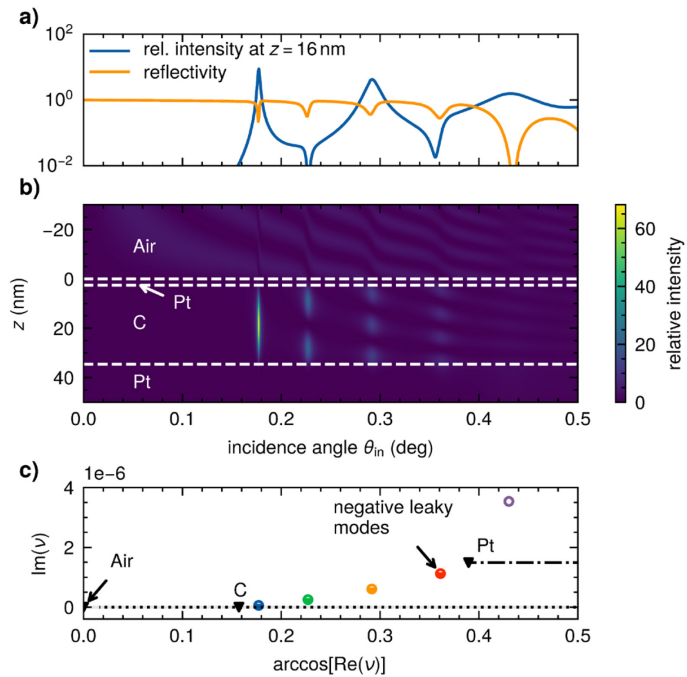

**Fig. 26 Intensity enhancement in a planar waveguide due to grazing-incidence illumination.** (**a**) The relative intensity at the center of the waveguide core shows pronounced maxima corresponding to the waveguide resonances (c). (**b**) Relative intensity as a function of incidence angle and vertical position, calculated via the Green's function of the system [420] assuming 11.4 keV photon energy (gold L$\beta$ emission). It also describes the angular spectrum of the field produced by an atom at z and observed in some plane above the platinum layer (Fig. 25b).

These works used the mode structure of the incident field (Fig. 25a). More recently, it was demonstrated that also the (incoherently) emitted fields are affected by the mode structure (Fig. 25b). Incoherent emissions, which are isotropic in a homogeneous environment, exhibit directionally structured emissions when placed into a waveguide environment. This has been reported in multilayer Bragg reflectors at the emission line of silicon, 1.7 keV [470,471]. Later, it was demonstrated in Ref. [472] with fluorescence from gold atoms and proposed as a time-resolved imaging technique with sub-nanometer spatial depth-resolution by the name X-ray waveguide fluorescence holography. At the same time, Vassholz and Salditt observed emission into waveguide modes, not only fluorescence from X-ray illumination but also characteristic radiation and bremsstrahlung due to irradiance with an electron beam (Fig. 27) [473]. The authors expect their discovery to lead to brighter X-ray sources based on multilayer anodes. In addition to intensity peaks at the resonant angles of the emissions leaking through the top cladding, they also observed pronounced oscillations in the field exiting the side boundary of the sample. These oscillations were later explained classically as interference between two resonant modes [420].

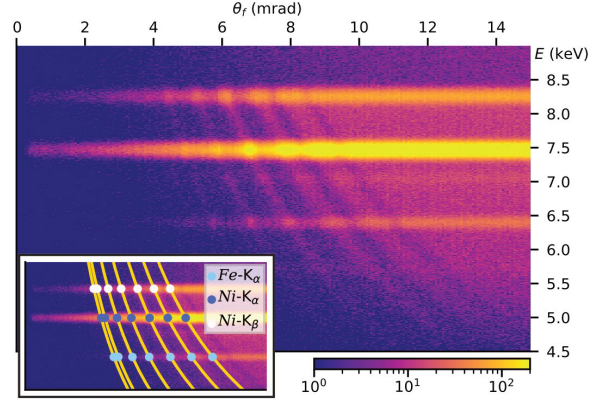

**Fig. 27** Structured emission of characteristic radiation and bremsstrahlung from a multilayer structure under electron irradiation, showing the modulation due to the waveguide modes. Reprinted with permission from M. Vassholz *et al.*, Sci. Adv. **7**, eabd5677 (2021) [473]. Licensed by a CC BY-NC 4.0 license.

### *6.5.2 Green's Function*

One powerful theoretical approach to model emitters in a layer structure, which is particularly suited for the present case, is via the (classical) electromagnetic Green's function, defined as the solution of

$$\{\nabla^2 + k^2 n(z)^2\} G(\mathbf{r}, \mathbf{r}', \omega) = \delta^{(3)}(\mathbf{r} - \mathbf{r}'). \tag{16}$$

where $k$ is the wavenumer and $n(z)$ is the refractive index, and $G(\mathbf{r}, \mathbf{r}', \omega)$ is the Green's function. It describes the field at $\mathbf{r}$ produced by an emitter at point $\mathbf{r}'$. We consider the scalar case, as polarization is mostly irrelevant for our purposes. Writing

$$G(x, y, z, z', \omega) = \int e^{i(q_x x + q_y y)} G(q_x, q_y, z, z', \omega) \frac{dq_x dq_y}{4\pi^2}, \tag{17}$$

considering the translational invariance in $x$ and $y$, a closed-form solution can be readily given for $G(q_x, q_y, z, z', \omega)$ [474]. A practical algorithm to compute solutions for arbitrary layer structures is given, for example, in Ref. [420], together with freely available code.

The Green's function models three situations. First, it describes the angular spectrum of the field radiated by an emitter in the waveguide core and observed in some plane above (Fig. 25b). Conversely, it also describes the field produced by a plane wave in grazing incidence as observed inside the waveguide – with the roles of $z'$ and $z$ reversed and $q_x = \cos\theta_{\text{in}}$ and $q_y = 0$ (Fig. 25a). Finally, it can also describe fields both produced and observed within the waveguide. To this end, a real-space solution to Eq. (16) can be given asymptotically in terms of resonant modes [420]. Using this approach, the ratio of the spontaneous emission rate $\Gamma_m$ into a planar waveguide mode compared to the total value, commonly referred to as the beta factor ($\beta$), can be estimated [420] as

$$\frac{\Gamma_m}{\Gamma_{\text{tot}}} \approx \lambda |u_m(z_0)|^2 \sim \frac{\lambda}{d_m}, \qquad (18)$$

where $u_m(z_0)$ is the mode profile from Eq. (15) evaluated at the emitter position $z_0$ and $d_m$ is the one-dimensional confinement of the mode. It is typically in the order of $d_m \gtrsim 10^2 \lambda$ for hard X-rays [425,473] because of the small contrast in refractive indices, resulting in $\Gamma_m/\Gamma_{\text{tot}} \lesssim 10^{-2}$. Since reflection under normal incidence is generally negligible for hard X-rays, a Purcell enhancement of the *total* emission rate is not expected and $\Gamma_{\text{tot}}$ is well approximated by its value in vacuum [420]. In this sense, we emphasize that hard X-ray waveguides generally are not classified as cavity resonators despite the contradictory use of the term in the literature.

### *6.5.3 Waveguides with Resonant Atoms*

Embedding resonant atoms in waveguides can produce a vast range of phenomena [475]. In addition to spatial patterns, atomic resonances also imprint an energetic or temporal structure into the emission. Collective effects such as superradiant speedup [476], shift of the resonance energy [477], collective oscillations between two ensembles [478], and many more [479–481] have been demonstrated.

One of the most successful platforms with resonant emitters is the Mössbauer nuclei discussed in Section 7.1 such as $^{57}$Fe. These nuclei have a low-lying isomeric state at transition energy $\hbar\omega = 14.4$ keV with a linewidth of only $\hbar\gamma = 4.7$ neV, corresponding to a lifetime of 141 ns. This is much shorter than the pulse duration produced by synchrotron light sources or X-ray free-electron lasers, typically 100 ps or shorter, which allows to stroboscopically excite the nuclei and observe the free decay as a function of time [482]. A second platform are the so-called "white line" resonances, dipole-like transitions close to the $L_3$ absorption edge of certain transition metals such as tantalum and tungsten. Here, a collective frequency shift and superradiance [478], control of core-hole lifetimes [483] and the observation of a flat line Fano profile [484] have been demonstrated. In all cases, the resonant materials can be deposited in the form of one or more thin films into the waveguide multilayer structure using thin-film deposition techniques such as magnetron sputtering.

The underlying interaction is similar for both nuclear and electronic resonances. The incident light drives an internal resonance of the atoms between some ground and excited states. Importantly, this process is coherent over the entire ensemble of atoms. Instead of emitting isotropically, the illumination coherently imprints its spatial phase so that the emission, or rather the scattered light, is concentrated in the direction of specular reflection (Fig. 25c). Since even modern synchrotron light sources provide only on the order of one photon per pulse within the resonance bandwidth of $^{57}$Fe, the degree of excitation is extremely low. While drastically higher photon numbers have been realized at X-ray free-electron lasers, no experiments with nuclear resonances in the non-linear regime have been reported. Nevertheless, a discussion of how such non-linear signatures might manifest has been started [485]. Recently, it was theoretically shown that optimized thin-film waveguides placed in tightly focused X-ray beams from free-electron laser could drive Mössbauer nuclei towards the nonlinear optics regime, although still significantly below inversion for present-day light sources [486,487]. For now, all experiments are well within the linear regime.

Theoretically, the atomic resonances can be accounted for by resonant modifications of either electric or magnetic susceptibility (depending on the kind of interaction) and absorbed into a frequency-dependent refractive index $n(\omega, z)$. This allows computing the reflectivity of any layer structure that contains resonant atoms using a transfer-matrix method [488], which also gives access to the field distribution inside the layer structure [489]. The propagation through structured samples can be simulated numerically using the (frequency-domain) finite-difference propagation techniques described earlier in this chapter. While being quantitatively accurate, this approach conceals the underlying physics, however. Changing the point of view, the coupled system of field and atomic excitations represents a fruitful platform for quantum-optical experiments (for a review, see [481]).

A system consisting of one or more films of resonant nuclei in a planar waveguide is equivalent to a Fabry-Pérot-type cavity resonator containing the atomic ensembles (Fig. 25d). We emphasize again that, from the perspective of a single atom, the waveguide itself is not a cavity resonator. This correspondence rather describes the infinitely extended ensemble of atoms in the translation-invariant waveguide. It allows to interpret the experiments in the language of cavity quantum electrodynamics (QED). To that end, several quantum theories have been formulated in the last decade, starting from few-mode theories [490–492], going to approaches based on macroscopic QED [492,493]. The latter are particularly interesting from the perspective of nano-optics, since they model the waveguide-mediated atom-atom interactions using the classical electromagnetic Green's function as defined in (3). The design and optimization of layer structures to obtain certain values for the quantum-optical observables has been studied in Ref. [494].

While the majority of experiments with resonant atoms in X-ray waveguides have been performed in grazing incidence reflection geometry, recently the aforementioned front-coupling geometry has gained some interest. Here, the cavity picture breaks down due to the finite length of the waveguide – or rather the presence of boundaries. Tapered waveguides in front-coupling geometry have been proposed to provide nanofocusing and drastically enhance the driving intensity for nuclei placed at the focal position [495]. This could facilitate exciting a significant fraction of nuclei and reach the non-linear regime using highly energetic X-ray free-electron laser (FEL) pulses. The theory of resonant propagation of X-rays in a waveguide containing resonant nuclei after front-coupled excitation has been worked out in detail within macroscopic QED, using the real-space mode-expansion of the electromagnetic Green's function [420,496]. In particular, the authors predict that the temporal emission pattern of a thin film of resonant material in a single-mode waveguide is equivalent to that of a homogeneous slab of resonant material with modified optical depth. The first experimental demonstration of resonant propagation after front-coupled excitation with a focussed beam of synchrotron radiation has been reported in Ref. [497], observing the predicted temporal emission pattern. In a third theoretical work, a similar waveguide structure is proposed as a gravitational sensor discussed in Section 7.1 [498] that harnesses the extremely narrow nuclear resonance of $^{45}$Sc (Fig. 28d).

# 7 Applications of X-rays and Free Electrons Using Nanophotonics

## 7.1 *Rabi Oscillation of X-rays for Measurements and Sensing*

Rabi oscillations [499] describe the coherent oscillation of the population between two states under an external oscillatory field. In the study of bond-length-dependent Rabi oscillations in the strong laser field dissociation of $H_2^+$, it was demonstrated that the interplay between bond stretching and Rabi frequencies significantly influences dissociation dynamics (Fig. 28a) [500]. The nuclear wave packet driven by Rabi oscillations can follow distinct pathways, either rolling outwards or looping between electronic states, each resulting in different kinetic energy releases of the ejected protons. Time-dependent Schrodinger equation simulations confirm that these dynamics yield rich proton energy spectra, providing deeper insights into the molecular dissociation process beyond the well-accepted resonant one-photon dissociation pathway. The study also highlights the importance of pulse length, showing that longer laser pulses enhance dynamic Rabi coupling, leading to more complex kinetic energy release distributions. Additionally, experimental exploration using the Cold-Target-Recoil-Ion-Momentum Spectrometer [501,502] technique involves analyzing dissociative fragments to validate the predicted Rabi oscillation-induced dissociation pathways. This comprehensive analysis of Rabi oscillations in a molecular system provides a detailed description of the underlying mechanisms and their implications for controlling molecular reactions and studying ultrafast molecular dynamics [500].

The detection of population inversions in Mössbauer nuclei can be enabled by utilizing Rabi oscillations and spectral interference [503,504]. The experiments with Mössbauer nuclei have been restricted to low-excitation regimes due to the low source brilliance and the sharpness of the spectral lines, limiting efficient population transfer and nuclear manipulation [490,491]. However, it has been proposed [505] that by employing pulsed excitation of the nuclei, the inversion of nuclear ensembles could be determined by detecting spectral symmetry flips (Fig. 28b), which occur at each half-cycle of Rabi oscillation, rather than relying on absolute intensity measurements. This method not only could improve the accuracy of detecting population inversions, but also could open up new possibilities for observing nonlinear light-matter interactions at X-ray frequencies [486,487,505].

Through manipulating the atomic electronic states and nuclear states, **Rabi oscillations** hold potential applications in quantum optics [506–508], precise spectroscopic techniques [509], and state-selective chemistry [510]. Resonance fluorescence induced by ultrafast and intense

X-ray pulses reveals Rabi flopping [511,512], characterized by repeated cycles of stimulated emission and absorption, which are observable in the fluorescence spectrum. This phenomenon provides insights into the coherent and nonlinear interactions between X-rays and atomic or ionic systems. Rabi flopping, driven by strong oscillatory fields, causes systems to transition repeatedly between ground and excited states. In the X-ray regime, the high-energy and short duration of pulses induce rapid transitions, highlighting coherent dynamics within the system. Probing Rabi flopping in fluorescence spectra allows for detailed analysis of quantum properties such as coherence times, external field influences, and electron correlation effects (Fig. 28c) [512].

The use of structured X-ray waveguides have been proposed to drive Rabi oscillations and enable precise quantum control of X-ray modes [498,513]. By leveraging nuclear resonant scattering and gravitational sensitivity, the system allows for coherent manipulation of X-rays, making it responsive to small changes in altitude and other external factors. The structured waveguide design introduces a method for generating higher-order X-ray modes and presents potential applications in X-ray interferometry and gravitationally sensitive X-ray optics. This approach offers a new pathway for integrating gravitational effects into practical X-ray technologies (Fig. 28d).

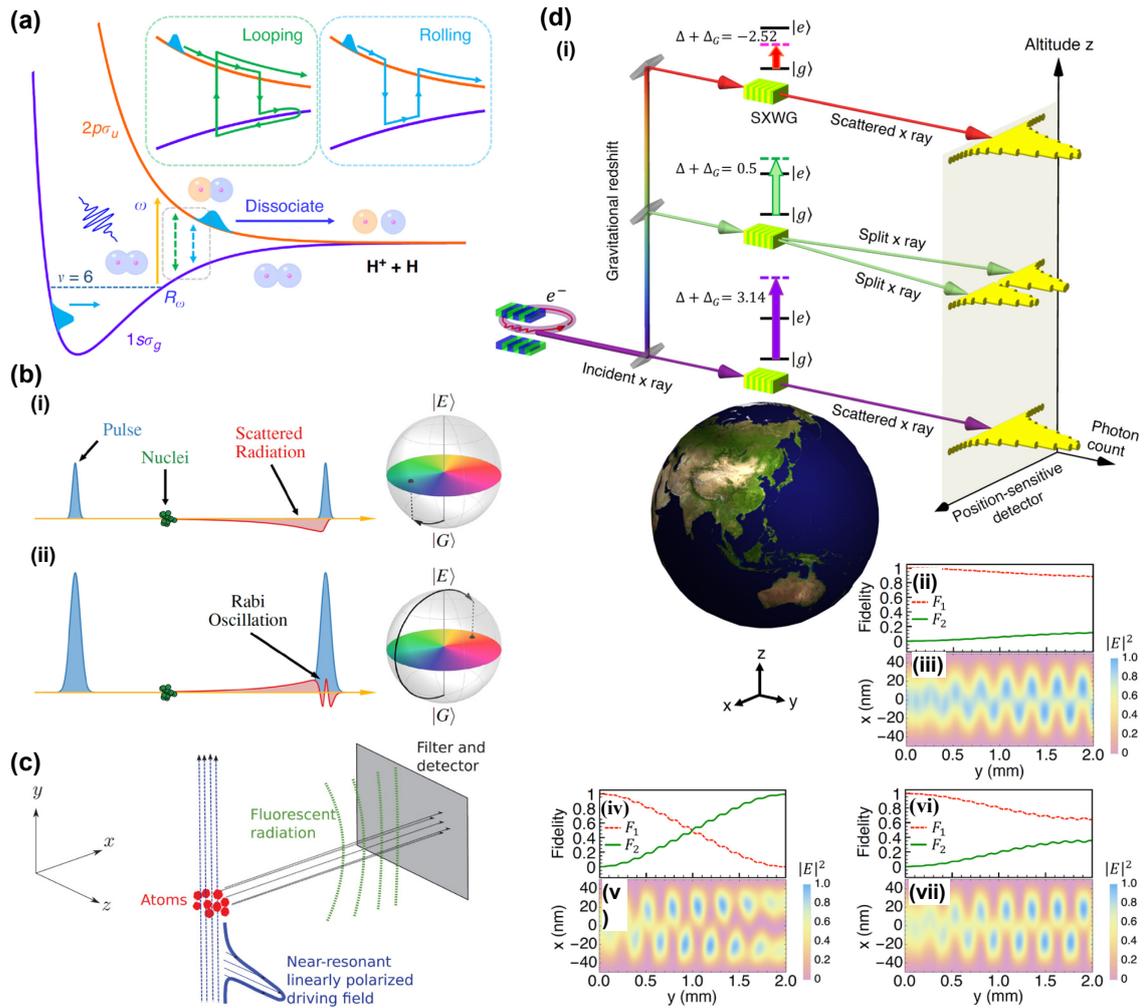

**Fig. 28 Exploring Rabi oscillations of X-rays in nuclear manipulation and nonlinear spectroscopy.** (**a**) Studies of bond-length-dependent Rabi oscillations in $H_2^+$ dissociation under strong laser fields. Reprinted with permission from S. Pan *et al.*, Light Sci. Appl. **12**, 35 (2023) **[500]**. Licensed by a Creative Commons Attribution 4.0 International License. (**b**) The illustration of detecting population inversions of Mössbauer nuclei using spectral interference. This could enable strong field manipulation of nuclear-level structures and observation of nonlinear light-matter interactions at X-ray frequencies. Reprinted with permission from K. P. Heeg *et al.*, arXiv:1607.04116 (2016) **[505]**. Licensed by under the arXiv Nonexclusive Distribution License. (**c**) Resonance fluorescence induced by ultrafast X-ray pulses, revealing Rabi flopping in the fluorescence spectrum, providing insights into coherent and nonlinear X-ray interactions with atomic systems. Reprinted with permission from S.M. Cavaletto *et al.*, Phys. Rev. A **86**, 033402 (2012) **[512]**. Copyright 2012 American Physical Society. (**d**) Exploration of using structured X-ray waveguides to drive Rabi oscillations for gravitational measurements. Reprinted with permission from S.-Y. Lee *et al.*, arXiv:2305.00613 (2023) **[498]**. Licensed by under the arXiv Nonexclusive Distribution License.

## 7.2 Quantum X-ray Imaging

Quantum-enhanced X-ray detection techniques exploit quantum correlations between photons to enhance the sensitivity and resolution, enabling the observation of subtle features inaccessible to classical methods. The advancement could particularly benefit medical imaging and non-destructive testing, where high precision is required without increasing radiation doses (Fig. 29a) [514]. Furthermore, leveraging higher-order intensity correlations of incoherently emitted light from free-electron lasers for imaging can achieve spatial resolutions close to or below the Abbe limit [515] (Fig. 29b). One desired application for quantum-enhanced X-ray detection is single-particle imaging, although it will be extremely challenging to implement [516].

Ghost imaging with X-ray photon pairs [517,518] is another innovative technique that utilizes quantum entanglement for high-resolution imaging. This method reduces radiation damage in medical imaging and enhances the resolution of structural characterization. By correlating the integrated transmitted intensity from a sample with the spatially resolved intensity measured in the empty beam, ghost imaging offers a promising approach for low-dose, high-precision imaging (Fig. 29c,d). Furthermore, lensless Fourier-transform ghost imaging with pseudothermal hard X-rays extends the capabilities of X-ray crystallography to noncrystalline samples. This technique, which measures second-order intensity correlation functions to obtain Fourier-transform diffraction patterns, does not require highly coherent X-ray sources. Thus, it can be implemented with laboratory X-ray sources, making it accessible and versatile for various applications [519].

The X-ray pairs are generally realized through the parametric down-conversion process [232–234] via nonlinear Bragg diffraction. This method generates entangled photon pairs from a high-flux X-ray beam, achieving detection rates from 0.1 pairs/sec to 4100 pairs/hour (Fig. 29e) [235–240]. Additionally, the use of two-dimensional detectors like the pnCCD for observing energy and momentum-correlated X-ray pairs significantly enhances imaging efficiency and reduces radiation levels. These detectors, offering high-efficiency measurements of X-ray photon pairs, are particularly beneficial for minimizing potential damage to delicate biological and material samples, and support advanced quantum optics experiments with high temporal and energy resolution [520].

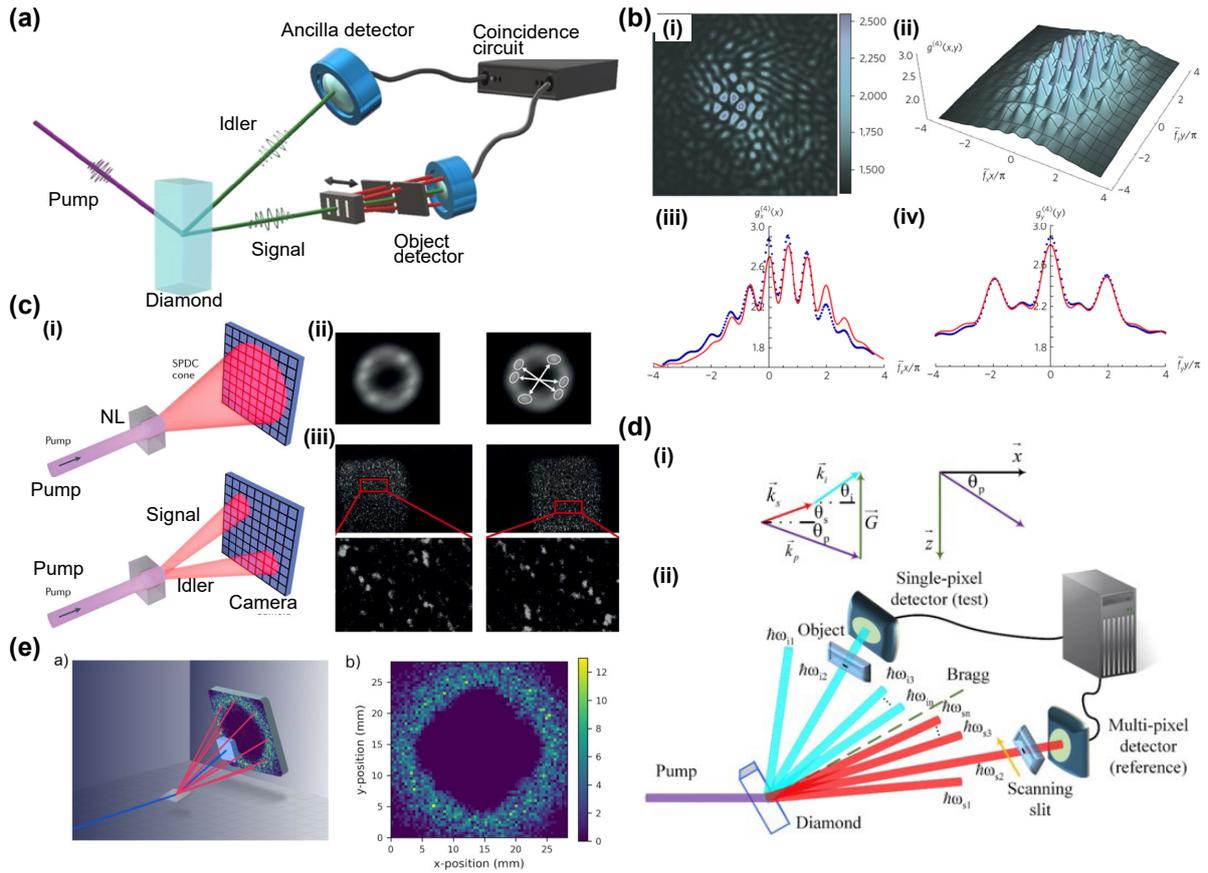

**Fig. 29 Imaging applications with quantum X-rays.** (**a**) Quantum-enhanced X-ray detection techniques use quantum correlations to improve the sensitivity and resolution of X-ray imaging, without increasing radiation dose. Reprinted with permission from S. Sofer *et al.*, Phys. Rev. X **9**, 031033 (2019) [514]. Licensed by a Creative Commons Attribution 4.0 International license. (**b**) Incoherently scattered light from free-electron lasers utilizes higher-order intensity correlations to achieve high spatial resolution imaging, providing structural information from incoherent processes. Reprinted with permission from R. Schneider *et al.*, Nat. Phys. **14**, 126–129 (2018) [515]. Copyright 2018 Springer Nature. (**c, d**) Ghost imaging with X-ray photon pairs leverages quantum entanglement to create high-resolution images without direct interaction with the imaging beam. Reprinted with permission from P.-A. Moreau *et al.*, Nat. Rev. Phys. **1**, 367–380 (2019) [518]. Copyright 2019 Springer Nature. Reprinted with permission from A. Schori *et al.*, Phys. Rev. A **97**, 063804 (2018) [517]. Copyright 2018 American Physical Society. (**e**) Spontaneous parametric down-conversion via nonlinear Bragg diffraction in diamond crystals generates entangled photon pairs from a high-flux X-ray beam. Reprinted with permission from J. C. Goodrich *et al.*, arXiv:2310.13078 (2023) [239]. Licensed by under the arXiv Nonexclusive Distribution License.

## 7.3 X-ray Focusing with Nanomaterials

X-ray focusing [419] with nanophotonics involves nanostructured materials[521,522] to manipulate X-rays, significantly enhancing the focusing efficiency and resolution. For example, one technique that benefits from nanoengineering is kinoform X-ray optics[98], which has the potential to focus X-rays with nearly 100% efficiency. The optics leverage the phase-shifting properties of kinoform structures to achieve high-resolution and high-efficiency focusing of X-rays. Recent studies have optimized the fabrication process using 3D printing technologies to create precise and intricate kinoform designs [523,524], enhancing the focusing capabilities of X-ray optics (Fig. 30a) [98]. Another significant advancement in the generation of X-ray nanobeams involves using reflective optics combined with speckle interferometry. Fig. 30b shows that through multilayer focusing mirrors, the width of the laser beam in a free-electron laser was reduced to a diameter of 6 nanometres [99].

The interactions of free electrons with van der Waals heterostructures offer more approaches for X-ray focusing directly from the source. In the theoretical proposal, free electrons generate focused X-ray beams by passing through the van der Waals heterostructure [33]. By engineering the van der Waals heterostructure through chirping the interlayer spacing, and controlling the electron energy, the X-rays can be focused. This proposal could help develop compact X-ray focusing devices with high spatial resolution and efficiency.

Achromatic X-ray lens has been developed by combining a focusing diffractive Fresnel zone plate (FZP) with a defocusing refractive lens. This combination leverages the differing dispersive properties of the FZP and refractive lens. Ref. [100] shows an experimental example with achromatic focusing over a broad energy range from 5.8 keV to 7.3 keV (Fig. 30c). The FZP focuses X-rays of varying energies at different focal planes, while the refractive lens counteracts this dispersion by defocusing the X-rays to compensate for chromatic effects. The result is a high-quality, consistent focal point across a wide range of energies. Experiments using scanning transmission X-ray microscopy and ptychography have demonstrated the achromat's capability to maintain image quality without the need for focal adjustments [100].

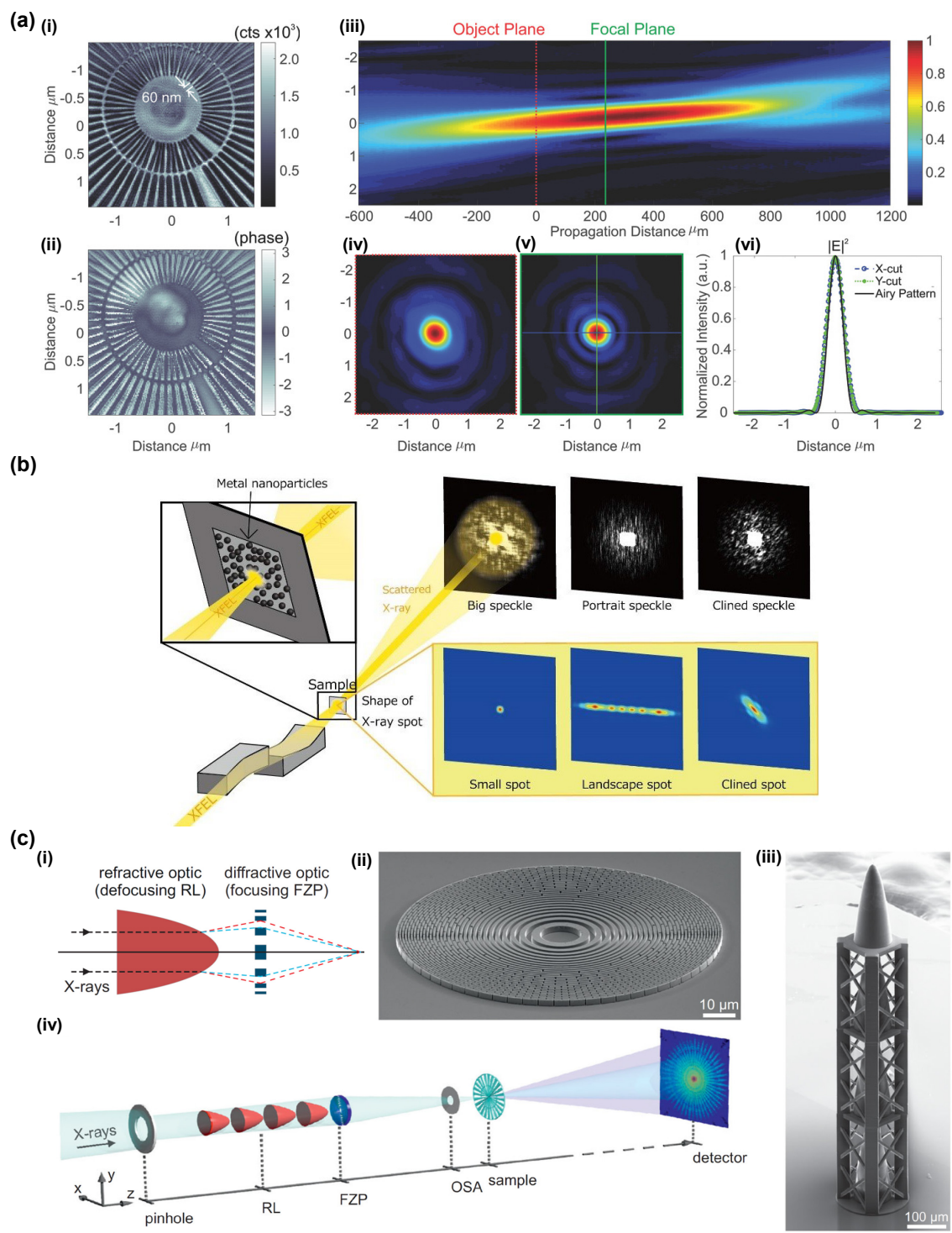

**Fig. 30 X-ray focusing with nanostructures**. (**a**) 3D nanoprinted kinoform structures significantly improve X-ray focusing efficiency from 0.9 to 1.5 keV, leveraging the phase-shifting properties of kinoform structures. Reprinted with permission from U.T. Sanli *et al.*, Adv. Mater. **30**, 1802503 (2018) [98]. Copyright 2018 Wiley-VCH. (**b**) Reflective optics combined with speckle interferometry allow for the generation of X-ray nanobeams with nanometer-scale diameters, enhancing the utility of X-ray free-electron lasers for high-resolution imaging and analysis. Reprinted with permission from T. Inoue *et al.*, J. Synchrotron Rad. **27**, 883–889 (2020) [99]. Licensed under a Creative Commons Attribution 4.0

International (CC BY 4.0) license. (**c**) An achromatic X-ray lens, combining a focusing diffractive Fresnel zone plate with a defocusing refractive lens, addresses chromatic aberration in X-ray optics, enabling consistent high-quality imaging over a broad energy range without the need for focal adjustments. Reprinted with permission from A. Kubec *et al.*, Nat. Commun. **13**, 1305 (2022) [100]. Licensed by a Creative Commons Attribution 4.0 International license.

### *7.4 Nanophotonic X-ray Scintillators*

A significant advancement in scintillator technology is the development of water-dispersible X-ray scintillators. Ref. [525] introduced $Tb_3^+$-doped $Na_5Lu_9F_{32}$ nanocrystals anchored on halloysite nanotubes [526,527]. Fig. 31a shows the innovative design that yields materials with exceptional water dispersibility and compatibility with polymer matrices, making them ideal for flexible substrates [528,529]. The nanocomposite demonstrates a remarkable steady-state X-ray light yield of 15,800 photons $MeV^{-1}$, indicative of high sensitivity and efficiency. The choice of $Tb_3^+$ doping enhances the scintillation properties due to its favorable energy levels for efficient energy transfer processes. The nanocrystals' integration with halloysite nanotubes provides structural stability and dispersibility, which are crucial for maintaining performance in diverse environments. The enhanced energy transfer between the doped nanocrystals and the host matrix results in efficient radiative recombination, making these scintillators highly effective for advanced X-ray imaging and radiation exposure monitoring technologies.

Nanomaterials have revolutionized scintillation technologies due to their tunable optical properties and high surface area-to-volume ratios. The implementation of nanomaterials in scintillators is typical for a computed tomography detector. These detectors typically consist of three layers: a scintillator that absorbs X-rays and emits visible light, photodiode arrays that convert this light into an electric current, and a substrate that provides structural support and facilitates signal processing, as shown in Fig. 31b. Ref. [530] provides a comprehensive review of advancements in all-inorganic perovskite nanocrystals, emphasizing their high light yield, fast decay times, and robust stability. These nanocrystals can incorporate various halide ions (Cl, Br, I) [531,532] to enable tunable emission wavelengths [532–534] and are suitable for high temporal resolution applications [533,535,536]. Their solution-processable synthesis allows integration into flexible films, facilitating scalable production of high-performance scintillators with tailored bandgaps and high quantum efficiency under X-ray excitation. Additionally, Ref. [533] demonstrated strong X-ray absorption and intense radioluminescence at visible wavelengths in perovskite nanocrystals, suitable for low-dose X-ray radiography (Fig. 31c). The defect-tolerant electronic properties enhance radioluminescence efficiency, making them ideal for high-resolution X-ray imaging.

By manipulating light at the nanometer scale, nanophotonics can significantly enhance scintillator performance. A comprehensive framework for scintillation in nanophotonics is presented in Ref. [104], showing how nanostructured materials can improve light extraction efficiency [537–541,103,542,106,102] and tailor emission spectra [80,543]. Nanoengineering the scintillator materials can enhance light-matter interactions and increase the probability of photon emission. Techniques such as plasmonic enhancement, where metallic nanostructures concentrate electromagnetic fields, can significantly boost the local density of states, enhancing radiative decay rates. Additionally, photonic crystals with a periodic dielectric structure can be designed to create photonic bandgaps, which control the flow of photons and improve light extraction efficiency [537–541,103,542,106,102] (Fig. 31d). This precise control at the nanoscale opens new possibilities for optimizing the performance of scintillators, making them more efficient and effective for high-performance applications. The application of photonic bandgap materials to control spontaneous emission rates and to channel emitted light more effectively is particularly noteworthy for improving scintillation efficiency.

Free electrons interacting with nanomaterials have also been harnessed to develop ultrafast and highly efficient scintillators. The fundamental principle is the nanoplasmonic Purcell effect, which enhances the radiative decay rate of excited states in scintillators by embedding them in plasmonic nanostructured environments [103,106,544]. The Purcell effect indicates the modification of the spontaneous emission rate of emitters placed in a resonant cavity or near plasmonic nanostructures [545]. Ref. [107] shows that by embedding scintillating materials within plasmonic structures, such as gold or silver nanoparticles (Fig. 31e), the local electromagnetic field is enhanced, leading to increased radiative recombination rates. This results in brighter and faster scintillation, which is particularly beneficial for applications requiring high temporal resolution. The research demonstrated that the use of plasmonic nanostructures could reduce decay times to the order of picoseconds, enhancing the temporal resolution and sensitivity of the scintillators. The implementation of nanoplasmonic structures to enhance light-matter interactions in scintillating materials paves the way for developing next-generation, ultrafast scintillation devices [546–557].

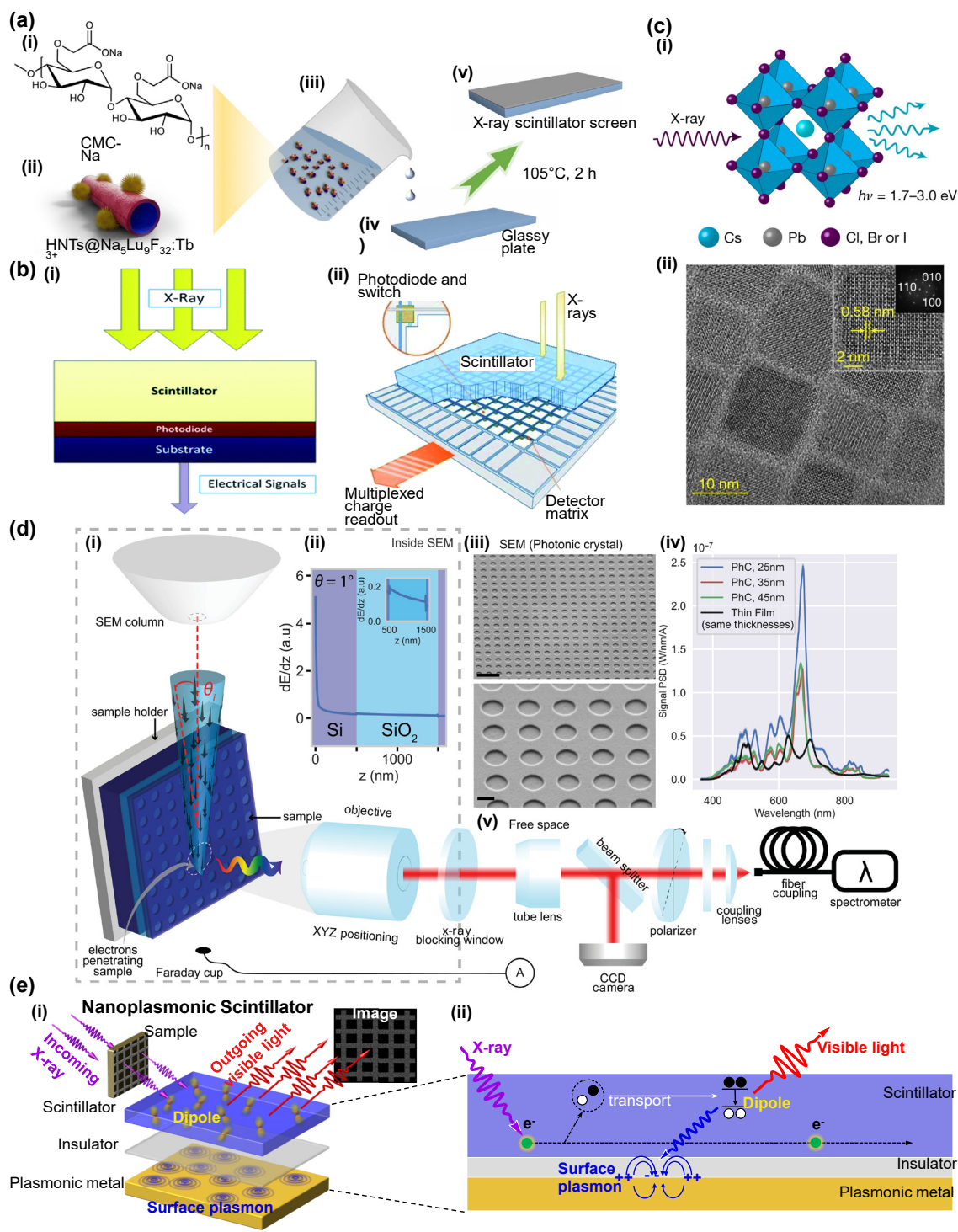

**Fig. 31 Advancements in X-ray Scintillators.** (**a**) Water-dispersible X-ray scintillators using $Tb^{3+}$-doped $Na_5Lu_9F_{32}$ nanocrystals anchored on halloysite nanotubes, demonstrating high sensitivity and efficiency. Reprinted with permission from H. Zhang *et al.*, Nat. Commun. **15**, 2055 (2024) [525]. Licensed by a Creative Commons Attribution 4.0 International license. (**b**) Schematic of a standard X-ray computed tomography detector. Reprinted with permission from M. Spahn *et al.*, NIM-A **731**, 57–63 (2013). Copyright 2013 Elsevier [536]. Copyright 2013 Elsevier. Reprinted with permission from E. Shefer *et al.*, Current Radiology Reports **1**, 76–91 (2013) [558]. Copyright 2013 Springer Nature. (**c**) Full-colour radioluminescence from $CsPbBr_3$ nanocrystal scintillators shown via X-ray-induced luminescence and transmission

electron microscope imaging. Reprinted with permission from Q. Chen *et al.*, Nature **561**, 88–93 (2018) [533]. Copyright 2018 Springer Nature. (**d**) Nanophotonics significantly enhances scintillator performance by improving light extraction efficiency and tailoring emission spectra through nanostructuring. Reprinted with permission from C. Roques-Carmes *et al.*, Science **375**, eabm9293 (2022) [104]. Copyright 2022 AAAS. (**e**) Nanoplasmonic Purcell effect significantly enhances the radiative decay rate in scintillators, leading to brighter and faster scintillation. Reprinted with permission from W. Ye *et al.*, Adv. Mater. **36**, 2309410 (2024) [107]. Copyright 2024 Wiley-VCH.

## 7.5 Quantum Information

### 7.5.1 Towards Superior Quantum Probes Based on Free Electrons

Quantum technologies such as quantum sensing, computing and communication rely on the ability to address and transfer quantum information between different systems. The most common building blocks for stationary qubits rely on bound electron systems such as atoms, solid state and artificial atoms like superconducting qubits. These qubits can be used as quantum probes, but they can also be probed by another type of quantum entity, a flying qubit, used for long-distance communication of quantum information. This role has been traditionally played by propagating photonic qubits. While technologies based on bound-electron qubits (in their various forms) are well established, they still suffer from fundamental limitations: having fixed transition frequencies, naturally-occurring emitters like atoms or defect centers are not spectrally tuneable; solid-state qubits such as quantum dots suffer from inhomogeneous broadenings; and artificial atoms such as superconducting qubits operate at cryogenic temperatures. Flying photon qubits are indeed excellent carriers of quantum information over long distances. However, as quantum probes, they are limited by the diffraction limit and by spectral mode matching (e.g. when the probed quantum system is integrated with a cavity, the photon wavepacket must be as temporally long as the cavity lifetime). These limitations make photons low-resolution and slow quantum probes, which can hinder the addressing of dense emitter arrays and limit operation rates.

The limitations of existing building blocks of quantum technologies motivate the pursuit of new qubit hardware. In this respect, this section will briefly outline the anticipated impact of free-electron quantum optics for overcoming these limitations using free-electron quantum probes. Free electrons can serve as new types of quantum emitters, quantum sensors, as well as flying ancillary qubits. They enjoy a broad spectral tunability, through the change of the electron velocity and design of the nanostructure, which dictate the frequency of the emitted photon through the phase-matching conditions. In addition, they can be focused well below the optical diffraction limit, to sub-nm scales, and operate at ultrafast rates, at the femtosecond and even attosecond scales. Free electrons can further enjoy high spatial and spectral coherence, allowing them to probe coherent quantum phenomena and resolve energy spectra down to the sub-meV level. Further, free electrons can operate at room temperature, and even display inherent nonlinearities and strong coupling at the single-electron and single-photon levels.

### 7.5.2   *Free Electrons as Carriers of Quantum Information*

The coherent modulation of free electrons sprouted ideas motivated by the free electrons' ability to carry quantum information as qubits. The first proposal of a free-electron qubit [559] employed the parity of the different electron energy levels $E_0 + n\hbar\omega$, $n$ being either even or odd. This property can be captured by the eigenstates of the hopping $b$ operators, the electron comb states [45,267] given as $|\text{comb}(\phi)\rangle = \sum_k e^{ik\phi}|E + \hbar\omega k\rangle$ with $b|\text{comb}(\phi)\rangle = e^{i\phi}|\text{comb}(\phi)\rangle$. These are non-normalizable states that approximately correspond to the attosecond electron pulse trains in the time domain [9,262,267]. The 2D Hilbert space of a free-electron qubit can in theory be spanned by even and odd combs [54,57], $|e\rangle = \sum_k |E + 2\hbar\omega k\rangle$, $|o\rangle = \sum_k |E + \hbar\omega(2k + 1)\rangle$, which satisfy $|e\rangle = b|o\rangle, |o\rangle = b|e\rangle$. In practice, this can be realized by a pair of normalizable states, where each of the states contains only odd or even electron energy harmonics. It was then shown that this subspace is closed under the semiclassical photon-induced near-field electron microscopy (PINEM) operations $S = \exp(gb - g^*b^\dagger)$ as well as free-space propagation (FSP), where the electron wavefunction accumulates a phase that is quadratic in the harmonic number, owing to free-space dispersion. The right combination of FSP and PINEM allows for a set of single-qubit gates. Another way to define the free-electron qubit in the energy domain is to use pairs of time and energy bins, as was recently demonstrated experimentally [560] by temporally shaping of the electron wavefunction using a laser in a transmission electron microscope.

Other manifestations of free-electron qubits (or most generally, qudits) harness the spatial degree of freedom. Path qubits can be realized using biprisms [561,562], electron beam splitters [563] and even encoding in the orbital angular momentum of free electrons [564]. Manipulation of electron path qubits entails progress in coherent electron beam splitters, guiders, mirrors, and interferometers [60,276,563,565–568]. In this respect, recent theoretical proposals have considered free-electron path and energy qubits as alternative ancilla qubits for quantum gates in cavity QED [57,67] .

### 7.5.3 *Quantum sensing and measurement*

Quantum technologies such as quantum sensing with free electrons harness the interaction, coherence and entanglement between free electron, bound electron, and photonic systems. Such advancements complement the existing platforms of electron microscopy that rely on the temporal and spatial coherence of free electron wavefunctions, such as PINEM and electron holography. PINEM is well known to allow for imaging the amplitude of optical modes in nanophotonics [19,20], polariton wavepacket dynamics [346], and attosecond phenomena [353], while electron holography has been used to image coherence lengths [569] and symmetries [280] of optical excitations, as well as magnetic phenomena such as circular dichroism [124].

The free-electron quantum probe has the potential to take these techniques a step further. Recently, PINEM was shown to permit the quantum coherent amplification and sensing of both the amplitude and phase of classical fields [355–357] and even interfere with the spontaneously emitted light from the same electron [302,570]. Similar advancements in quantum electron holography offer interaction-free measurements [561] and coherent Lorentz microscopy of optical fields [571]. Beyond sensing of classical fields, free electron quantum probes have been predicted to enable quantum optical detection of photon statistics [22,40,270], quantum optical coherence [46,302], enhanced sensing of light-matter coupling [301,359,572–574], measurement of quantum correlations of matter [304], and even inducing quantum correlations across different platforms[48,53], all made possible with ultrafast time scales and deep subwavelength resolution.

This promising research direction was stimulated by the work of Gover and Yariv [118], who considered firstly a semiclassical model of a modulated free electron beam. When the modulation frequency is set equal to the transition frequency of a two-level bound electron system the electrons are interacting with, it was shown that a resonant interaction may occur [118]. Following that work, several other studies, including extensions to the fully quantum regime [47,48,358,575,576], showed that free electrons could be used to manipulate, control, and read out bound electron qubits [47,359,576]. The free-electron probe is particularly appealing for such a task, since its high spatial and temporal resolution can increase the addressing capabilities of individual qubits. Further exciting possibilities were theoretically investigated, such as inducing entanglement between distant bound electron systems using free electrons [48], driving of large-scale quantum correlations leading to superradiance [304,306], quantum interference between light emission and atomic excitations [301,359,572,574] with applications for quantum sensing of strong coupling [359].

### 7.5.4 Quantum light sources and quantum gates

Free electrons can further function as alternative emitters for quantum optics, specifically when integrated into cavity QED platforms. This is particularly exciting since some of these quantum states, for example, cat, GKP, and large-number Fock states, are hard to generate in the optical (or other) domains with current cavity QED platforms. Interestingly, free electrons can also be used to coherently control existing cavity QED systems and enable multi-qubit quantum gates for quantum information processing [57,67].

Quantum light states such as Fock states [18,36,37,45,50], cat and GKP states [54,55,57], and squeezed states[577] can be generated by shaped free electrons interacting with optical cavities. For higher electron energies, protocols for quantum light generation involve a probabilistic process where the successful generation of the state is heralded by the detection of the free electron energy [45,49,50,56]. For lower electron energies, it is predicted that strong quantum recoil effects can introduce an inherent nonlinearity to the emission process, which transforms the free electron emitter into a saturable emitter [38,52,58,60,64,65,273]. In the limit of a strong nonlinearity, the free electron turns into an effective two-level system, offering a completely new platform for the Jaynes-Cummings model [58,60,65,273]. Quantum gates controlled by free electrons were proposed employing the strong coupling in cavity QED which enables a polariton blockade mechanism [67]. The free electron path qubits act as ancilla qubits mediating single and two-qubit universal gate sets for the cavity polariton qubits. Similar ideas were proposed for quantum gates controlling photonic GKP qubits [57] that are promising for quantum error correction in continuous variable quantum computation.

## 8   Conclusion

In this review, we have explored recent advancements in exploiting quantum nanophotonics to investigate and control free-electron-light interactions. Nanostructures not only enhance these interactions but also highlight the significance of quantum features of photons and electrons under less stringent conditions than previously studied. For instance, the quantum recoil phenomenon has been experimentally demonstrated with semi-relativistic electrons interacting with van der Waals materials in the soft X-ray regime. Similar phenomena in Compton scattering require ultra-strong fields, relativistic electrons, and high-energy photons. The experiment leverages the unique properties of nanomaterials to achieve comparable results under more accessible conditions. Most importantly, it demonstrates the significance of quantum mechanics in the investigation of free electrons interacting with nanomaterials.

Moreover, the wave nature of electrons is evident in their interactions with nanomaterials, providing an additional degree of freedom to control quantum electrodynamic processes through electron wavefunction engineering. By manipulating engineered electron wavefunctions, free-electron radiation can be enhanced and controlled, and quantum light can be generated through interactions between these tailored electrons and nanomaterials. The current advancements have been limited to low-energy photons, but the techniques could be applied to high-energy photons, such as extreme ultraviolet (XUV) photons and X-rays.

The advancements in nanophotonics have facilitated the development of novel X-ray optics through innovative nanostructural and atomic designs. Notably, compact and coherent X-ray sources have emerged from the interactions of free electrons with nanomaterials, enabling the creation of customized X-ray sources with specific spatial and spectral properties directly at the source, thereby eliminating the need for additional optical components.

Furthermore, nanophotonics offers a promising platform for generating and manipulating ultrashort electron pulses, unveiling novel aspects of quantum electrodynamics in the ultrashort pulse regime. For instance, attosecond electron pulses can be produced through the interaction of femtosecond light with tip emitters, where the nanoscale apex of the tip facilitates the excitation of ultrafast electrons within sub-optical cycles. Recent theoretical and experimental demonstrations of heralded electron sources have opened new avenues for exploring quantum light generation using these electrons in the X-ray regime. Additionally, high-harmonic generation benefits significantly from nanomaterials; engineered nanostructures and structured surfaces could enhance high-harmonic generation yield, reduce intensity thresholds, and enable the focusing of emitted X-ray and extreme ultraviolet rays to small spots.

With the advances in nanophotonics, X-ray technologies are increasingly finding significant applications across various fields. Nanophotonics facilitates the development of advanced optical elements that enhance the utility of X-rays, particularly in applications like coherent nanoscale imaging. Furthermore, leveraging the quantum properties of X-rays has the potential to significantly improve measurement techniques, paving the way for innovations in sensing and quantum imaging.

**Author Contributions**

X. S. contributed to Chapters 1-4 and 8. W. W. L. contributed to Sections 7.1-7.4 and organized the manuscript. A. K. contributed to Section 2.3 and Section 7.5. L. M. L. and T. S. contributed to Chapter 6. A.G. contributed Chapter 5. L. W. W. W. and L. J. W contributed to Chapter 1. All authors participated in the discussion and review of the paper. L.J.W. conceived the idea and supervised the review.

**Acknowledgment:**

This project was partially supported by the Ministry of Education, Singapore, under its AcRF Tier 2 program (award no. MOE-T2EP50222-0012) and the National Research Foundation (Project ID NRF2020-NRF-ISF004-3525). A.K. is supported by the Urbanek-Chodorow postdoctoral fellowship by the Department of Applied Physics at Stanford University, the Zuckerman STEM leadership postdoctoral program, the VATAT-Quantum fellowship by the Israel Council for Higher Education, and the Viterbi fellowship of the Technion – Israel Institute of Technology. L.M.L. acknowledges funding by the Max Planck School of Photonics and by the DFG through Project No. 432680300 (SFB 1456-C03). X.S. is supported in part by a fellowship of the Israel Council for Higher Education and by the Technion's Helen Diller Quantum Center.